\documentclass[prd,twocolumn,notitlepage,superscriptaddress,nofootinbib]{revtex4-1}

\usepackage{amsmath}
\usepackage{amsfonts}
\usepackage{graphicx}
\usepackage[utf8]{inputenc}
\usepackage{slashed}
\usepackage{multirow}
\usepackage[normalem]{ulem}

\usepackage{hyperref}
\hypersetup{colorlinks, linkcolor = [rgb]{0,0.0,0.75}, citecolor = [rgb]{0,0.0,0.75}, urlcolor = [rgb]{0,0.0,0.75}}

\makeatletter
\g@addto@macro\bfseries{\boldmath}
\makeatother

\newcommand{\MSb}{\overline{\text{MS}}}
\newcommand{\LambdaQCD}{\Lambda_{\textrm{QCD}}}
\newcommand{\numax}{\nu_{\textrm{max}}}
\newcommand{\zmax}{z_{\textrm{max}}}

\begin{document}

\title{Continuum limit of parton distribution functions \\ from the pseudo-distribution approach on the lattice}

\author{Manjunath Bhat}
\affiliation{Faculty of Physics, Adam Mickiewicz University, ul.\ Uniwersytetu Pozna\'nskiego 2, 61-614 Poznań, Poland}
\author{Wojciech Chomicki}
\affiliation{Faculty of Physics, Adam Mickiewicz University, ul.\ Uniwersytetu Pozna\'nskiego 2, 61-614 Poznań, Poland}
\author{Krzysztof~Cichy}
\affiliation{Faculty of Physics, Adam Mickiewicz University, ul.\ Uniwersytetu Pozna\'nskiego 2, 61-614 Poznań, Poland}
\author{Martha~Constantinou}
\affiliation{Temple University, 1925 N.\ 12th Street, Philadelphia, PA 19122-1801, USA}
\author{Jeremy R.~Green}
\affiliation{School of Mathematics and Hamilton Mathematics Institute, Trinity College, Dublin 2, Ireland}
\author{Aurora~Scapellato}
\affiliation{Temple University, 1925 N.\ 12th Street, Philadelphia, PA 19122-1801, USA}

\date{\today}

\begin{abstract}
Precise quantification of the structure of nucleons is one of the crucial aims of hadronic physics for the coming years.
The expected progress related to ongoing and planned experiments should be accompanied by calculations of partonic distributions from lattice QCD.
While key insights from the lattice are expected to come for distributions that are difficult to access experimentally, it is important that lattice QCD can reproduce the well-known unpolarized parton distribution functions (PDFs) with full control over systematic uncertainties.
One of the novel methods for accessing the partonic $x$-dependence is the pseudo-distribution approach, which employs matrix elements of a spatially-extended nonlocal Wilson-line operator of length  $z$. In this paper, we address the issue of discretization effects, related to the necessarily nonzero value of the lattice spacing $a$,  which start at first order in $a$ as a result of the nonlocal operator.
We use twisted mass fermions simulated at three values of the lattice spacing, at a pion mass of 370 MeV, and extract the continuum limit of isovector unpolarized PDFs.
We also test, for the first time in the pseudo-distribution approach, the effects of the recently derived two-loop matching. Finally, we address the issue of the reliability of the extraction with respect to the maximal value of $z$.
\end{abstract}

\maketitle

\section{Introduction}

\noindent Existence of the internal partonic structure of nucleons was discovered over fifty years ago in groundbreaking experiments at Stanford.
Since then, data from deep inelastic scattering (DIS) and other high-energy processes were used to probe this structure extensively.
Yet many aspects of the rich internal dynamics of the nucleon and other hadrons, such as polarized and multidimensional observables, remain elusive.
This continues to be a very active area of research, with new planned or ongoing experiments, designed specifically to probe different aspects of hadronic structure.
Significant amounts of new experimental data are expected from e.g.\ the COMPASS++/AMBER experiment at CERN \cite{Adams:2018pwt}, the 12 GeV upgrade of Jefferson Lab's \mbox{CEBAF} accelerator \cite{Dudek:2012vr,Burkert:2018nvj} and the recently approved Electron-Ion Collider at Brookhaven National Laboratory \cite{NAP25171,AbdulKhalek:2021gbh}.
The experimental effort should be supplemented by theoretical developments in multiple areas: phenomenological models, perturbative QCD, and nonperturbative first-principles calculations.
The appropriate tool for the latter is lattice QCD (LQCD), a nonperturbative formulation of QCD allowing for quantitative predictions from first principles.

The standard theoretical description of quantum field theories involves perturbation theory.
However, in the case of QCD, the low-energy properties cannot be accounted for in perturbation theory.
A common way of overcoming this restriction is the factorization framework in which cross sections are separated into short- and long-distance parts:  the former are treated perturbatively, while the latter are parametrized in terms of partonic distributions.
Such distributions can be evaluated by performing global fits to experimental data.
Obviously, this is viable only under the condition of sufficient abundance of such experimental input.
The largest body of data concerns the case of nucleon's unpolarized parton distribution functions (PDFs), for which thousands of measurements impose stringent constraints over nearly the whole $x$-dependence.
Additionally, the fits are performed by several independent groups (see, e.g., Refs.~\cite{Martin:2009iq,Accardi:2016qay,Alekhin:2017kpj,Ball:2017nwa}), allowing for comparison and assessment of the influence of some necessarily introduced assumptions.

In contrast, once one considers polarized observables, already the case of longitudinal polarization of the parton and the nucleon (helicity PDFs) is significantly less constrained (see, e.g., Refs.~\cite{deFlorian:2009vb,Nocera:2014gqa,Ethier:2017zbq}).
In the case of transversely polarized quarks in a transversely polarized nucleon, experimental data are hardly enough to get any quantitative knowledge of transversity PDFs \cite{Lin:2017stx,Radici:2018iag}, although including more experimental data improves the extraction precision \cite{Cammarota:2020qcw}.
The extraction can also be augmented by an additional input, e.g.\ the tensor charge calculated on the lattice \cite{Lin:2017stx}.
Even less can be inferred for the three-dimensional nucleon structure, which is quantified in terms of generalized parton distributions (GPDs) \cite{Ji:1996ek,Radyushkin:1996nd,Diehl:2003ny,Ji:2004gf,Belitsky:2005qn} and transverse-momentum dependent PDFs (TMD PDFs or TMDs) \cite{Collins:1981uk,Collins:1981uw,Boer:2011fh,Accardi:2012qut,Angeles-Martinez:2015sea}.
Access to them is not possible in standard DIS: GPDs require exclusive processes such as deeply virtual Compton scattering (DVCS)~\cite{Ji:1996nm} and deeply virtual meson production~\cite{Favart:2015umi}, while TMDs require processes such as Drell-Yan~\cite{Qiu:2000hf,DAlesio:2014mrz,Bacchetta:2017gcc} and semi-inclusive DIS~\cite{Boer:2011fh,Accardi:2012qut}.
The amount of data available from such experiments is much smaller, leading to insufficient constraining power to fully quantify the three-dimensional structure.
No global fits have so far been performed in the whole DVCS kinematic domain to extract GPDs, but first attempts at extracting GPD-related observables are under way \cite{Kumericki:2016ehc,dHose:2016mda,Moutarde:2019tqa}.
Similarly, while there is insufficient data for full mapping of TMDs, first fits are being performed \cite{Bacchetta:2017gcc,Bertone:2019nxa,Hautmann:2020cyp}, particularly for the unpolarized case. 
Although significant progress is expected from the above mentioned new experimental setups, it would be invaluable to complement it with first-principle lattice investigations.

Being a genuinely nonperturbative approach, LQCD can, in principle, fill the gap coming from the lack of access to low-energy properties in perturbation theory.
A potentially crucial restriction of LQCD for calculating partonic distributions is its Euclidean metric, prohibiting direct access to them.
However, indirect access is still possible and different approaches to it are intensely investigated in the last years.
This present surge of studies was initiated by seminal papers of Ji \cite{Ji:2013dva,Ji:2014gla}.
He proposed that while light-front correlations are inaccessible in Euclidean spacetime, lattice-calculable spatial correlations in a boosted hadron can be used to define alternative observables that can be appropriately ``translated'' to the desired distributions.
There are several lattice observables that are well-suited to extract partonic distributions from them.
Apart from being calculable on the lattice, these observables need to have the same infrared structure as their light-front counterparts and be renormalizable.
While Ji's proposal of quasi-distributions sparked intense studies, earlier approaches also existed \cite{Liu:1993cv,Aglietti:1998ur,Detmold:2005gg,Braun:2007wv}, some of them revived in the last years, and new ones were put forward \cite{Chambers:2017dov,Radyushkin:2017cyf,Radyushkin:2019mye,Ma:2014jla,Ma:2017pxb}.
Similarly to the standard phenomenological approach, all these methods make use of factorization at the stage of relating the lattice data to light-front distributions.
The different lattice approaches were all subject to broad theoretical and practical studes, see, e.g., Refs.~\cite{Lin:2014zya,Alexandrou:2015rja,Chen:2016utp,Alexandrou:2016jqi,Chambers:2017dov,Alexandrou:2017huk,Orginos:2017kos,Ishikawa:2017faj,Ji:2017oey,Radyushkin:2018cvn,Alexandrou:2018pbm,Chen:2018fwa,Alexandrou:2018eet,Liu:2018uuj,Karpie:2018zaz,Bhattacharya:2018zxi,Zhang:2018diq,Li:2018tpe,Braun:2018brg,Sufian:2019bol,Karpie:2019eiq,Liu:2019urm,Alexandrou:2019lfo,Wang:2019tgg,Chen:2019lcm,Izubuchi:2019lyk,Cichy:2019ebf,Joo:2019jct,Radyushkin:2019owq,Joo:2019bzr,Son:2019ghf,Ma:2019agv,Green:2020xco,Chai:2020nxw,Lin:2020ssv,Braun:2020ymy,Joo:2020spy,Bhat:2020ktg,Bhattacharya:2020xlt,Zhang:2020gaj,Bhattacharya:2020jfj, Bhattacharya:2021boh,Zhang:2020dbb,Fan:2020cpa,Alexandrou:2020zbe,Alexandrou:2020uyt,Bringewatt:2020ixn,Liu:2020rqi,DelDebbio:2020rgv,Alexandrou:2020qtt,Liu:2020krc,LatticePartonCollaborationLPC:2021xdx,Detmold:2021uru,Fan:2021bcr,Karpie:2021pap,Karthik:2021sbj,Alexandrou:2021oih,Li:2021wvl,Bhattacharya:2021moj,Egerer:2021ymv,HadStruc:2021wmh,Shanahan:2021tst,Alexandrou:2021bbo,Detmold:2021qln,CSSMQCDSFUKQCD:2021lkf,HadStruc:2021qdf,Balitsky:2021qsr,Chirilli:2021euj,Balitsky:2021cwr,Gao:2021dbh,Bhattacharya:2021oyr,Xu:2022krn,LPC:2022ibr,Chou:2022drv} and the reviews \cite{Cichy:2018mum,Ji:2020ect,Constantinou:2020pek,Cichy:2021lih,Cichy:2021ewm}.

Quasi-distributions \cite{Ji:2013dva} are defined as Fourier transforms of Euclidean matrix elements (MEs) of boosted hadrons with an operator insertion of a spatially separated quark-antiquark pair connected by a Wilson line and with a Dirac structure that determines the type of the distribution. 
Exactly the same MEs can be used to define another generalization of light-front distributions, dubbed pseudo-distributions \cite{Radyushkin:2016hsy,Radyushkin:2017cyf,Radyushkin:2017lvu,Radyushkin:2017sfi,Radyushkin:2018cvn,Radyushkin:2018nbf,Radyushkin:2019owq,Radyushkin:2019mye}.
With $z$ denoting the vector describing the position of the Wilson line and $p$ being the hadron's 4-momentum,
the difference between the two approaches consists in the Fourier transform being either in $|z|$ at fixed $p_z$ (quasi) or in $p\cdot z$ at fixed $|z|$ (pseudo).
The Lorentz-invariant product $p\cdot z$ is often called the ``Ioffe time''.
However, the key difference comes at the stage of factorization, performed in momentum space (quasi) or in coordinate space (pseudo).
In practice, in the quasi-distribution method, renormalized MEs are first subjected to reconstruction of the $x$-dependence, i.e.\ coordinate-space MEs are an input to a procedure that brings them to momentum space of Bjorken-$x$ fractions.
The ensuing functions are called quasi-distributions and they are then subjected to a factorization-based matching procedure, that, in turn, ``translates'' the spatial correlations that they express to light-front correlations that define physical distributions like PDFs and GPDs.
In the pseudo-distribution approach, renormalized MEs, called pseudo-ITDs (Ioffe time distributions), are usually at this stage subjected to a matching procedure, leading to light-front ITDs.
These functions are then input to the $x$-dependence reconstruction.
A Fourier transform of pseudo-ITDs can also be done prior to matching, leading to the so-called pseudo-PDFs, however the factorization is still in coordinate space, at short distances.  
These differences between the two approaches have far-reaching consequences and imply possibly very different systematic effects, although in the end, the physical distributions from both methods should coincide.

In this paper, we investigate discretization effects, which are particularly important for nonlocal operators.
We use the pseudo-distribution approach to extract unpolarized isovector PDFs of the nucleon in the continuum limit.
We use three ensembles of twisted mass gauge field configurations at lattice spacings $a\approx0.064,\,0.082$ and $0.093$ fm, at a non-physical pion mass of around 370 MeV.
This setup has been used to determine unpolarized and helicity PDFs within the quasi-distribution approach \cite{Alexandrou:2020qtt}, at a fixed nucleon boost of 1.8 GeV.
Here, we supplement the lattice data of Ref.~\cite{Alexandrou:2020qtt} with additional three to four nucleon momenta to cover the full range of Ioffe times required in the pseudo-PDF method and to utilize the standard ratio scheme renormalization, where the divergences are canceled by taking a ratio with respect to zero-boost MEs.

The outline of the paper is as follows.
In Sec.~\ref{sec:theory}, we recall some theoretical principles of pseudo-PDFs along with their practical aspects.
Our lattice setup is outlined in Sec.~\ref{sec:lattice}.
The results of our study are described in Sec.~\ref{sec:results}.
Finally, Sec.~\ref{sec:summary} concludes and discusses future prospects.

\section{Theoretical setup and analysis techniques}
\label{sec:theory}
We refer the reader to the review of Ref.~\cite{Radyushkin:2019owq} for an extensive discussion on the theoretical principles and properties of pseudo-distributions and summarize here only the main aspects.

\subsection{Euclidean matrix elements}
Euclidean correlations that underlie quasi- and pseudo-PDFs of the nucleon are described by bare MEs, $\mathcal{M}_\Gamma(P,z)$, of the form
\begin{equation}
\label{eq:bare}
  \left\langle P,s' \middle| \mathcal{O}_\Gamma(0,z) \middle| P,s \right\rangle
  = \mathcal{M}_\Gamma(P,z) \bar u(P,s') \Gamma u(P,s),
\end{equation}
where $u(P,s)$ is a spinor corresponding to a Euclidean 4-momentum $P$ and spin $s$.
The bare non-local operator is
\begin{equation}
  \mathcal{O}_\Gamma(x,z) \equiv \bar\psi(x) \Gamma \tau_3 W(x,x+z) \psi(x+z),
\end{equation}
with $x$ and $z$ denoting position 4-vectors, the latter being the displacement between light-quark doublets $\psi$ and $\bar{\psi}$, which are connected by a Wilson line $W$ that maintains gauge invariance.\footnote{We employ here the opposite convention to the one commonly used in the quasi-PDF literature, wherein $\bar{\psi}$ rather than $\psi$ is displaced by $z$. This implies the opposite sign of the imaginary part of MEs.}
The Pauli matrix $\tau_3$ corresponds to the isovector flavor combination $u-d$ for which all results of this work are obtained.
In the following, we choose $z=(0,0,0,z_3)$ and $P=(P_0,0,0,P_3)$ and henceforth, $z$ will denote the length of the Wilson line.
The Dirac structure $\Gamma$ determines the type of the accessed PDF and here, we choose $\Gamma=\gamma_0$ to consider unpolarized PDFs\footnote{The other choice of $\Gamma=\gamma_3$ leads to slower convergence \cite{Radyushkin:2016hsy} and to mixing with the twist-3 scalar operator for non-chiral lattice fermions \cite{Constantinou:2017sej,Green:2017xeu,Chen:2017mie}. }
and drop the index $\Gamma$ in $\mathcal{M}$ below.

The above MEs, $\mathcal{M}(P,z)$, can be viewed as functions of the Wilson line length $z$ and the Ioffe time $\nu\equiv P_3z$, giving rise to the notion of ITDs mentioned above.
From now on, we will, thus, use the notation $\mathcal{M}(\nu,z)$ to refer to these objects.
The lattice-calculated ITDs contain the standard logarithmic divergence and, at non-zero $z$, additionally a power divergence induced by the Wilson line.
These divergences have been shown to be multiplicatively renormalizable to all orders in perturbation theory \cite{Ishikawa:2017faj,Ji:2017oey} and they can be removed through a ratio with a ME of the same operator at $P_3=0$. We employ a double ratio that also involves $z=0$ MEs to cancel additional systematics and ensure exact normalization of the charge~\cite{Orginos:2017kos}, 
\begin{equation}
\label{eq:reduced}
\mathfrak{M}(\nu,z) \equiv \frac{\mathcal{M}(\nu,z)\,/\,\mathcal{M}(\nu,0)}{\mathcal{M}(0,z)\,/\,\mathcal{M}(0,0)}.
\end{equation}
The renormalized MEs, $\mathfrak{M}(\nu,z)$, are referred to as reduced- or pseudo-ITDs.
Apart from serving the purpose of renormalization, the above ratio can be plausibly conjectured to remove some systematic effects \cite{Orginos:2017kos}.
In particular, this concerns discretization effects and $\mathcal{O}(z^2\LambdaQCD^2)$ higher-twist effects (HTEs), which are likely similar in the numerator and the denominator of the ratio.
We note that the above prescription defines a nonperturbative renormalization scheme and $1/z$ is a kinematic scale that suppresses higher-twist contributions, analogous to the momentum transfer in deep inelastic scattering.
As a ratio, $\mathfrak{M}$ is renormalization group invariant. 
However, its leading-twist contribution is related by factorization to the PDF at scale $\mu$ via the dimensionless product $z^2 \mu^2$ (see below). Therefore, under the leading-twist approximation, the dependence of $\mathfrak{M}$ on $z$ is governed by Dokshitzer-Gribov-Lipatov-Altarelli-Parisi (DGLAP) evolution.
\vspace*{0.1mm}\\

\subsection{Matching to light-front correlations}
Reduced ITDs are Euclidean observables that can be ``translated'' to their light-front counterparts via a perturbative matching procedure.
We will denote the light-front (matched) ITDs by $Q(\nu,\mu)$, where the $\MSb$ renormalization scheme is chosen with the renormalization scale denoted by $\mu$.
Matched ITDs are related to light-front PDFs, $q(x,\mu)$, by a Fourier transform:
\begin{equation}
\label{eq:FT}
q(x,\mu)=\frac{1}{2\pi}\int_{-\infty}^\infty d\nu' \, e^{-i\nu'x} Q(\nu',\mu).
\end{equation}

The relevant one-loop matching formulae were derived in Refs.~\cite{Radyushkin:2018cvn,Zhang:2018ggy,Izubuchi:2018srq,Radyushkin:2018nbf} and recently, the formalism was extended to two loops \cite{Li:2020xml}.
In this work, we apply it for the first time to actual lattice data.
Ref.~\cite{Li:2020xml} provides a factorization relation for the unpolarized matrix element renormalized in a generic scheme (denoted below with the superscipt ''R'').
For the case of interest here, i.e.\ the $\Gamma=\gamma_0$ Dirac structure and with the standard relativistic normalization of states, this relation reads:\footnote{We insert an additional factor 2 in the denominator of the right-hand side of Eq.~(\ref{eq:fact}) to reconcile the convention in Ref.~\cite{Li:2020xml} (in which in Eq.~(\ref{eq:A}) $iA(0,z,\mu)=2$ at tree-level) with our Eq.~(\ref{eq:bare}).}\vspace*{-2mm}
\begin{equation}
\label{eq:fact}
\mathcal{M}(\nu,z)^{\rm R}=\frac{1}{2R(z,\mu)}\int_{-1}^1 dx \, q(x,\mu) iA(x\nu,z,\mu),
\end{equation}
where $\mu$ is the renormalization/factorization scale and the factorization is valid up to $\mathcal{O}(z^2\Lambda_{\rm QCD}^2)$ higher-twist corrections.
$R(z,\mu)$ is the conversion factor between the chosen renormalization scheme and the $\MSb$ scheme, in which the light-cone PDF $q(x,\mu)$ is expressed.
$A(x\nu,z,\mu)$ is the perturbatively calculable matching kernel, available to two loops:
\begin{widetext}
\begin{eqnarray}
\label{eq:A}
&&\!\!\!\!\!iA(x\nu,z,\mu)= 2e^{ix\nu}+\frac{\alpha_s C_F}{\pi}\,\sum_{i=0}^1 L^i \left( a_{i10}^{(1)}e^{ix\nu} + \int_0^1 \!\! du \, a_{i11}^{(1)}(u)\left(e^{ixu\nu}\!-e^{ix\nu}\right)\right)\\ 
&+&\frac{\alpha_s^2}{\pi^2}\left[\sum_{i=0}^2 \sum_{j=1}^3 L^i C_j^{(2)} a_{ij0}^{(2)} e^{ix\nu}
+\sum_{i=0}^2 \sum_{j=1}^3 L^i C_j^{(2)} \!\!\int_0^1 \!\! du\, a_{ij1}^{(2)} \left(e^{ixu\nu}\!-e^{ix\nu}\right) 
+\sum_{i=0}^1  L^i C_4^{(2)} \!\!\int_{-1}^0 \!\! du\, a_{i42}^{(2)}(u) \left(e^{ixu\nu}\!-e^{ix\nu}\right)\right],\nonumber
\end{eqnarray}
where $L=\ln(z^2\mu^2/4)+2\gamma_E$ and the gauge group factors are $C_1^{(2)}=C_F^2$, $C_2^{(2)}=C_FC_A$, $C_3^{(2)}=n_f C_F T_F$ and $C_4^{(2)}=C_F^2-C_FC_A/2$, with $n_f$ the number of quark flavors.
The one-loop coefficients are $a_{010}^{(1)}=5/2$, $a_{110}^{(1)}=3/2$, $a_{011}^{(1)}(u)=(u^2-4u+1-4\ln(1-u))/(1-u)$ and $a_{111}^{(1)}(u)=(u^2+1)/(u-1)$.
Explicit expressions for the two-loop functions $a_{ijk}^{(2)}$ are lengthy and are given in the supplemental material of Ref.~\cite{Li:2020xml}.

To get the appropriate expression in the double ratio scheme, one can form a suitable ratio of the right-hand sides of Eq.~(\ref{eq:fact}), in which the conversion factors $R(z,\mu)$ cancel and one uses the normalization condition of $q(x,\mu$):
\begin{equation}
\label{eq:fact2}
\mathfrak{M}(\nu,z)=\int_{-\infty}^\infty dx \, q(x,\mu) \mathcal{A}(x\nu,z,\mu),
\end{equation}
where the kernel $\mathcal{A}(x\nu,z,\mu)=A(x\nu,z,\mu)/A(0,z,\mu)$ and the limits of the integration can be extended to infinity.
Writing:
\begin{equation}
iA(x\nu,z,\mu)=2e^{ix\nu}+\frac{\alpha_s}{\pi}A^{(1)}(x\nu,z,\mu)+\frac{\alpha_s^2}{\pi^2}A^{(2)}(x\nu,z,\mu)
\end{equation}
and Taylor-expanding $\mathcal{A}(x\nu,z,\mu)$ to $\mathcal{O}(\alpha_s^2)$, one arrives at:
\begin{eqnarray}
\label{eq:Anu}
\mathcal{A}(x\nu,z,\mu)=e^{ix\nu}&+&\frac{\alpha_s}{2\pi}\left(A^{(1)}(x\nu,z,\mu)-e^{ix\nu}A^{(1)}(0,z,\mu)\right)\\
&+&\frac{\alpha_s^2}{2\pi^2}\left(A^{(2)}(x\nu,z,\mu)-\frac{A^{(1)}(x\nu,z,\mu)A^{(1)}(0,z,\mu)}{2}-e^{ix\nu}A^{(2)}(0,z,\mu)+e^{ix\nu}\frac{(A^{(1)}(0,z,\mu))^2}{2}\right).\nonumber
\end{eqnarray}

Next, we derive the matching formulae in coordinate space, that transform the reduced ITDs into light-cone ITDs.
This procedure can be split into the evolution part, which takes the ITDs defined at different scales $1/z$ to a common scale $1/z'\mu$, yielding evolved ITDs $\mathfrak{M}'(\nu,\mu)$, and the matching part, leading finally to $Q(\nu,\mu)$.
First, we demonstrate that the above matching reproduces the known one-loop formulae \cite{Radyushkin:2018cvn,Zhang:2018ggy,Izubuchi:2018srq,Radyushkin:2018nbf} used in previous work.
Taking the explicit form of the one-loop functions $a_{i11}^{(1)}(u)$ given below Eq.~(\ref{eq:A}) and plugging in Eq.~(\ref{eq:FT}), one obtains:
\begin{eqnarray}
\mathfrak{M}(\nu,z)&=&Q(\nu,\mu)\\
&+&\frac{\alpha_s C_F}{4\pi^2} \int \!\! dx 
\int_0^1 \!\! du\, \left( \frac{u^2-4u+1-4\ln(1-u)}{1-u} - \frac{u^2+1}{1-u} \ln\frac{z^2\mu^2e^{2\gamma_E}}{4}\right)\left(e^{iux\nu}\!-\!e^{ix\nu}\right)\,\int_{-\infty}^\infty \!\! d\nu' \, e^{-i\nu'x} Q(\nu',\mu),\nonumber
\end{eqnarray}
where the contributions from the terms containing the coefficients $a_{i10}^{(1)}$ cancel between $A^{(1)}(x\nu)$ and $e^{ix\nu}A^{(1)}(0)$.
Using the integral representation of the Dirac delta, the exponentials in the above equation lead to $Q(u\nu,\mu)$ and $Q(\nu,\mu)$.
To arrive at the form of Ref.~\cite{Radyushkin:2018cvn} that we used in our previous study \cite{Bhat:2020ktg}, we rearrange the terms to have $\ln(z^2\mu^2/4)e^{2\gamma_E+1}$, leading finally to:
\begin{equation}
\label{eq:1L}
\mathfrak{M}(\nu,z)=Q(\nu,\mu) 
+\frac{\alpha_s}{\pi} \int_0^1  du\,\, C^{(1)}(u,z,\mu)\left(Q(u\nu,\mu)-Q(\nu,\mu)\right),
\end{equation}
with
\begin{equation}
C^{(1)}(u,z,\mu)=\frac{C_F}{2}\left(L^{(1)}(u) + B^{(1)}(u) \ln\frac{z^2\mu^2e^{2\gamma_E+1}}{4}\right),
\vspace*{-3mm}
\end{equation}
\begin{equation}
B^{(1)}(u)=\frac{1+u^2}{u-1},
\vspace*{-3mm}
\end{equation}
\begin{equation}
L^{(1)}(u) = 4 \frac{\ln(1-u)}{u-1} - 2(u-1).
\end{equation}

A similar procedure at the two-loop level, invoking additionally cancellations between $A^{(2)}(x\nu)$ and $e^{ix\nu}A^{(2)}(0)$ in Eq.~(\ref{eq:Anu}) and redefinition of the squared logarithm, leads to the two-loop matching, 
\begin{equation}
\label{eq:2L}
\mathfrak{M}(\nu,z)\!=\!Q(\nu,\mu) + \frac{\alpha_s}{\pi} \!\int_0^1 \!\! du\, C^{(1)}(u,z,\mu)\left(Q(u\nu,\mu)-Q(\nu,\mu)\right)
+ \frac{\alpha_s^2}{\pi^2} \int_{-1}^1  du\, C^{(2)}(u,z,\mu)\left(Q(u\nu,\mu)-Q(\nu,\mu)\right),
\end{equation}
\begin{equation}
C^{(2)}(u,z,\mu) = \frac{1}{2} \left(L^{(2)}(u)+B_1^{(2)}(u)\ln\frac{z^2\mu^2e^{2\gamma_E+1}}{4}+B_2^{(2)}(u)\ln^2\frac{z^2\mu^2e^{2\gamma_E+1}}{4}\right),
\end{equation}
\begin{eqnarray}
\label{eq:L2}
&& L^{(2)}(u) = \left\{
\begin{aligned}
& C_F^2 \left(a_{011}^{(2)}(u)-a_{111}^{(2)}(u)+a_{211}^{(2)}(u)-\frac{1}{2}a_{011}^{(1)}(u)+\frac{1}{2}a_{111}^{(1)}(u)\right) + C_FC_A \left(a_{021}^{(2)}(u)-a_{121}^{(2)}(u)+a_{221}^{(2)}(u)\right) & u>0\,,\\
& + n_fC_FT_F \left(a_{031}^{(2)}(u)-a_{131}^{(2)}(u)+a_{231}^{(2)}(u)\right) \\
& \left( C_F^2-\frac{1}{2}C_F C_A \right)\left(a_{042}^{(2)}(u)-a_{142}^{(2)}(u)\right) & u<0\,,
\end{aligned}\right.
\end{eqnarray}
\begin{eqnarray}
\label{eq:B21}
&& B_1^{(2)}(u) = \left\{
\begin{aligned}
& C_F^2 \left(a_{111}^{(2)}(u)-2a_{211}^{(2)}(u)-\frac{3}{4}a_{011}^{(1)}(u)+\frac{1}{4}a_{111}^{(1)}(u)\right) + C_FC_A \left(a_{121}^{(2)}(u)-2a_{221}^{(2)}(u)\right) & u>0\,,\\
& + n_fC_FT_F \left(a_{131}^{(2)}(u)-2a_{231}^{(2)}(u)\right) \\
& \left( C_F^2-\frac{1}{2}C_F C_A \right)a_{142}^{(2)}(u) & u<0\,,
\end{aligned}\right.
\end{eqnarray}
\begin{eqnarray}
\label{eq:B22}
&& B_2^{(2)}(u) = \left\{
\begin{aligned}
& C_F^2 \left(a_{211}^{(2)}(u)-\frac{3}{4}a_{111}^{(1)}(u)\right) + C_FC_A a_{221}^{(2)}(u) + n_fC_FT_F a_{231}^{(2)}(u) & u>0\,,\\
& 0 & u<0\,.
\end{aligned}\right.
\end{eqnarray}

One can check that the matching relation (\ref{eq:2L}) can be inverted in the following way:
\begin{eqnarray}
\label{eq:inv}
Q(\nu,\mu;z)&=&\mathfrak{M}(\nu,z)-\frac{\alpha_s}{\pi} \int_0^1  du\, C^{(1)}(u)\left(\mathfrak{M}(u\nu,z)-\mathfrak{M}(\nu,z)\right)
-\frac{\alpha_s^2}{\pi^2} \int_{-1}^1  du\, C^{(2)}(u)\left(\mathfrak{M}(u\nu,z)-\mathfrak{M}(\nu,z)\right)\nonumber\\
&+& \frac{\alpha_s^2}{\pi^2} \int_0^1  du\, C^{(1)}(u) \int_0^1  du'\, C^{(1)}(u')  \left(\mathfrak{M}(uu'\nu,z)-\mathfrak{M}(u\nu,z)-\mathfrak{M}(u'\nu,z)+\mathfrak{M}(\nu,z)\right),
\end{eqnarray}
\end{widetext}
which is equivalent to Eq.~(\ref{eq:2L}) up to $\mathcal{O}(\alpha_s^3)$ effects.
We use this equation to calculate light-cone ITDs, $Q(\nu,\mu)$, splitting the procedure into two parts: 
\begin{enumerate}
\item $B^{(1)}(u)$ (one-loop) and $B_i^{(2)}(u)$ (two-loop) kernels -- evolution of $z$-dependent reduced ITDs to a common scale $1/z'\equiv\mu$, taken to be 2 GeV -- yielding evolved ITDs, $\mathfrak{M}'(\nu,\mu;z)$,
\item $L^{(1)}(u)$ (one-loop) and $L^{(2)}(u)$ (two-loop) kernels -- matching and scheme conversion at fixed $\mu$ -- yielding the final light-cone ITDs, $Q(\nu,\mu;z)$.
\end{enumerate}
Note that at this stage, the evolved and matched ITDs keep track of the initial scale $1/z$ of reduced ITDs.
This will allow us to check the expected independence of $z$ after the matching procedure.
In practice, this will provide an important criterion for the maximal $z$ that can be used for the reconstruction of PDFs.
In principle, $z$ is limited to the perturbative regime -- thus, it should not exceed $\mathcal{O}(0.2-0.3)$ fm, at which point uncontrolled $\mathcal{O}(z^2\LambdaQCD^2)$ HTEs may be enhanced. 
However, a significant part of HTEs is expected to cancel when forming the double ratio and moreover, the practically achievable level of precision is limited.
Thus, it is plausible to extend the maximal value of $z$ in the reconstruction, $\zmax$, to a value such that $Q(\nu,\mu;z)$ is independent of $z$, considering ITDs coming from different combinations of $(P_3,z)$, but corresponding to the same Ioffe time $P_3z$.
Having established the empirical value of such $\zmax$, we will drop the $z$-argument of evolved and matched ITDs and ITDs from different combinations of $(P_3,z)$ will be averaged over.

\begin{table*}[t!]
\begin{center}
\begin{tabular}{|c|ccccc|cccc|}
\hline
Ensemble & $(L/a)^3\times(T/a)$ & $L$ [fm] & $m_\pi L$ & $a$ [fm] & $t_s/a$ & $P_3$ & $P_3$ [GeV] & $N_{\rm conf}$ & $N_{\rm meas}$\\ 
\hline
\multirow{4}{*}{A60} & \multirow{4}{*}{$24^3\times48$} & \multirow{4}{*}{2.2} & \multirow{4}{*}{4.1} & \multirow{4}{*}{$0.0934$} & \multirow{4}{*}{10} & 0 & 0 & 630 & 2520\\
 & & & & & & $2\pi/L$ & 0.55 & 630 & 2520\\
 & & & & & & $4\pi/L$ & 1.11 & 1260 & 10080\\
 & & & & & & $6\pi/L$ & 1.66 & 1260 & 40320\\
\hline
\multirow{5}{*}{B55} & \multirow{5}{*}{$32^3\times64$} & \multirow{5}{*}{2.6} & \multirow{5}{*}{5.0} & \multirow{5}{*}{$0.0820$} & \multirow{5}{*}{12} & 0 & 0 & 458 & 1832\\
 & & & & & & $2\pi/L$ & 0.47 & 458 & 1832\\
 & & & & & & $4\pi/L$ & 0.94 & 915 & 7320\\
 & & & & & & $6\pi/L$ & 1.42 & 1830 & 29280\\
 & & & & & & $8\pi/L$ & 1.89 & 1829 & 58528\\
\hline
\multirow{4}{*}{D45} & \multirow{4}{*}{$32^3\times64$} & \multirow{4}{*}{2.1} & \multirow{4}{*}{3.9} & \multirow{4}{*}{$0.0644$} & \multirow{4}{*}{15} & 0 & 0 & 630 & 2520\\
 & & & & & & $2\pi/L$ & 0.60 & 630 & 2520\\
 & & & & & & $4\pi/L$ & 1.20 & 1259 & 10072\\
 & & & & & & $6\pi/L$ & 1.80 & 1259 & 40288\\
\hline
\end{tabular}
\caption{Parameters of gauge field configurations ensembles used in this work. We give the ensemble label, the lattice size ($(L/a)^3\times(T/a)$), the spatial lattice extent ($L$) in fm, the value of the product $m_\pi L$, the lattice spacing ($a$) in fm, the source-sink separation in lattice units ($t_s/a$) and the values of the available nucleon boosts ($P_3$ as multiples of $2\pi/L$ and in GeV), together with the numbers of emplyed configurations ($N_{\rm confs}$) and the number of measurements ($N_{\rm meas}$, with 4, 8, 16 or 32 measurements per configuration).}
\label{tab:lat}
\end{center}
\end{table*}

\subsection{Reconstruction of momentum-space distributions}
PDFs are formally related to matched ITDs by Eq.~(\ref{eq:FT}).
However, this equation assumes an infinite range of continuous Ioffe times, while lattice evaluations provide only a discrete set of matched ITDs, truncated at some Ioffe time implied by the available $\zmax$. 
This generic limitation for lattice determinations of partonic functions was discussed in detail in Ref.~\cite{Karpie:2019eiq}.
As in our previous work \cite{Bhat:2020ktg}, we will follow three ways of reconstructing the light-cone distributions:
\begin{enumerate}
\item naive Fourier transform, i.e.\ using a discretized version of Eq.~(\ref{eq:FT}),
\item Backus-Gilbert (BG) method \cite{BackusGilbert,Karpie:2018zaz} (see the appendix for details),
\item reconstruction with an ansatz for the light-cone PDF \cite{Joo:2019jct}.
\end{enumerate}
In each of these ways, additional assumptions are provided that fill the gap between discrete lattice data and continuous distributions.
In the naive Fourier transform, this assumption is most severe -- the ITDs are assumed to be zero beyond $\zmax$ and no further criterion is used.
In the BG method, there is the model-independent assumption of maximizing the stability of the reconstructed distribution with respect to variation of the data within their errors, which alleviates to some extent the data missing beyond $\zmax$.
One can show that the naive Fourier transform yields a convolution of the PDF with a sinc-type kernel of width $(P_3 \zmax)^{-1}$; likewise, the BG-reconstructed $q(x)$ is an integral over nearby points $x'$ of the product of the true PDF with a computable $x$-dependent smearing kernel \cite{Alexandrou:2020qtt,BackusGilbert}.
Finally, ansatz reconstruction assumes a functional form of the PDF, analogously to procedures used in global fits of experimental collider data.

It is convenient to consider separately the real and imaginary parts of ITDs.
The former are related to the valence distribution, $q_v=q-\bar{q}$,
\begin{equation}
\label{eq:ReQ}
{\rm Re}\,Q(\nu,\mu^2) = \int_0^1 dx \cos(\nu x) q_v(x,\mu).
\end{equation}
The imaginary part is related to the distribution $q_{v2s}\equiv q_v+2\bar{q}=q+\bar{q}$,
\begin{equation}
\label{eq:ImQ}
{\rm Im}\,Q(\nu,\mu^2) = \int_0^1 dx \sin(\nu x) q_{v2s}(x,\mu^2).
\end{equation}
Combining $q_v$ and $q_{v2s}$, we will also present results for $q=q_v+\bar{q}$ and for the antiquark (sea quark) PDF, $q_s\equiv\bar{q}$.
In the ansatz reconstruction, we will use the simplest plausible functional form capturing the limiting behaviors for small and large $x$,
\begin{equation}
\label{eq:ansatz}
q(x) = N x^\alpha (1-x)^\beta,
\end{equation}
with fitting parameters $\alpha,\,\beta$.
For the valence PDF, normalized to 1, $N=1/B(\alpha+1,\beta+1)$, $B(x,y)=\Gamma(x)\Gamma(y)/\Gamma(x+y)$ being the Euler beta function and $\Gamma(x)$ the gamma function. 
In turn, the normalization of the distribution $q_{v2s}$, $N$, is an additional fitting parameter.
The fits minimize the $\chi^2$ function,
\begin{equation}
\label{eq:chi2}
\chi^2=\sum_{\nu=0}^{\numax}\frac{Q(\nu,\mu)-Q_f(\nu,\mu)}{\sigma_Q^2(\nu,\mu)},
\end{equation}
where $\sigma_Q^2(\nu,\mu)$ is the statistical error of $Q(\nu,\mu)$ and $Q_f(\nu,\mu)$ is the cosine/sine Fourier transform of the fitting ansatz, for the $q_v$ and $q_{v2s}$ case, respectively.
Such fits define the fitted ITDs, $Q_f(\nu,\mu)$, which are continuous functions of the Ioffe time.
Note the fits depend on the maximum Ioffe time, $\numax$, which is determined by the maximal length of the Wilson line, $\zmax$, and the maximum nucleon boost.

\begin{figure*}[t!]
\begin{center}
    \includegraphics[width=\textwidth]{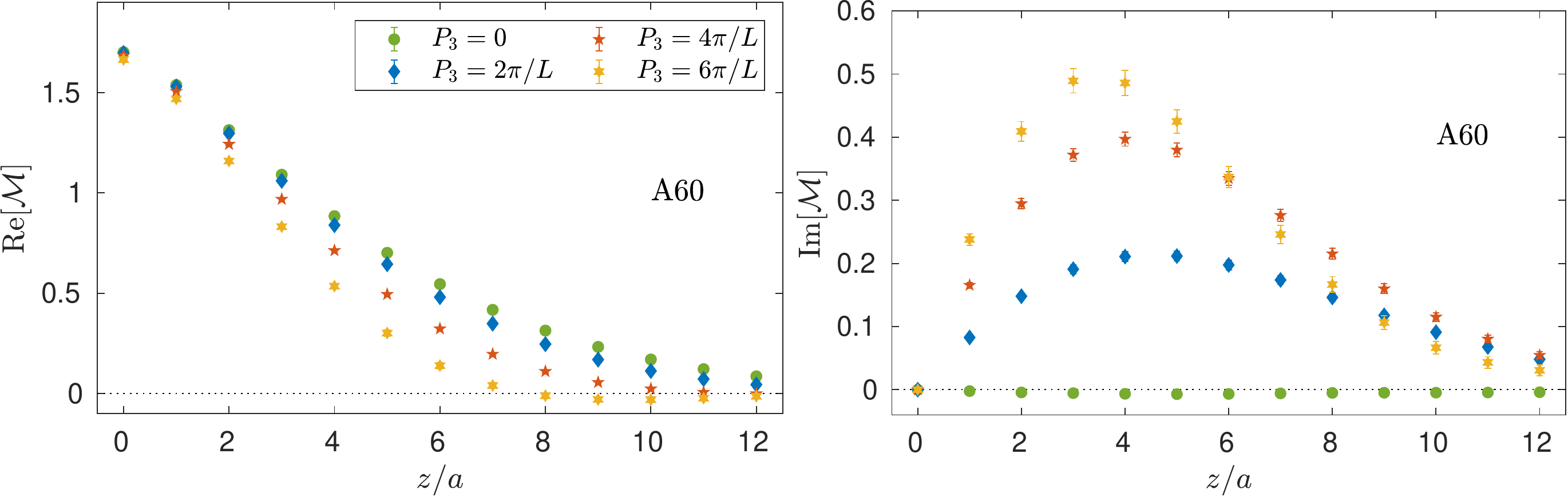}
    \includegraphics[width=\textwidth]{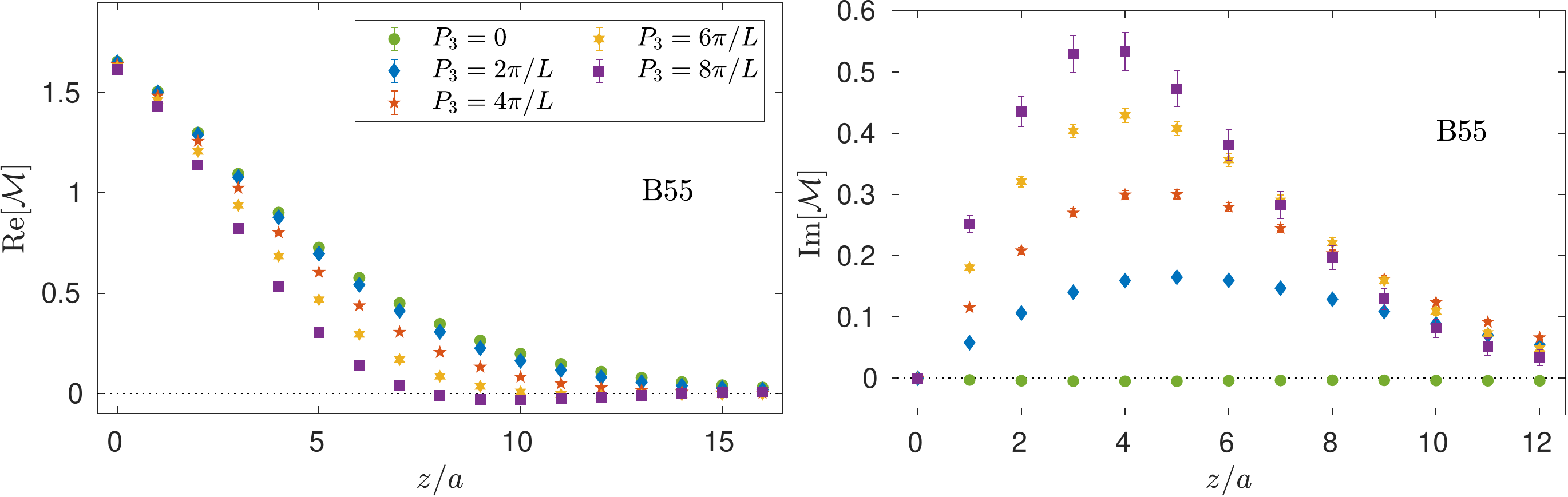}
    \includegraphics[width=\textwidth]{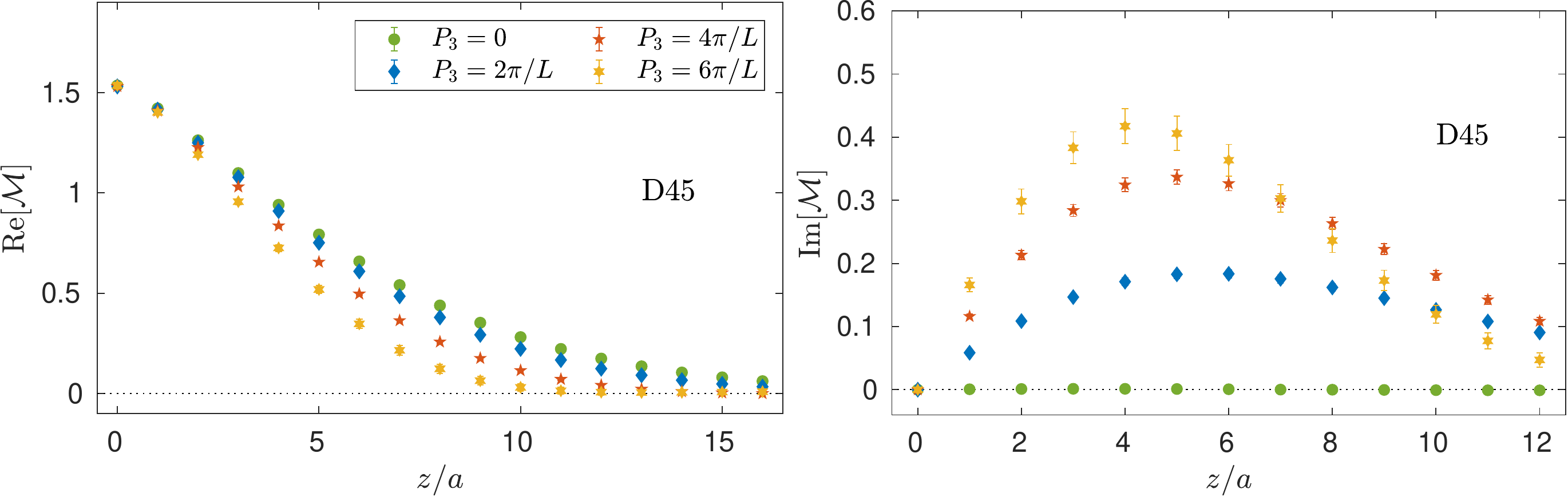}
\end{center}
\vspace*{-0.5cm} 
\caption{Real (left) and imaginary (right) part of bare matrix elements, $\mathcal{M}(\nu,z)$, at different values of $P_3$ for the ensembles A60 (top), B55 (middle) and D45 (bottom). The different symbols correspond to: $P_3=0$ (green circles), $P_3=2\pi/L$ (blue rhombuses), $P_3=4\pi/L$ (red 5-stars), $P_3=6\pi/L$ (yellow 6-stars) and $P_3=8\pi/L$ (purple squares; only B55).}
\label{fig:bare}
\end{figure*}

\begin{figure*}[t!]
\begin{center}
    \includegraphics[width=\textwidth]{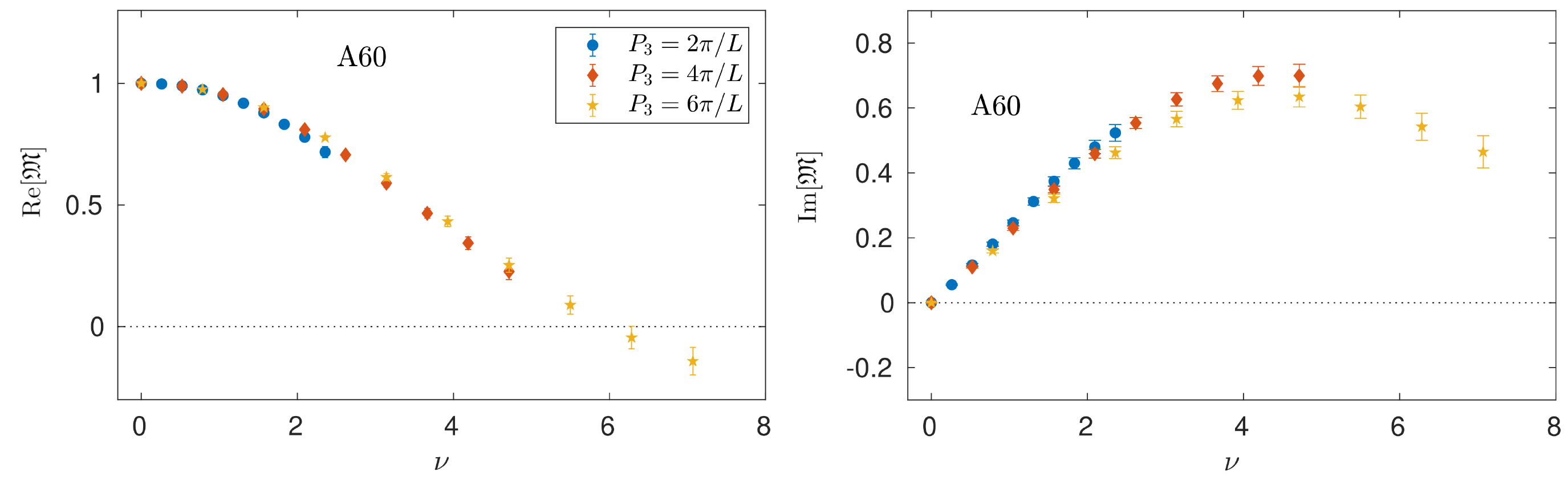}
    \includegraphics[width=\textwidth]{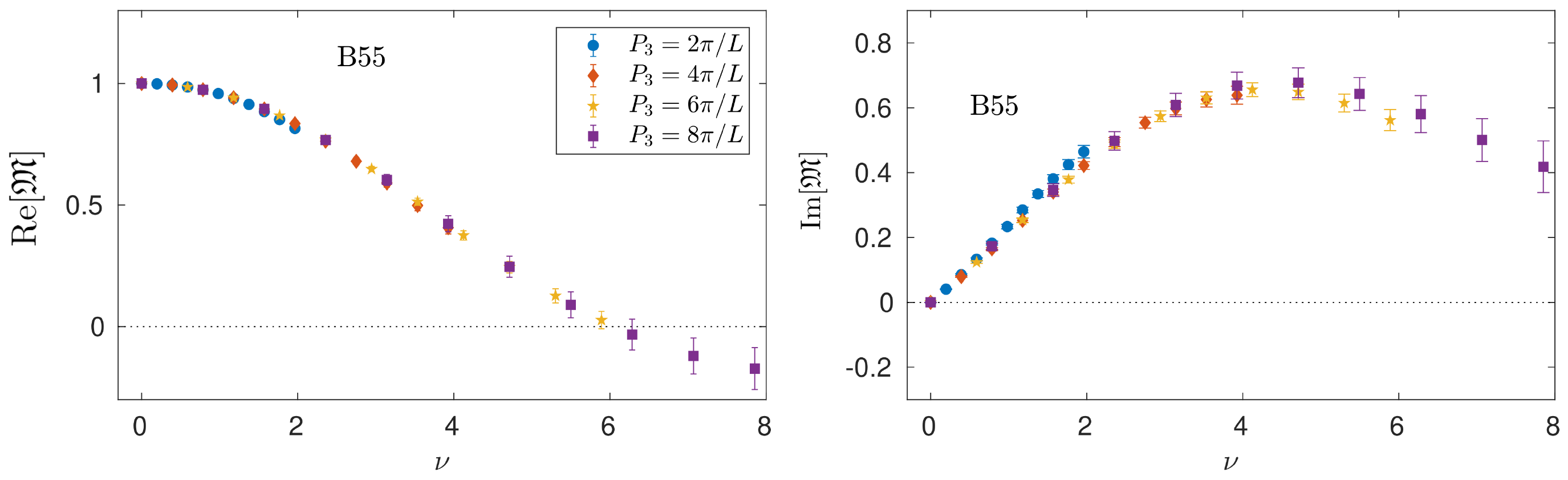}
    \includegraphics[width=\textwidth]{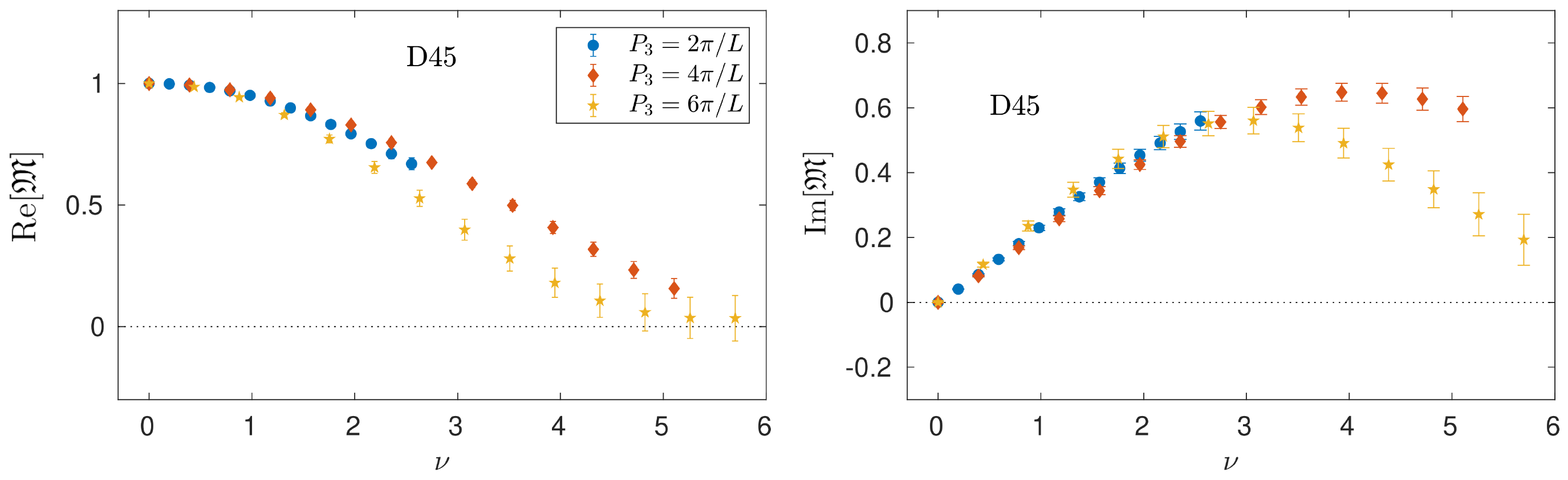}
\end{center}
\vspace*{-0.5cm} 
\caption{Real (left) and imaginary (right) part of reduced ITDs, $\mathfrak{M}(\nu,z)$, at different values of $P_3$ for the ensembles A60 (top), B55 (middle) and D45 (bottom). The different symbols correspond to: $P_3=2\pi/L$ (blue circles), $P_3=4\pi/L$ (red rhombuses), $P_3=6\pi/L$ (yellow stars) and $P_3=8\pi/L$ (purple squares; only B55).}
\label{fig:reduced}
\end{figure*}
\begin{figure*}[t!]
\begin{center}
    \includegraphics[width=\textwidth]{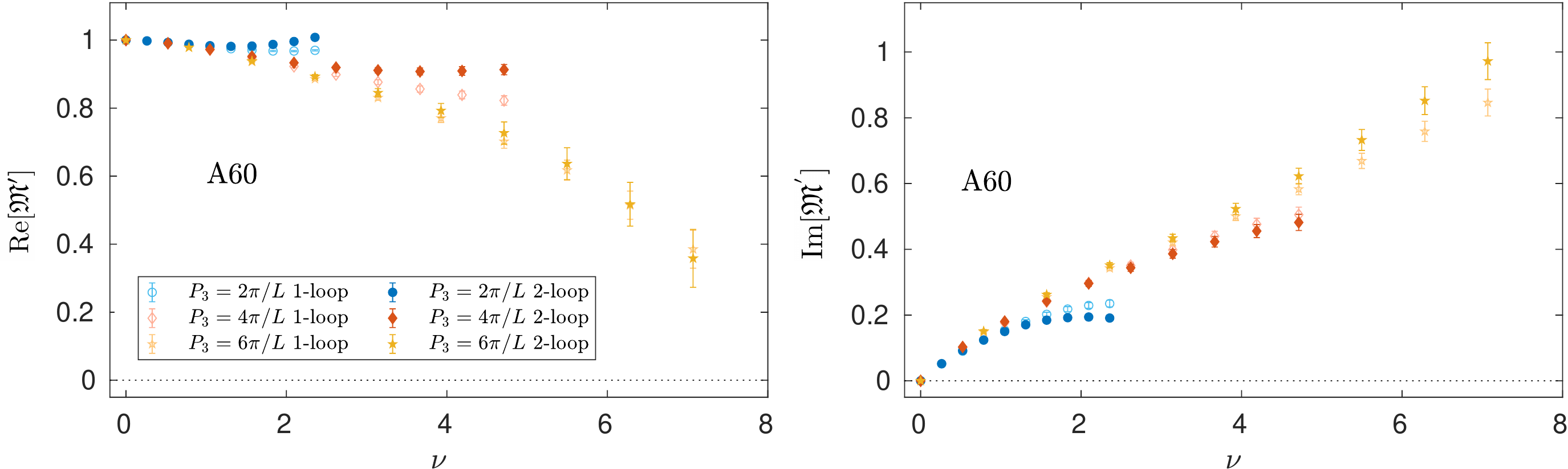}
    \includegraphics[width=\textwidth]{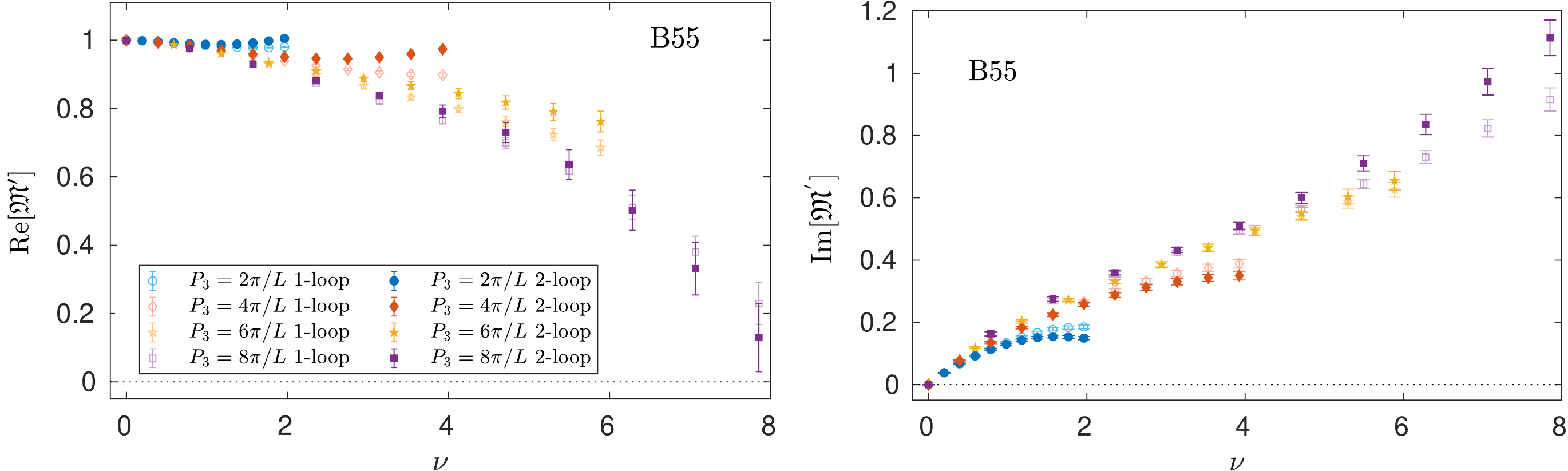}
    \includegraphics[width=\textwidth]{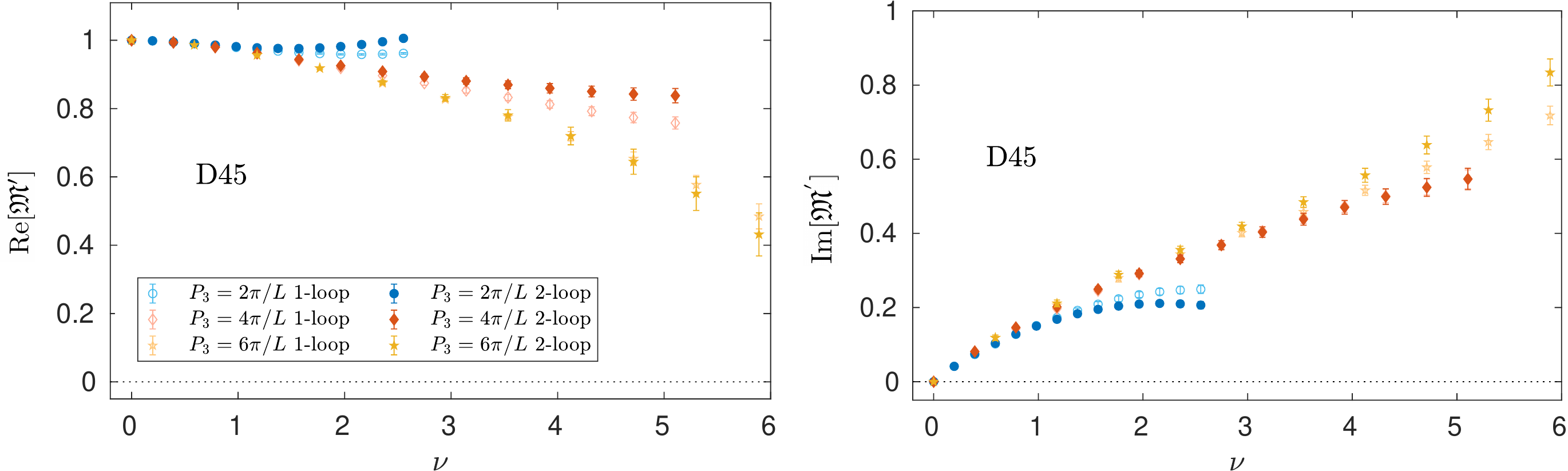}
\end{center}
\vspace*{-0.5cm} 
\caption{Real (left) and imaginary (right) part of ITDs evolved to a common scale of 2 GeV, $\mathfrak{M}'(\nu,z,1/z'=2{\rm\,GeV})$, at different values of $P_3$ for the ensembles A60 (top), B55 (middle) and D45 (bottom). The different symbols correspond to: $P_3=2\pi/L$ (blue circles), $P_3=4\pi/L$ (red rhombuses), $P_3=6\pi/L$ (yellow stars) and $P_3=8\pi/L$ (purple squares; only B55). The open lighter-color markers show the one-loop effect and the closed darker-color ones the two-loop effect.}
\label{fig:evolved}
\end{figure*}

\begin{figure*}[t!]
\begin{center}
    \includegraphics[width=\textwidth]{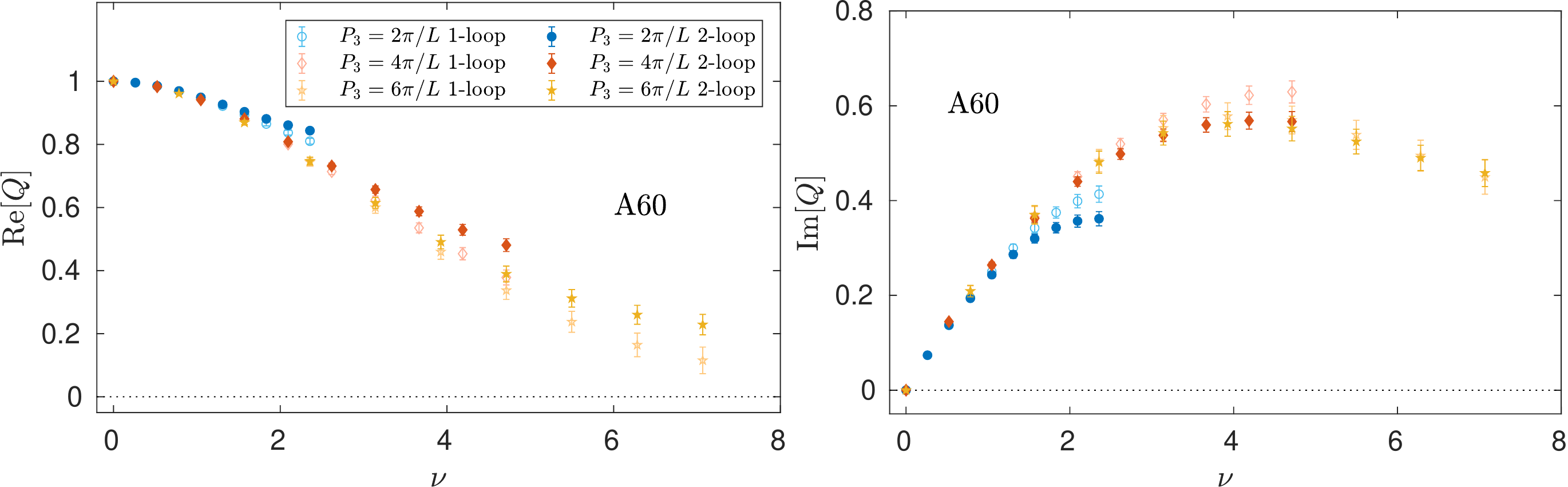}
    \includegraphics[width=\textwidth]{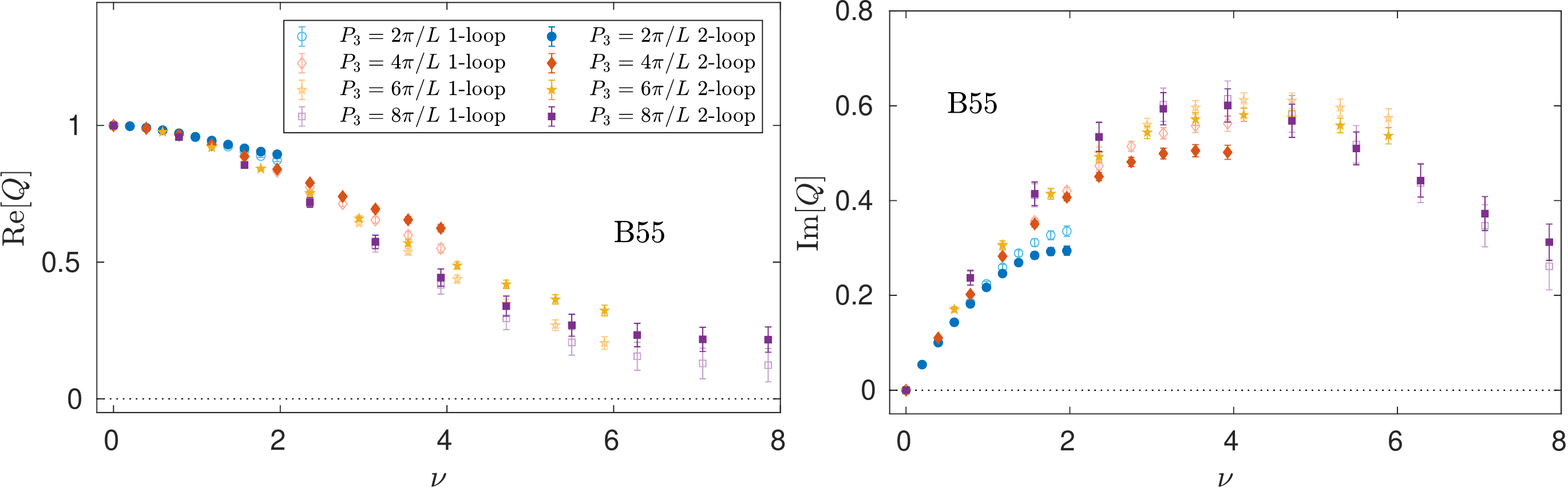}
    \includegraphics[width=\textwidth]{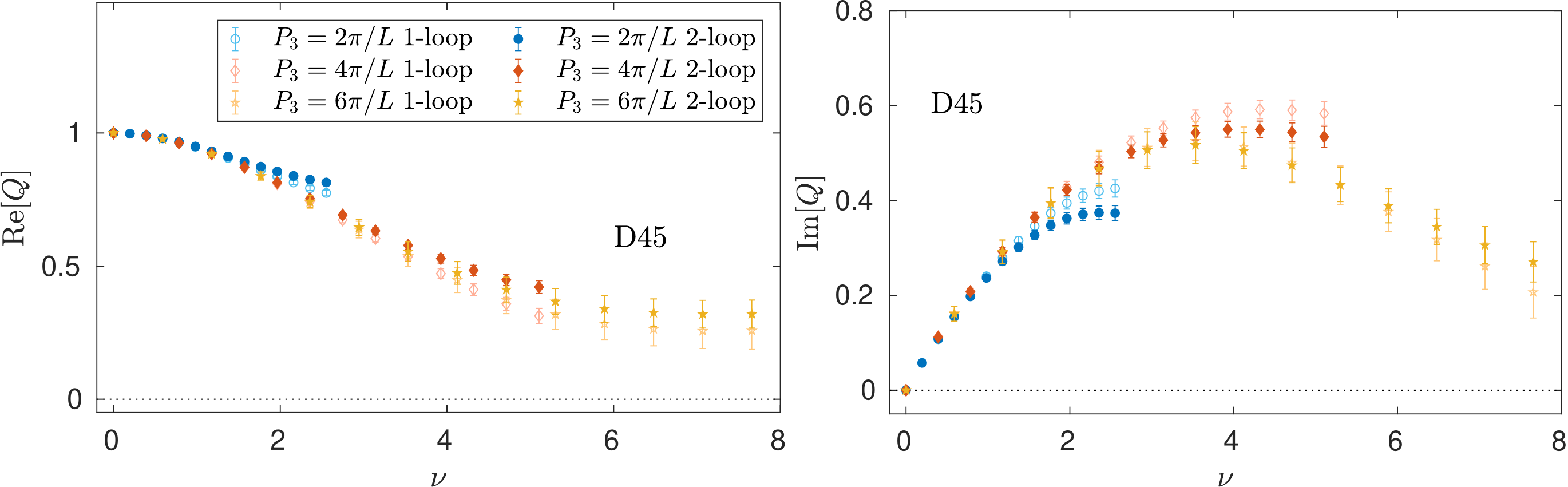}
\end{center}
\vspace*{-0.5cm} 
\caption{Real (left) and imaginary (right) part of matched ITDs, $Q(\nu,z,\mu=2{\rm\,GeV})$, at different values of $P_3$ for the ensembles A60 (top), B55 (middle) and D45 (bottom). The different symbols correspond to: $P_3=2\pi/L$ (blue circles), $P_3=4\pi/L$ (red rhombuses), $P_3=6\pi/L$ (yellow stars) and $P_3=8\pi/L$ (purple squares). The open lighter-color markers show the one-loop effect and the closed darker-color ones the two-loop effect.}
\label{fig:matched}
\end{figure*}

\begin{figure*}[p!]
\begin{center}
    \includegraphics[width=\textwidth]{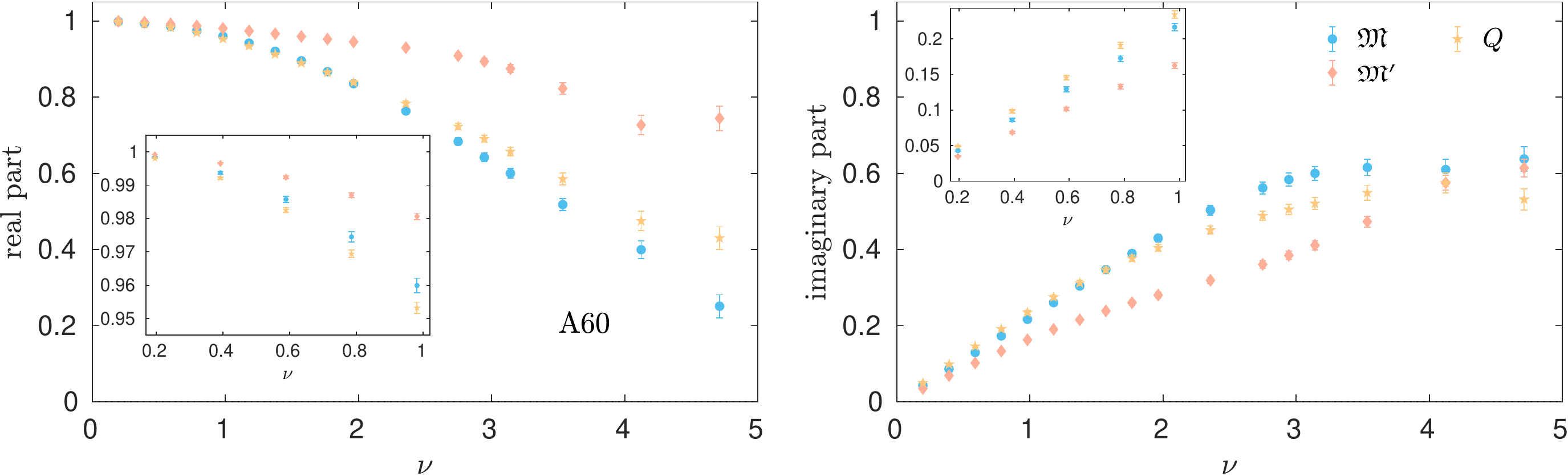}
    \includegraphics[width=\textwidth]{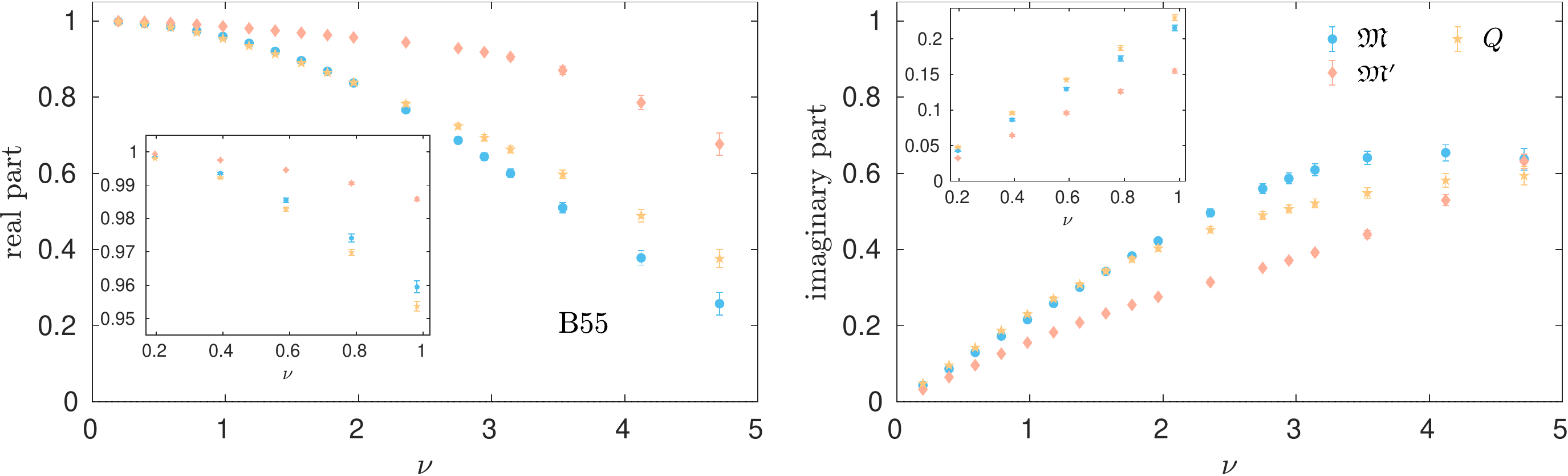}
    \includegraphics[width=\textwidth]{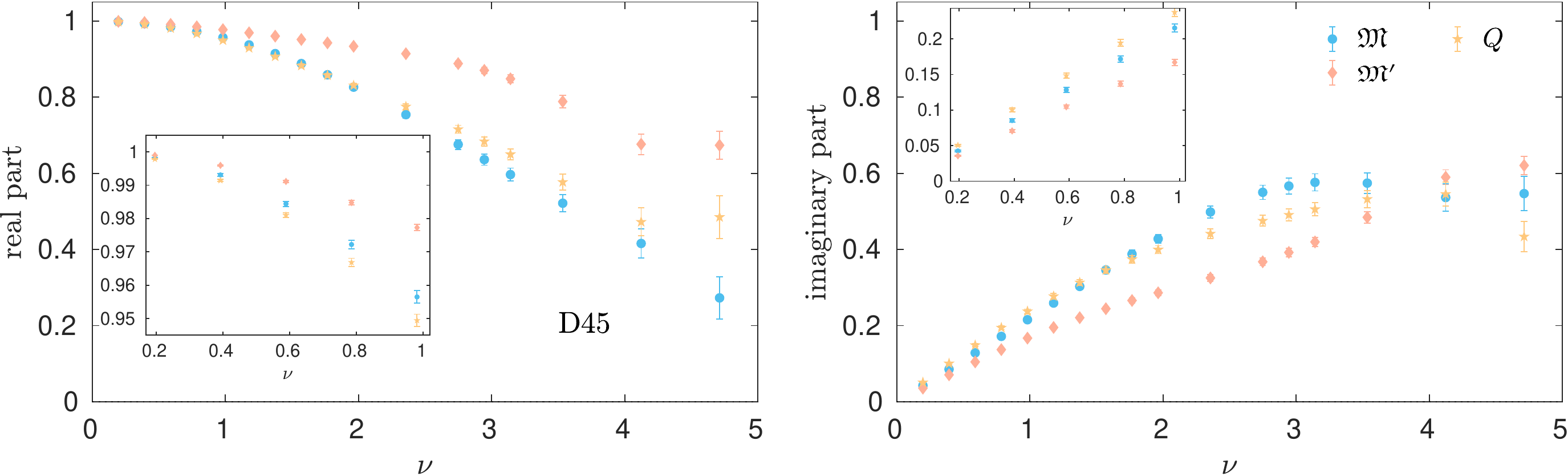}
    \includegraphics[width=\textwidth]{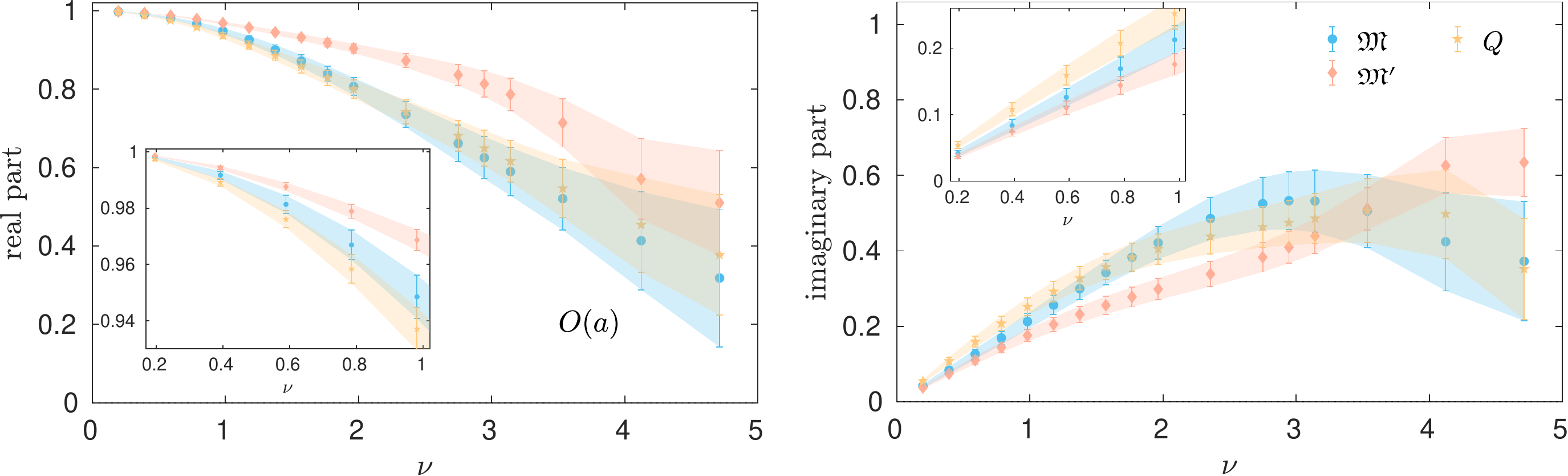}
\end{center}
\vspace*{-0.5cm} 
\caption{Real (left) and imaginary (right) part of reduced (blue circles), evolved (red rhombuses) and matched (yellow stars) ITDs, with the same Ioffe times corresponding to different combinations $(P_3,z)$ averaged over, including Wilson line lengths $z\leq\zmax=0.5$ fm. Shown are Ioffe times corresponding to the discrete values attainable for B55 and D45; the data for A60 (with different values of the boost in lattice units) are interpolated with fourth order polynomials. From top to bottom: A60, B55, D45, continuum limit from $\mathcal{O}(a)$ extrapolation (evolved and matched ITDs from two-loop formulae). The insets zoom in the small-$\nu$ behavior.}
\label{fig:ITDs}
\end{figure*}

\section{Lattice setup}
\label{sec:lattice}
In this work, we study the continuum limit of lattice-extracted ITDs and the resulting unpolarized PDFs.
For each of the employed three lattice spacings, $a=0.0644,\,0.0820,\,0.0934$ fm, we use the lattice data of Ref.~\cite{Alexandrou:2020qtt} pertaining to nucleon boosts of around 1.8 GeV and the Dirac structure $\gamma_0$.
These data were used for an analogous continuum limit study in the quasi-distribution framework.
For quasi-PDFs, one can only use data of a sufficiently large momentum, for which contact with the light-cone frame can be attained.
In turn, pseudo-PDFs can utilize all nucleon boosts, including small ones, thus leading to precisely extracted ITDs at small Ioffe times.
To take advantage of this fact, we supplement the data of Ref.~\cite{Alexandrou:2020qtt} with all intermediate nucleon boosts (including zero boost used to form the reduced ITDs) by performing additional calculations.
In Ref.~\cite{Alexandrou:2020qtt}, all data were produced employing five steps of stout smearing \cite{Morningstar:2003gk} applied to the Wilson line of the non-local operator.
Here, we also consider data obtained without stout smearing, checking the independence of the results concerning this aspect.

The computational techniques are the same as in Ref.~\cite{Alexandrou:2020qtt} and we refer to this paper for more details, discussing here only the main aspects.
The used ensembles of gauge field configurations were generated by the European Twisted Mass Collaboration \cite{Baron:2010bv}, the predecessor of the current Extended Twisted Mass Collaboration (ETMC).
They have two degenerate light flavors of maximally twisted mass fermions \cite{Frezzotti:2000nk,Frezzotti:2003ni} with masses corresponding to a pion mass of around 370 MeV and the strange and charm quarks with near-physical values of the mass.
The gluonic part of the action is Iwasaki-improved \cite{Iwasaki:1985we}.
The parameters of our calculations are given in Tab.~\ref{tab:lat}.
While twisted mass fermions yield automatic $\mathcal{O}(a)$-improvement of physical observables when tuned to maximal twist, the evaluated non-local matrix elements do not belong to this category and thus, the improvement holds only at $z=0$.
Consequently, the ITDs calculated in this work have $\mathcal{O}(a)$ leading discretization effects.
Thus, our continuum extrapolations are performed assuming $\mathcal{O}(a)$ fitting ansatzes, but we check also the alternative $\mathcal{O}(a^2)$ ones for comparison.
The latter may be plausible when including relatively small $z$ values, for which remnants of automatic $\mathcal{O}(a)$-improvement may be present.
Also, it was shown in Ref.~\cite{Green:2020xco} that maximal twist can remove some of the $\mathcal{O}(a)$ contributions and some reduction of these can also ensue in the double ratio that defines reduced ITDs.
We also remark that all statistical analyses are performed using 1000 bootstrap samples generated by reshuffling the original data.

\section{Results}
\label{sec:results}
\subsection{ITDs}
The lattice input to the determination of PDFs via the pseudo-distribution approach are bare MEs, which we show in Fig.~\ref{fig:bare} for all our ensembles and for all employed nucleon boosts, with five steps of stout smearing.
We note that the $z=0$ matrix element is independent of the nucleon boost and yields 1 upon multiplication with the appropriate scale- and scheme-independent normalization factor $Z_V$, reflecting vector current conservation.
At non-zero $z$, the real part decays to zero faster as the boost increases.
The imaginary part vanishes for $z=0$ at any boost and for all $z$'s in the zero-momentum case, within uncertainties.
For $z>0$, it becomes more pronounced with increasing boost, with its maximum moving towards smaller Wilson line lengths.

In Fig.~\ref{fig:reduced}, we show reduced ITDs (five iterations of stout smearing), formed according to Eq.~(\ref{eq:reduced}), as a function of the Ioffe time.
For sufficiently small Ioffe times, all nucleon boosts yield ITDs consistent with the ones for other values of $P_3$.
Such ITDs are defined at different scales $1/z$, which suggests that the scale dependence is relatively small.
This residual scale dependence is expected to be further reduced after the matching.

Evolved and matched ITDs at the level of separate ensembles are shown in Fig.~\ref{fig:evolved} and Fig.~\ref{fig:matched}, respectively.
Here, we decompose the perturbative evolution and matching procedure into one-loop and two-loop parts.
The two-loop effect in the evolution is much smaller than the one-loop one, but its relevance is increasing at larger Ioffe times.
However, one needs to keep in mind that the perturbative procedure becomes unreliable at large distances $z$.
Nevertheless, the two-loop effect is clearly statistically significant for ITDs with Wilson line lengths $z\gtrsim0.4-0.5$ fm and the evolved ITDs are substantially different than reduced ITDs.
The second, $L$-kernel part of the perturbative matching acts in the opposite direction, bringing matched ITDs closer to reduced ones and making the two-loop effects statistically significant only for ITDs originating from $z\gtrsim0.5-0.6$ fm.

Fig.~\ref{fig:matched} allows us also to establish the aformentioned practical criterion for the value of $\zmax$ in the PDF reconstruction procedure.
On the one hand, the coordinate-space factorization should include only ITDs at perturbative values of $z$, with large-$z$ ITDs contaminated by potentially uncontrollable HTEs.
On the other hand, the double ratio is likely subtracting a part of the HTEs and the remaining ones may be well below our statistical precision for values of $z$ extending beyond the perturbative regime.
Inspecting Fig.~\ref{fig:matched}, we observe that ITDs at small Ioffe times are independent on the nucleon boost at which they have been obtained as long as the product $P_3z$ is the same.
The $P_3$-dependence of such equal-$\nu$ ITDs starts to set in for ITDs obtained at values of $z$ larger than around $0.5$ fm for the real part and already around $0.3$ fm for the imaginary part (we note that the two-loop correction slightly decreases this value), indicating statistical significance of HTEs for such large lengths of the Wilson line and the breakdown of coordinate-space factorization.
Below, we keep these values in mind when reconstructing the PDFs.
Particularly the value for the imaginary part is rather low, close to the expectation for the validity of perturbation theory.
This provides a clue that the reconstruction of distributions involving antiquarks (i.e.\ using the imaginary part of ITDs) may be more difficult for the lattice.
Thus, we will look at PDFs reconstructed with three different values of $\zmax$: $0.3$ fm, $0.5$ fm and $0.7$ fm.

\begin{figure*}[p!]
\begin{center}
    \includegraphics[width=\textwidth]{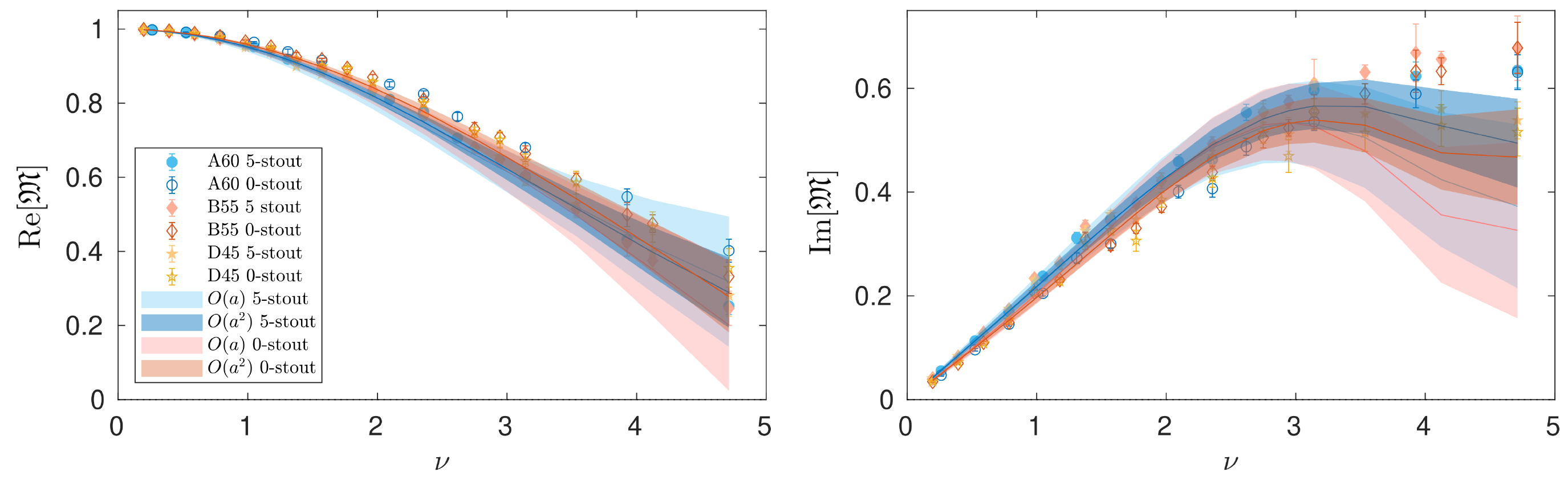}    
\end{center}
\vspace*{-0.5cm} 
\caption{Real (left) and imaginary (right) part of reduced ITDs, $\mathfrak{M}(\nu,z)$, with the same Ioffe times corresponding to different combinations $(P_3,z)$ averaged over, $z\leq\zmax=0.5$ fm. Shown are data points for the ensembles A60 (blue circles), B55 (red rhombuses) and D45 (yellow stars). The red/blue bands correspond to data obtained with 0/5 steps of stout smearing of the Wilson line entering the non-local operator, with the lighter/darker color pertaining to $\mathcal{O}(a)$/$\mathcal{O}(a^2)$ extrapolation to the continuum.}
\label{fig:reduced_nuavg}
\end{figure*}

\begin{figure*}[p!]
\begin{center}
    \includegraphics[width=\textwidth]{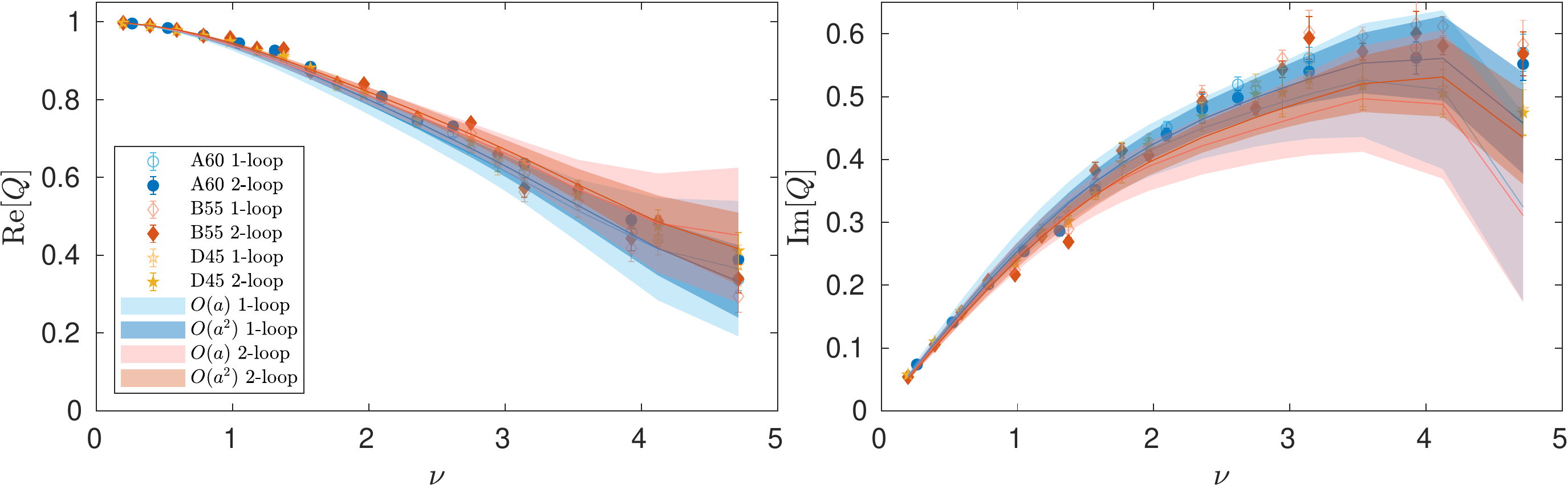}    
\end{center}
\vspace*{-0.5cm} 
\caption{Real (left) and imaginary (right) part of matched ITDs, $Q(\nu,\mu=2{\rm\,GeV})$, with the same Ioffe times corresponding to different combinations $(P_3,z)$ averaged over, $z\leq\zmax=0.5$ fm. Shown are data points for the ensembles A60 (blue circles), B55 (red rhombuses) and D45 (yellow stars). The blue bands correspond to matching peformed at one-loop/two-loop order, with the lighter/darker color pertaining to $\mathcal{O}(a)$/$\mathcal{O}(a^2)$ extrapolation to the continuum.}
\label{fig:matched_nuavg}
\end{figure*}

\begin{figure*}[p!]
\begin{center}
    \includegraphics[width=\textwidth]{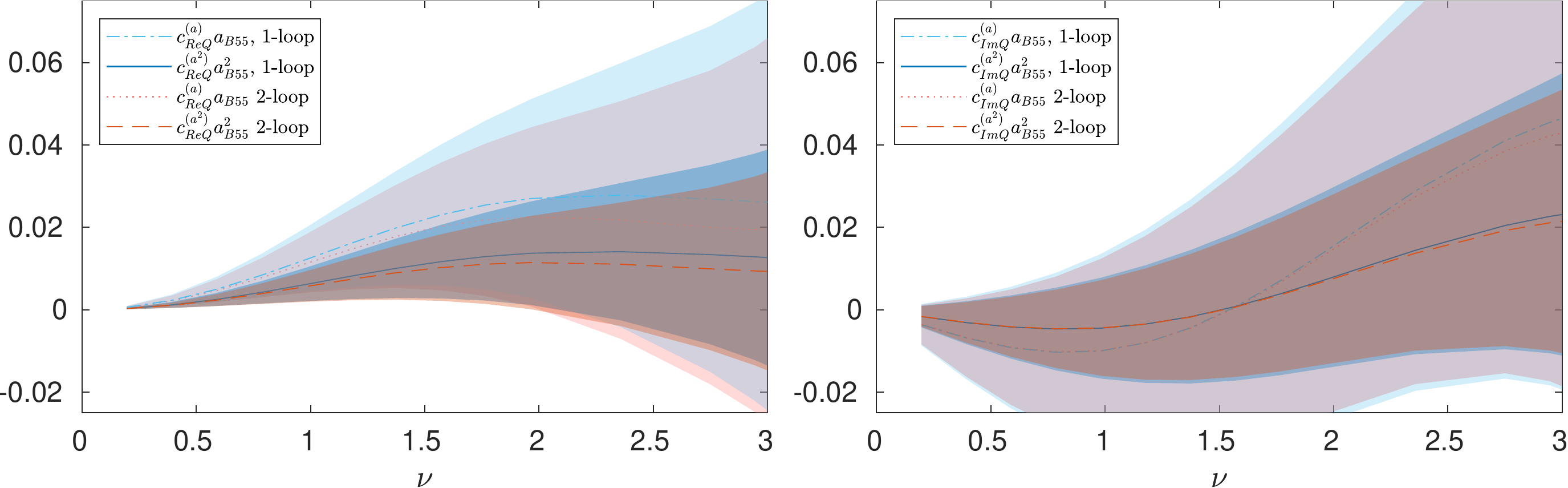}    
\end{center}
\vspace*{-0.5cm} 
\caption{Real (left) and imaginary (right) part of cut-off effects coefficients $c_{{\rm Re\,}Q}^{(a^i)}$ and $c_{{\rm Im\,}Q}^{(a^i)}$ of the $\mathcal{O}(a^i)$ continuum limit extrapolations of 1-loop and 2-loop matched ITDs, multiplied by the $i$-th power of our middle lattice spacing, $a_{\rm B55}=0.082$ fm.
}
\label{fig:matched_cutoff}
\end{figure*}

For a given value of $\zmax$, all data corresponding to $z>\zmax$ are dismissed and ITDs corresponding to the same Ioffe times that originate from different combinations of $(P_3,\,z)$ are averaged.
A comparison of such $\nu$-averaged reduced, evolved and matched ITDs (with two-loop formulae) at a finite lattice spacing is shown in the three upper rows of Fig.~\ref{fig:ITDs}, for $\zmax=0.5$ fm.
We show ITDs at Ioffe times corresponding to the discrete values attainable for B55 and D45, i.e.\ $2\pi n/32$, with $n$ integer.
For A60 ($L/a=24$), ITDs are interpolated to these discrete values by fitting fourth-order polynomials to the $\nu$-dependence.
We note such polynomials provide very good description of the $\nu$-dependence in the entire range of Ioffe times.
As mentioned above, the effect of matching is opposite to the one of evolution, with effects of the former almost canceling the latter.
In the real part, at small/large Ioffe times, matched ITDs are below/above reduced ones, with exact cancellation of evolution and matching occuring around $\nu=1.8$.
The behavior is exactly the opposite in the imaginary part, with small/large-$\nu$ matched ITDs above/below reduced ones.

At this stage, we are ready to perform continuum limit extrapolations from our three lattice spacings.
The fitting ansatz takes the form:
\begin{equation}
\mathcal{I}(a)=\mathcal{I}(0)+c_{\mathcal{I}}^{(a^i)}\,a^{i}, 
\end{equation}
where $\mathcal{I}(a)$ is the considered ITD (reduced or matched at one/two-loop order, either real or imaginary part) at lattice spacing $a$ and $c_{\mathcal{I}}^{(a^i)}$ is the slope of the leading discretization effects linear in $a^i$, with $i=1,\,2$.
The extrapolations are performed always at fixed Ioffe times being integer multiples of $2\pi/32$.

The bottom row of Fig.~\ref{fig:ITDs} compares our reduced, evolved and matched ITDs in the $\mathcal{O}(a)$ continuum limit, again showing $\zmax=0.5$ fm.
Clearly, the errors are inflated in the continuum limit, decreasing the significance of differences between reduced and matched ITDs.
In fact, at all Ioffe times, these differences become statistically insignificant, with largest ones slightly exceeding 1-$\sigma$ (imaginary part at small values of $\nu$).

In the continuum limit, we also address the issue of the potential influence of the number of stout smearing iterations on our results.
In Fig.~\ref{fig:reduced_nuavg}, we show the $\nu$-averaged reduced ITDs of the three ensembles with zero or five steps of stout smearing, together with their continuum limit extrapolations, performed both with an $\mathcal{O}(a)$ and $\mathcal{O}(a^2)$ fitting ansatz ($\zmax=0.5$ fm).
This reveals that there is significant dependence of ITDs on the number of stout iterations for the separate ensembles, particularly at large Ioffe times.
However, the results in the continuum limit are fully compatible between zero and five stout steps.
This holds in the whole considered range of Ioffe times with extrapolations linear in $a$.
In the case of $a^2$ extrapolations, the difference between 0-stout and 5-stout is mildly statistically significant in narrow ranges around $\nu=2$ (real part) or $\nu=1$ (imaginary part).
Since we know that $\mathcal{O}(a)$ effects are bound to be present, this suggests that these effects may be enhanced around these Ioffe times and supports the need for the $\mathcal{O}(a)$-improvement of the underlying MEs.
We also note the compatibility of continuum results obtained with extrapolations linear in $a$ and $a^2$, for the whole considered range of Ioffe times.
This is a consequence of the relative smallness of discretization effects, with reduced ITDs compatible between ensembles for almost all values of $\nu$.
Obviously, the longer $\mathcal{O}(a)$-extrapolation inflates the errors much more significantly.
Overall, it is clear that ITDs obtained with different numbers of stout smearing iterations differ only by $\mathcal{O}(a)$ cutoff effects.
Hence, below we concentrate exclusively on the slightly more precise case of five stout steps.

Fig.~\ref{fig:matched_nuavg} presents the dependence of $\nu$-averaged matched ITDs, again at the level of single ensembles and the $\mathcal{O}(a)$ and $\mathcal{O}(a^2)$ continuum limits.
In this case, we compare results from one- and two-loop matching.
As observed above in Fig.~\ref{fig:matched}, the two-loop effects for $z\lesssim0.5-0.6$ fm are smaller than statistical uncertainties already for separate ensembles.
Thus, the continuum limits with inflated errors are also consistent.
Again, cutoff effects are almost invisible at this level of precision.
The slope $c_{\mathcal{I}}^{(a^i)}$ is statistically insignificant for all Ioffe times, being at most 1-1.5-$\sigma$ away from zero at $\nu\lesssim2$ in the real part.
This is depicted in Fig.~\ref{fig:matched_cutoff}, where we plot $c_{\mathcal{I}}^{(a^i)}a^i_{\rm B55}$, where $a_{\rm B55}=0.082$ fm is our middle lattice spacing.
Thus, this quantity can be interpreted as the difference of the fitted B55 result and the continuum limit value.

\begin{figure*}[t!]
\begin{center}
    \includegraphics[width=\textwidth]{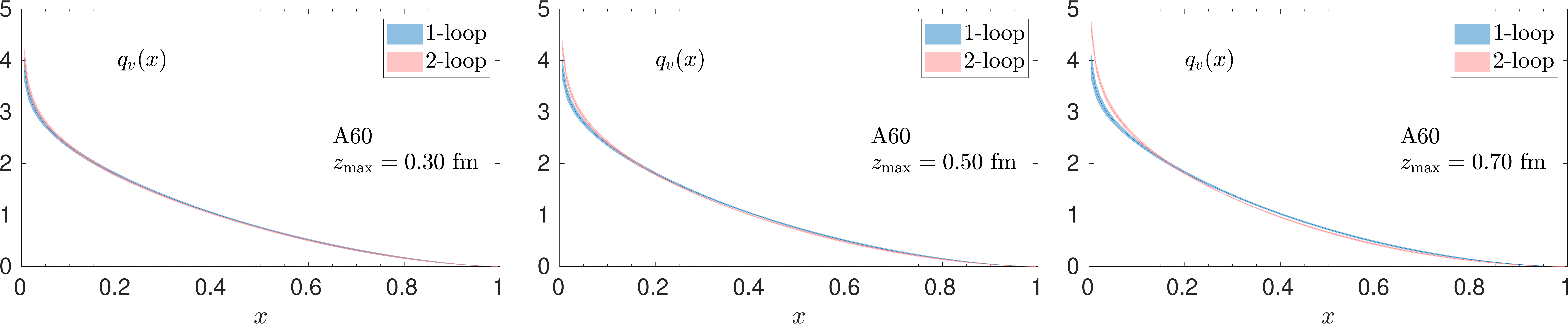}    
    \includegraphics[width=\textwidth]{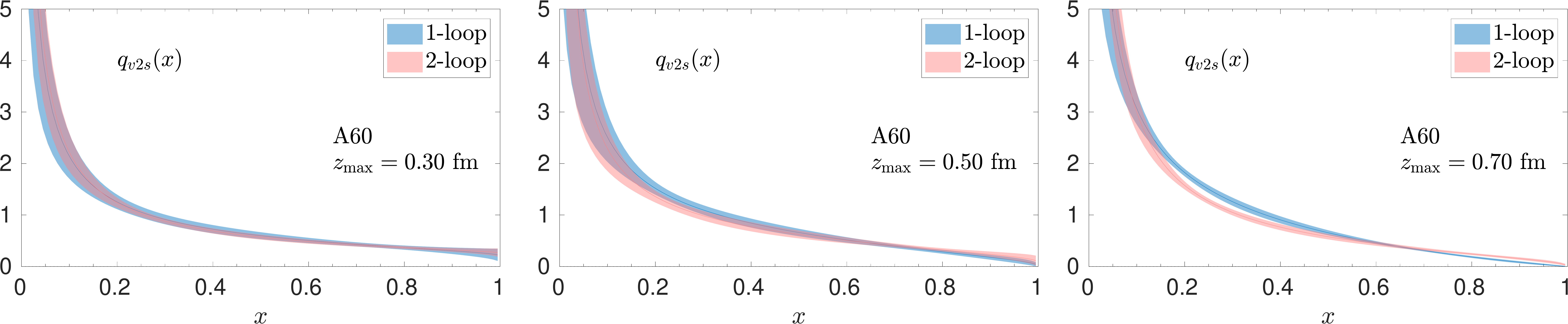}
    \includegraphics[width=\textwidth]{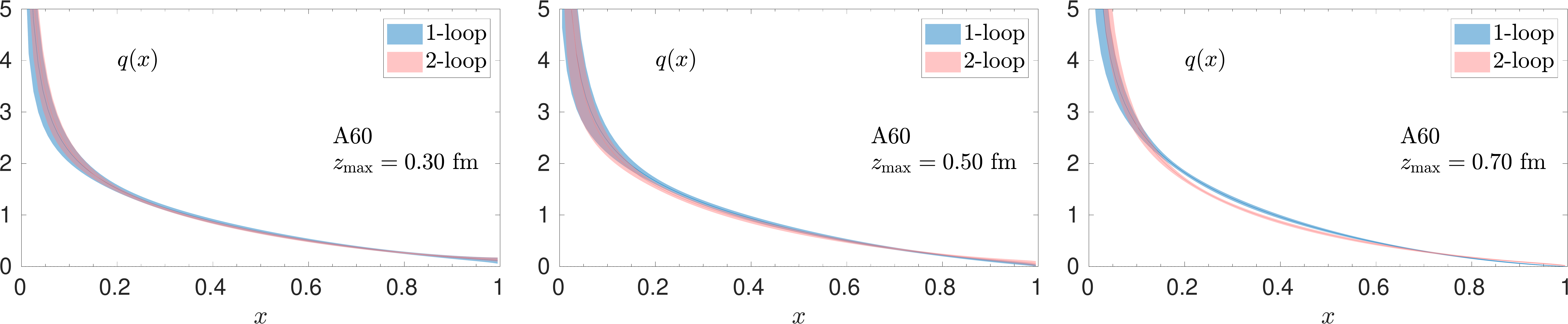}
    \includegraphics[width=\textwidth]{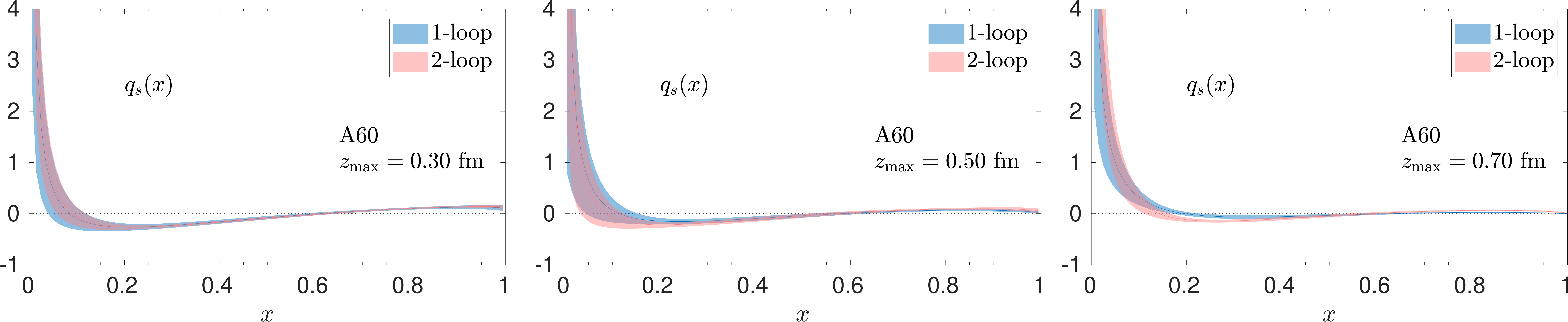}
\end{center}
\vspace*{-0.5cm} 
\caption{PDFs for ensemble A60 from fitting ansatz reconstruction. From top to bottom: $q_v$, $q_{v2s}=q_v+2\bar{q}$, $q=q_v+\bar{q}$, $q_s=\bar{q}$. The left/middle/right column shows results with $\zmax=0.3/0.5/0.7$ fm, respectively. Results from one- and two-loop matching are compared in each plot.}
\label{fig:A60}
\end{figure*}

\begin{figure*}[t!]
\begin{center}
    \includegraphics[width=\textwidth]{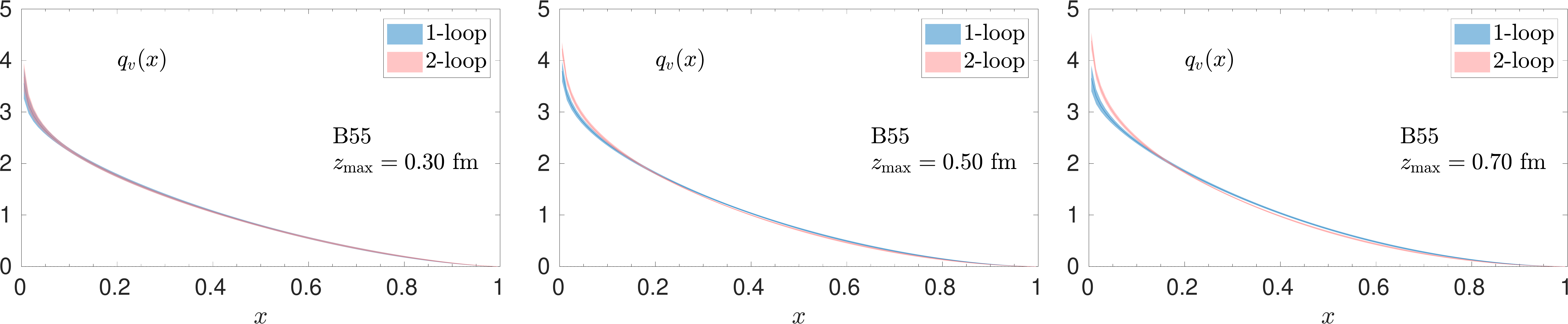}    
    \includegraphics[width=\textwidth]{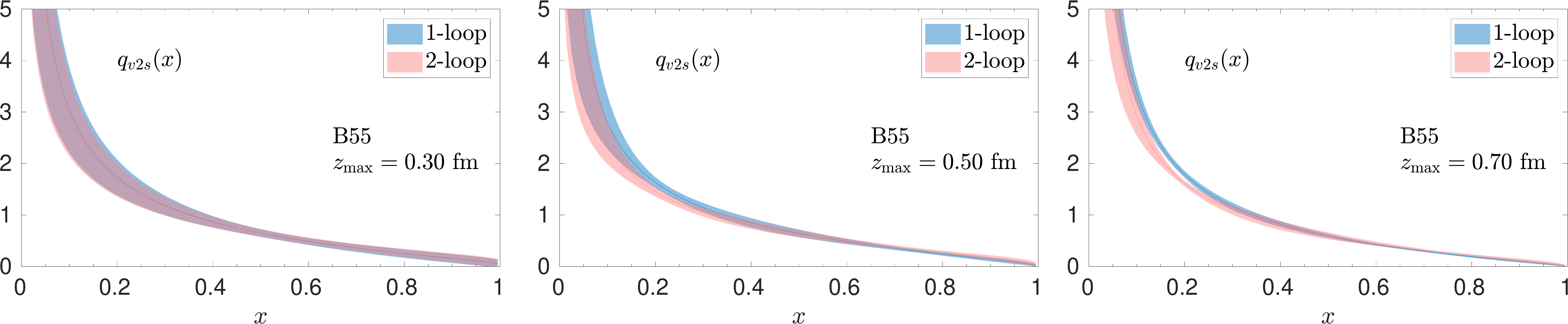}
    \includegraphics[width=\textwidth]{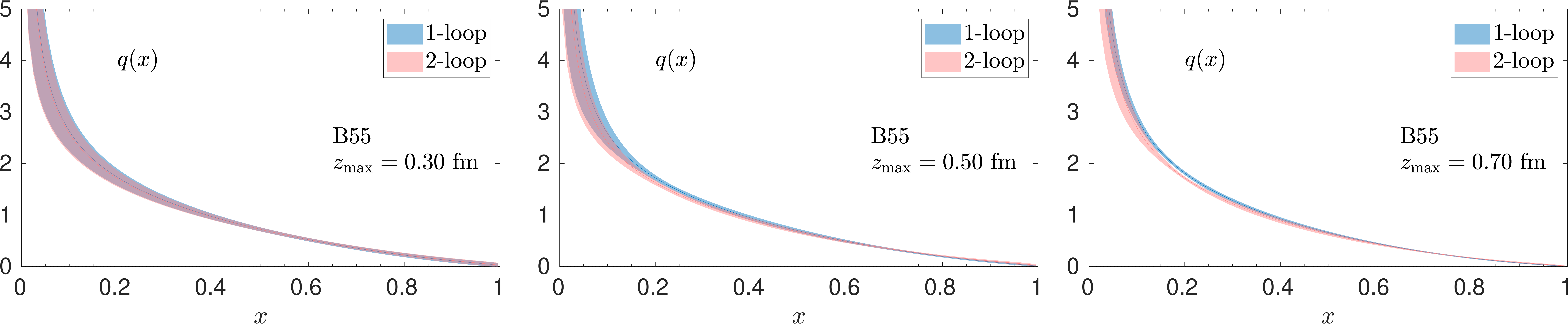}
    \includegraphics[width=\textwidth]{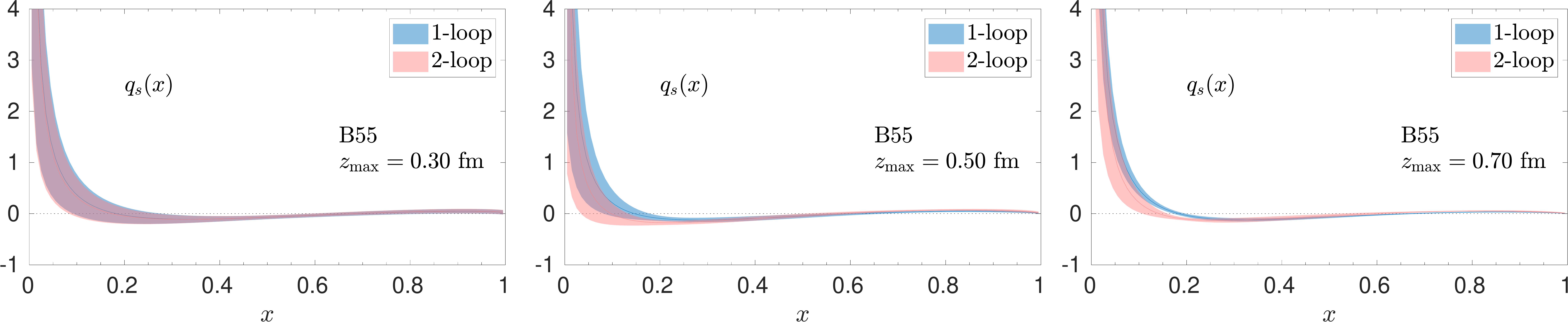}
\end{center}
\vspace*{-0.5cm} 
\caption{PDFs for ensemble B55 from fitting ansatz reconstruction. From top to bottom: $q_v$, $q_{v2s}=q_v+2\bar{q}$, $q=q_v+\bar{q}$, $q_s=\bar{q}$. The left/middle/right column shows results with $\zmax=0.3/0.5/0.7$ fm, respectively. Results from one- and two-loop matching are compared in each plot.}
\label{fig:B55}
\end{figure*}

\begin{figure*}[t!]
\begin{center}
    \includegraphics[width=\textwidth]{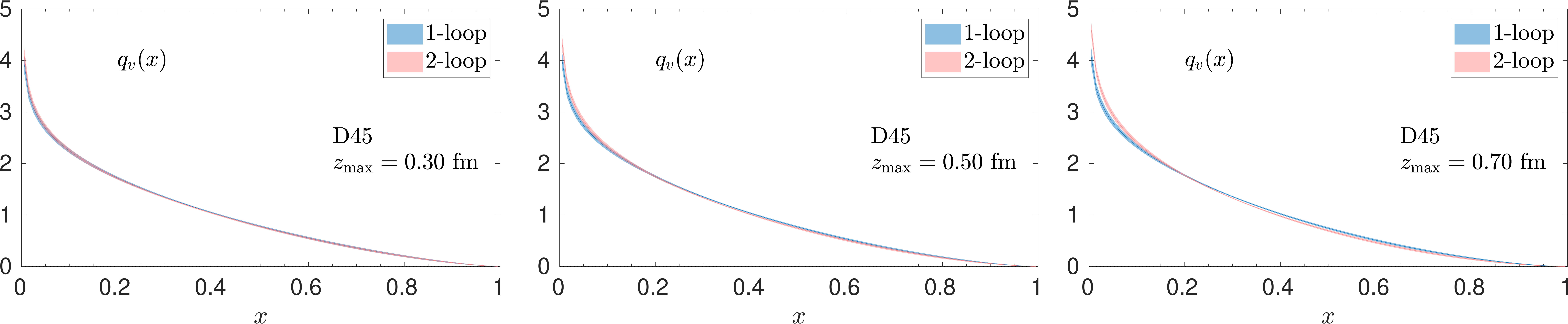}    
    \includegraphics[width=\textwidth]{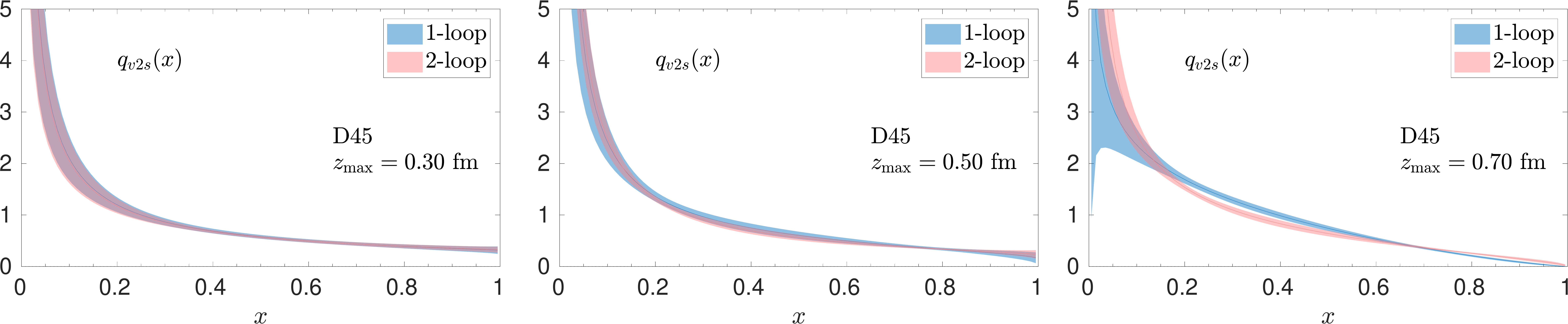}
    \includegraphics[width=\textwidth]{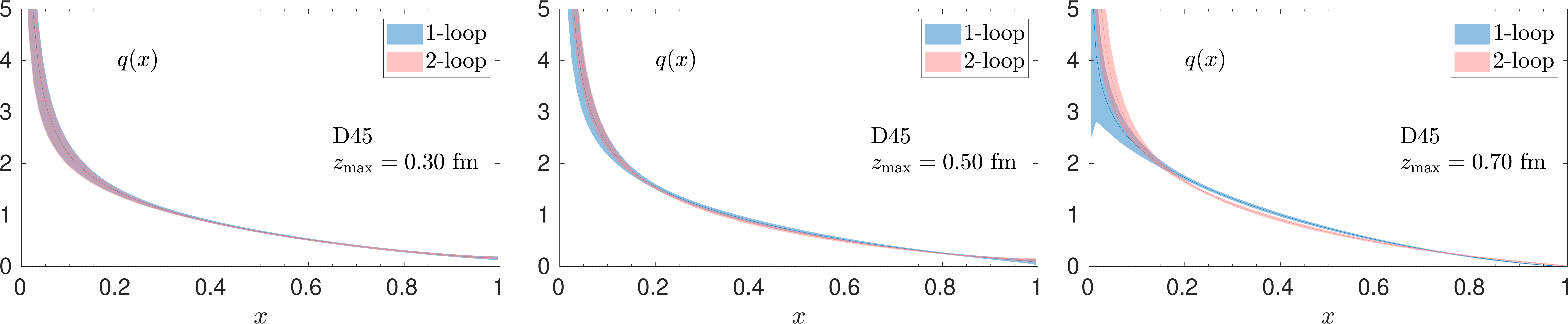}
    \includegraphics[width=\textwidth]{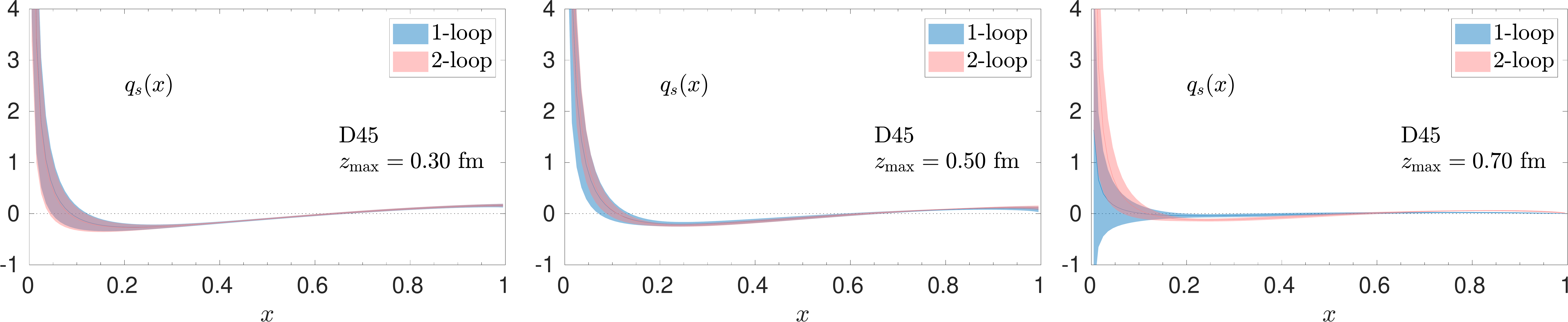}
\end{center}
\vspace*{-0.5cm} 
\caption{PDFs for ensemble D45 from fitting ansatz reconstruction. From top to bottom: $q_v$, $q_{v2s}=q_v+2\bar{q}$, $q=q_v+\bar{q}$, $q_s=\bar{q}$. The left/middle/right column shows results with $\zmax=0.3/0.5/0.7$ fm, respectively. Results from one- and two-loop matching are compared in each plot.}
\label{fig:D45}
\end{figure*}

\subsection{PDFs}
Now, we move on to reconstruction of PDFs from our matched ITDs.
The criterion of restricting lattice data to MEs including only rather low Wilson line lengths has an important consequence for the choice of the preferred reconstruction method.
Namely, ITDs restricted to these small $z$'s, combined with our nucleon boosts of up to around 1.8 GeV, allow us to explore Ioffe time range of up to e.g.\ around 4.7 at $\zmax=0.5$ fm.
As implied by Fig.~\ref{fig:matched_nuavg}, this means that neither the real nor the imaginary part of ITDs has yet decayed to zero. In the naive Fourier reconstruction method, this leads to a sharp cutoff of ITDs, which are taken as identically equal to zero for larger Ioffe times.
The BG approach addresses the inverse problem with a model-independent mathematical assumption, but does not provide the data missing beyond $\zmax$ either.
In turn, the fitting ansatz reconstruction supplements the data by assuming a certain model parametrization reflecting the expected small- and large-$x$ behavior, which implicitly models the missing large-$\nu$ region of the ITD.
While the implied model dependence is non-ideal from the point of view of achieving genuine first-principle results, in practice it is inevitable at this stage of lattice calculations.
The conclusion that the quality of lattice data needs to improve is well-known, see e.g.~Refs.~\cite{Cichy:2018mum,Cichy:2021lih} for extensive discussions, and it should be understood as being able to obtain robust data at larger nucleon boosts.
The latter is essential in both quasi- and pseudo-distribution approaches, by allowing these methods to make reliable contact with the light-cone frame and/or exploring the full range of Ioffe times.

Thus, we first present our PDF reconstructions from fitting, at the level of separate ensembles.
All four kinds of considered PDFs are shown in Figs.~\ref{fig:A60},\ref{fig:B55},\ref{fig:D45}, for ensembles A60, B55 and D45, respectively.
In each plot, we compare results from one- and two-loop matching and we show $\zmax=0.3$ fm (left columns), $\zmax=0.5$ fm (middle columns) and $\zmax=0.7$ fm (right columns).
Starting with the valence distribution, $q_v$, we note that it is reconstructed with very good statistical precision, with errors of order 1-2\% for a wide kinematic range.
Consistently with the size of the two-loop correction up to $z\approx0.5$ fm, the two-loop-matched PDFs are consistent with their one-loop counterparts for $\zmax=0.3$ fm and $0.5$ fm, with the two-loop correction visible at $\zmax=0.7$ fm and small $x\lesssim0.1$.
Distributions involving the imaginary part of ITDs, employing 3-parameter fits, have larger relative errors.
The additional fitting parameter, the normalization $N$, produces a less-constrained model, leading to much worse precision of the extracted PDFs, with errors for a large range of $x$ at the level of 10-20\% for the case of $q_{v2s}$, 5-10\% for $q$ and over 25\% for the suppressed $\bar{q}$.
The two-loop correction affects a somewhat wider $x$-range, $0.2\lesssim x\lesssim0.4$, with the effect at smaller $x$ obscured by the large errors.

\begin{figure*}[t!]
\begin{center}
    \includegraphics[width=\textwidth]{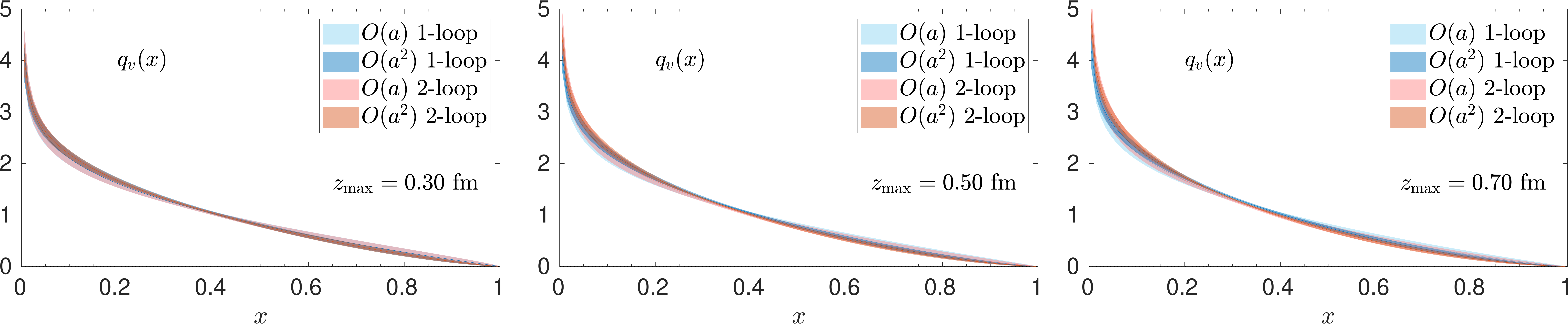}    
    \includegraphics[width=\textwidth]{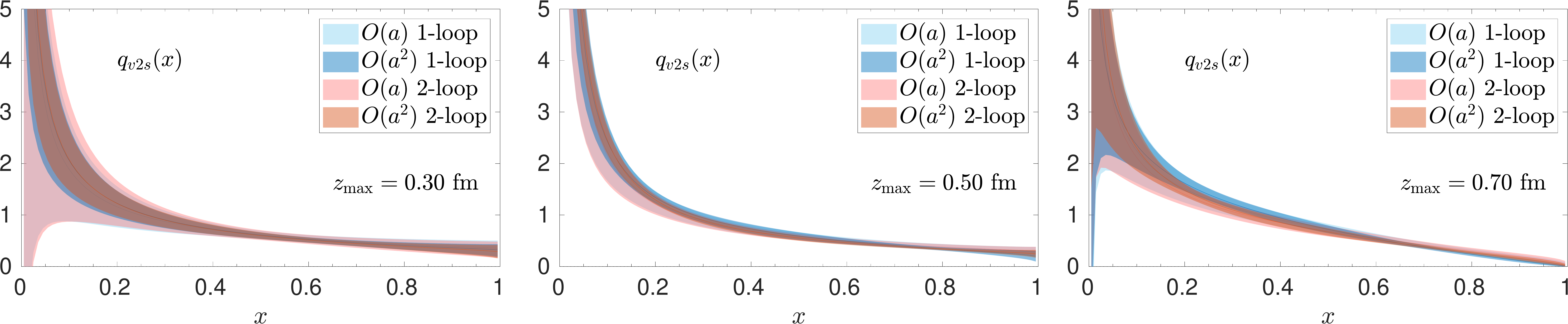}
    \includegraphics[width=\textwidth]{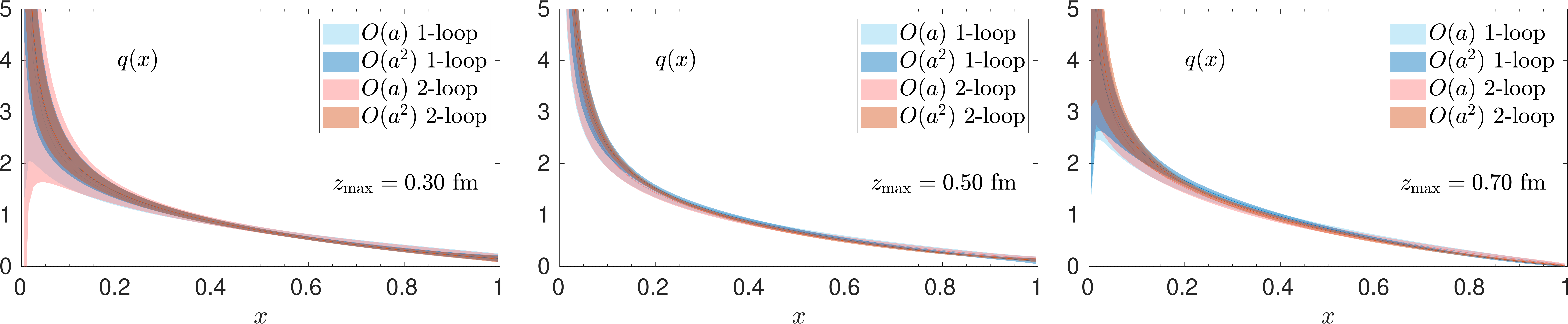}
    \includegraphics[width=\textwidth]{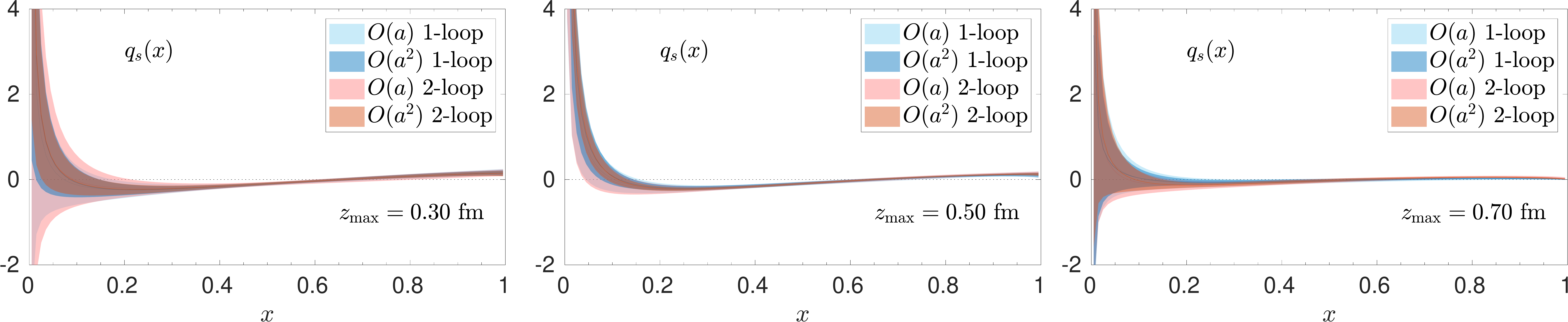}
\end{center}
\vspace*{-0.5cm} 
\caption{Continuum-extrapolated PDFs from fitting ansatz reconstruction. From top to bottom: $q_v$, $q_{v2s}=q_v+2\bar{q}$, $q=q_v+\bar{q}$, $q_s=\bar{q}$. The left/middle/right column shows results with $\zmax=0.3/0.5/0.7$ fm, respectively. Results from one- and two-loop matching as well as from $\mathcal{O}(a)$ and $\mathcal{O}(a^2)$ extrapolations are compared in each plot.}
\label{fig:cont}
\end{figure*}

\begin{figure*}[p!]
\begin{center}
    \includegraphics[width=0.49\textwidth]{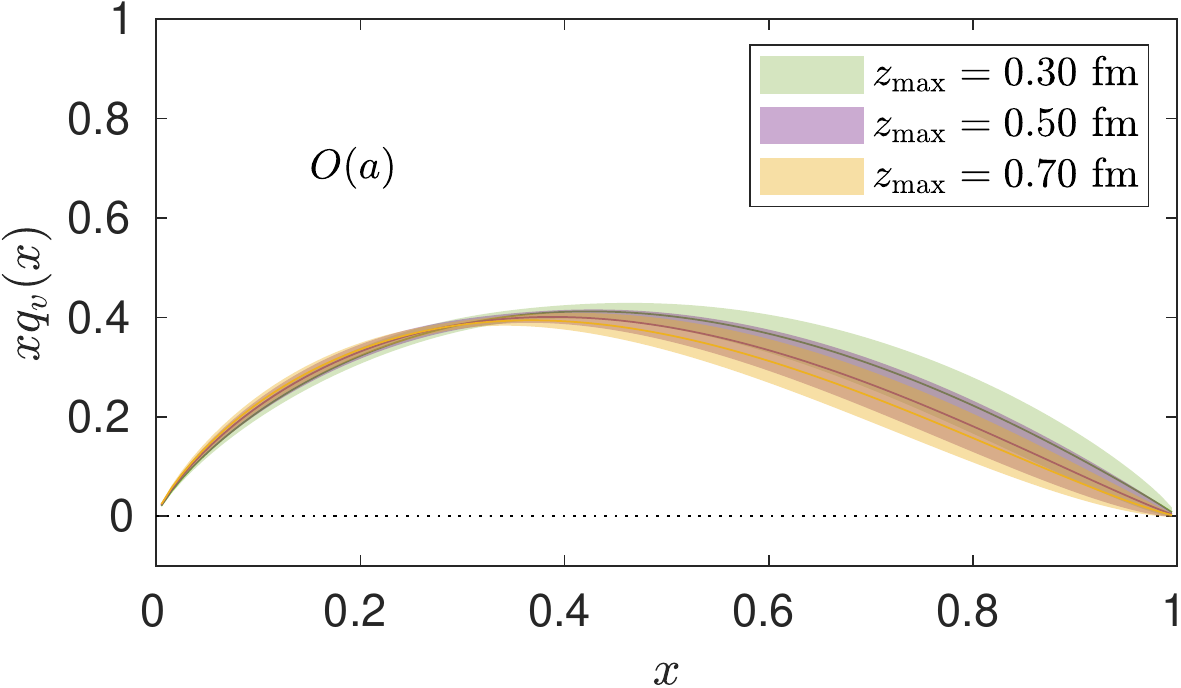}
    \includegraphics[width=0.49\textwidth]{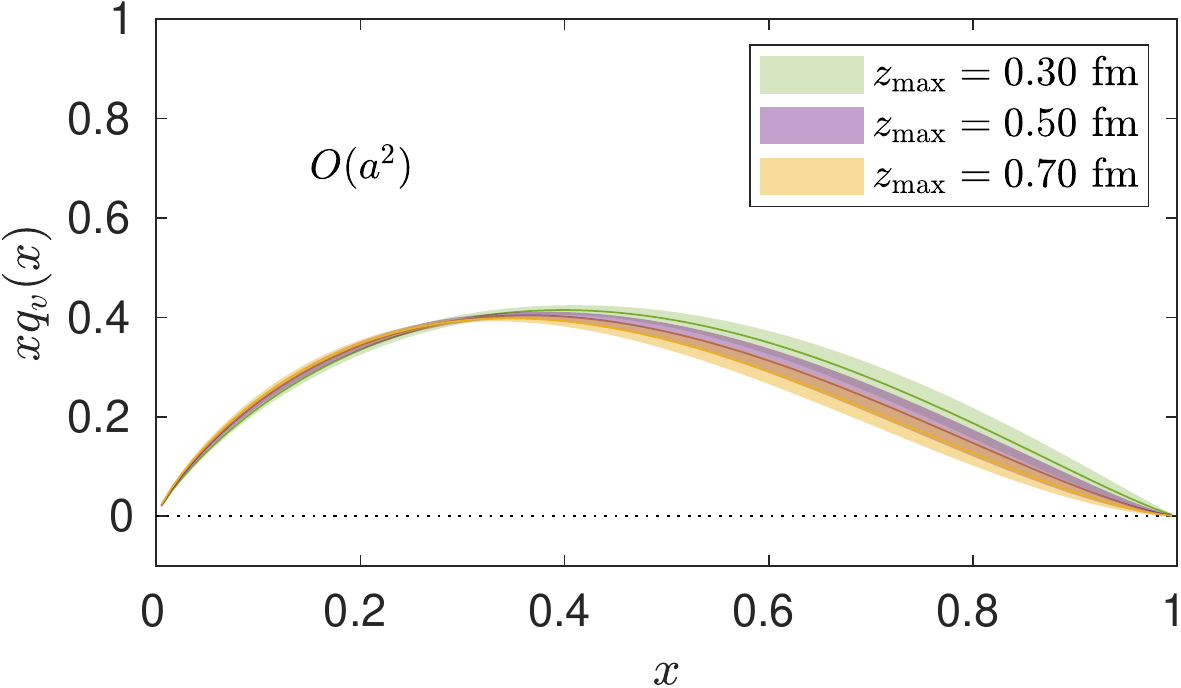}    
    \includegraphics[width=0.49\textwidth]{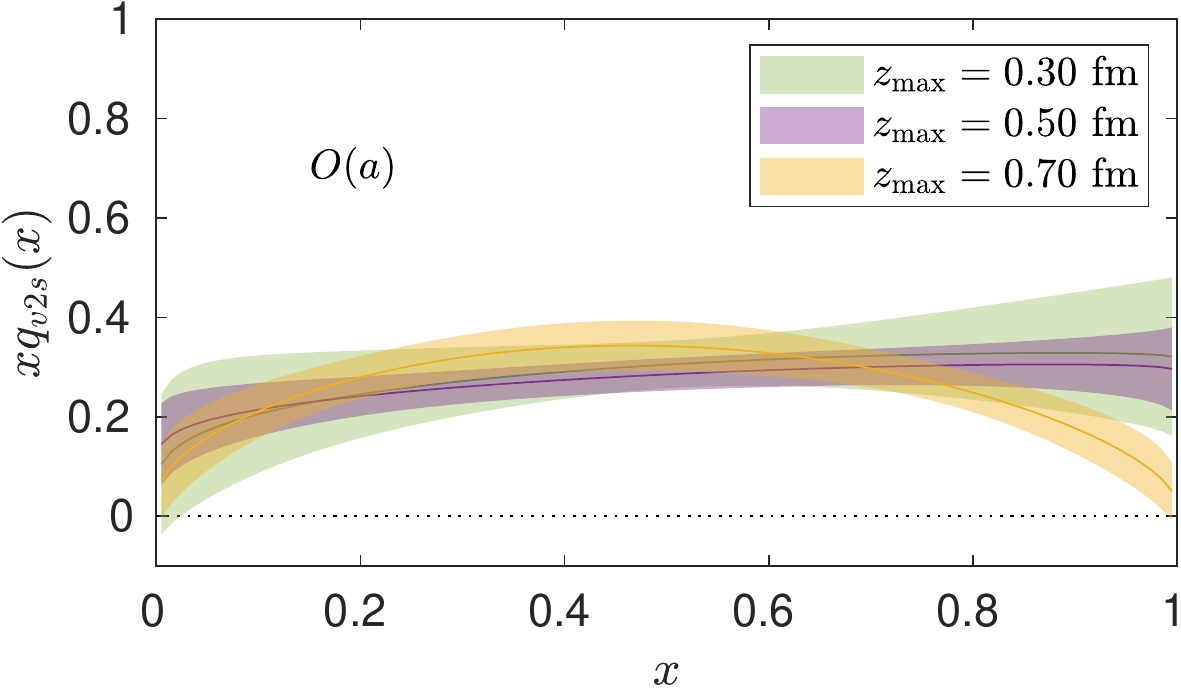}
    \includegraphics[width=0.49\textwidth]{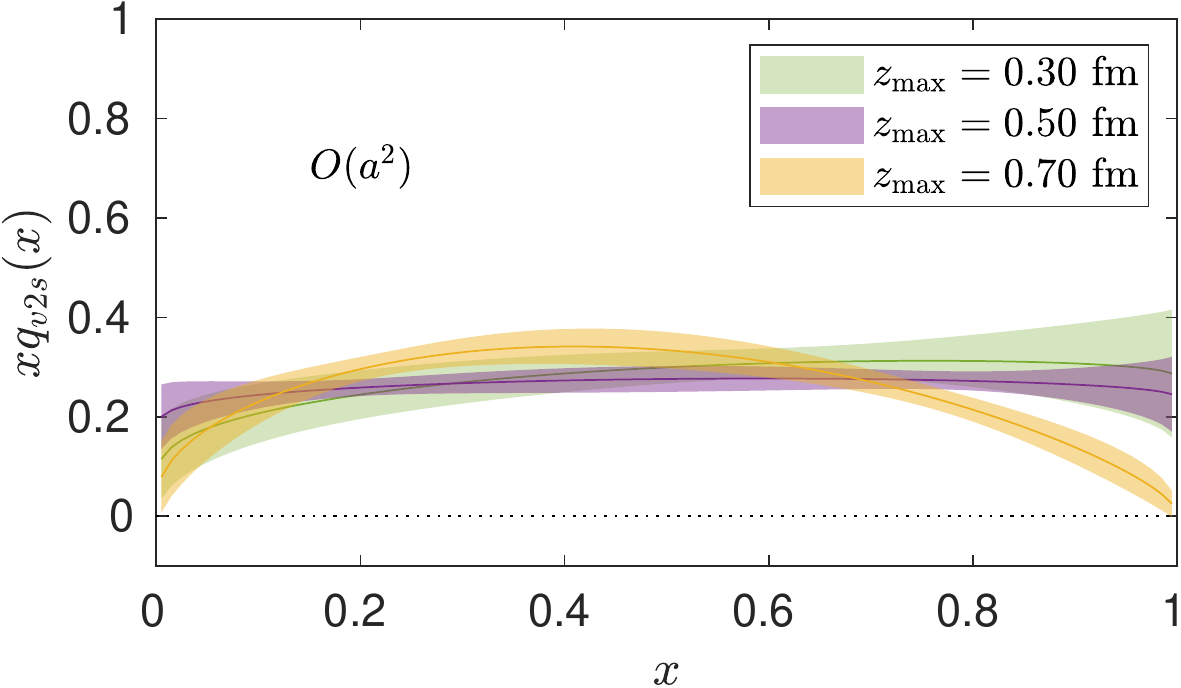}
    \includegraphics[width=0.49\textwidth]{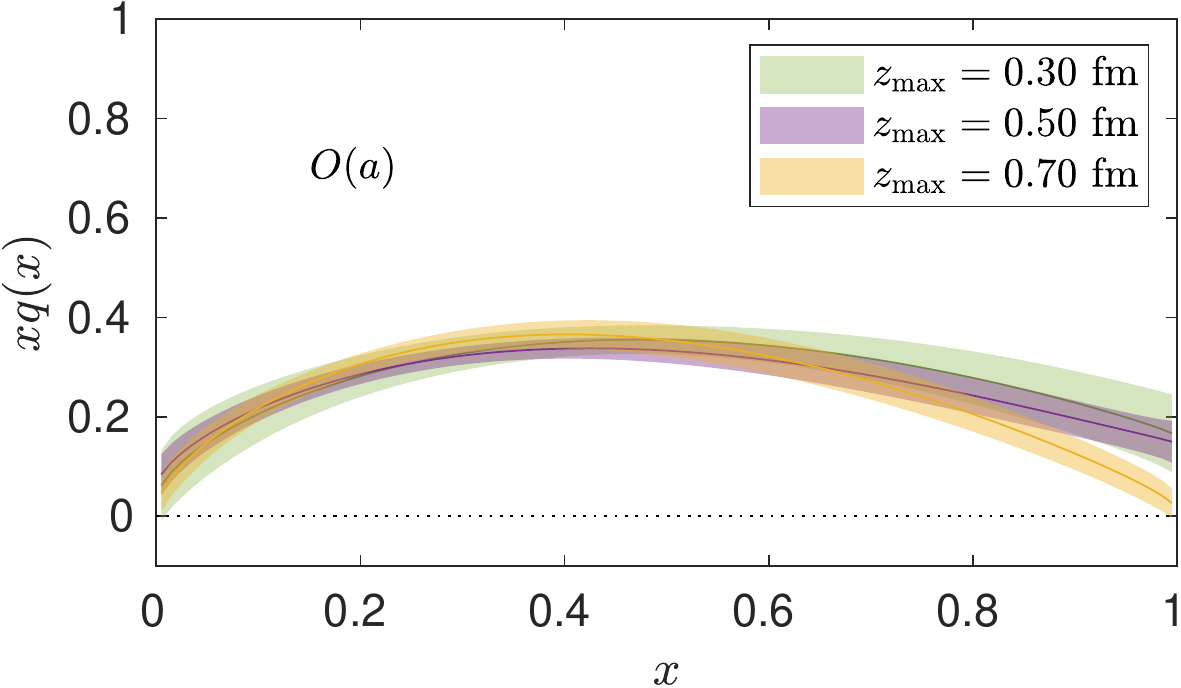}
    \includegraphics[width=0.49\textwidth]{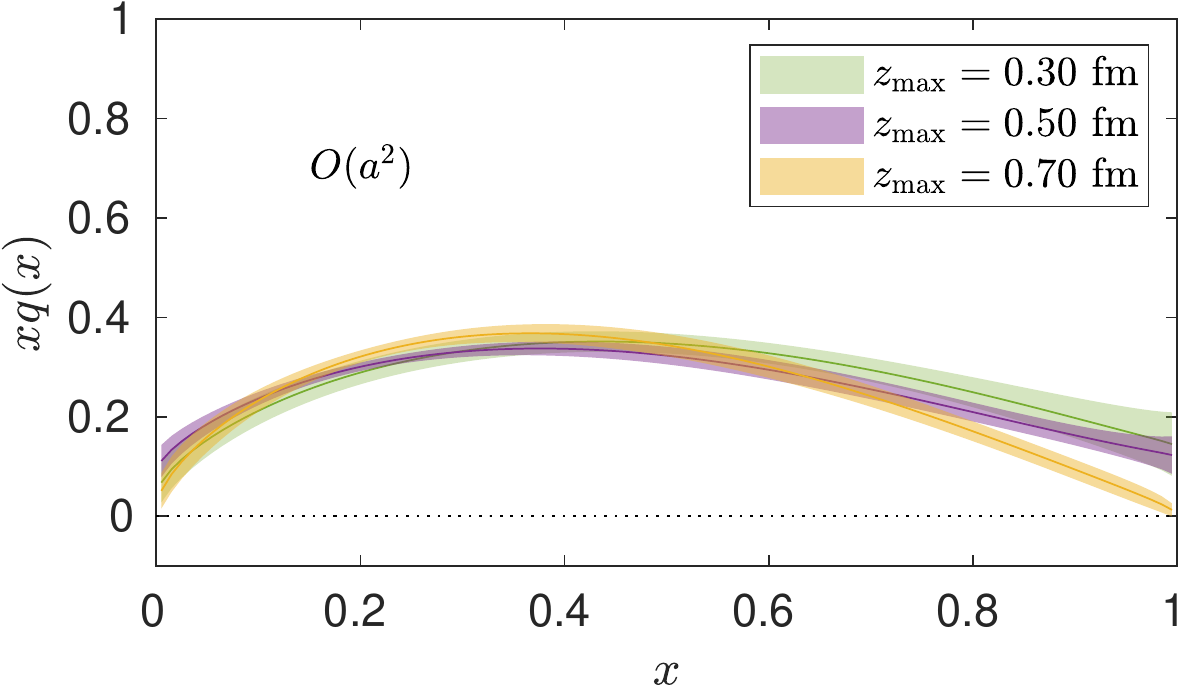}
    \includegraphics[width=0.49\textwidth]{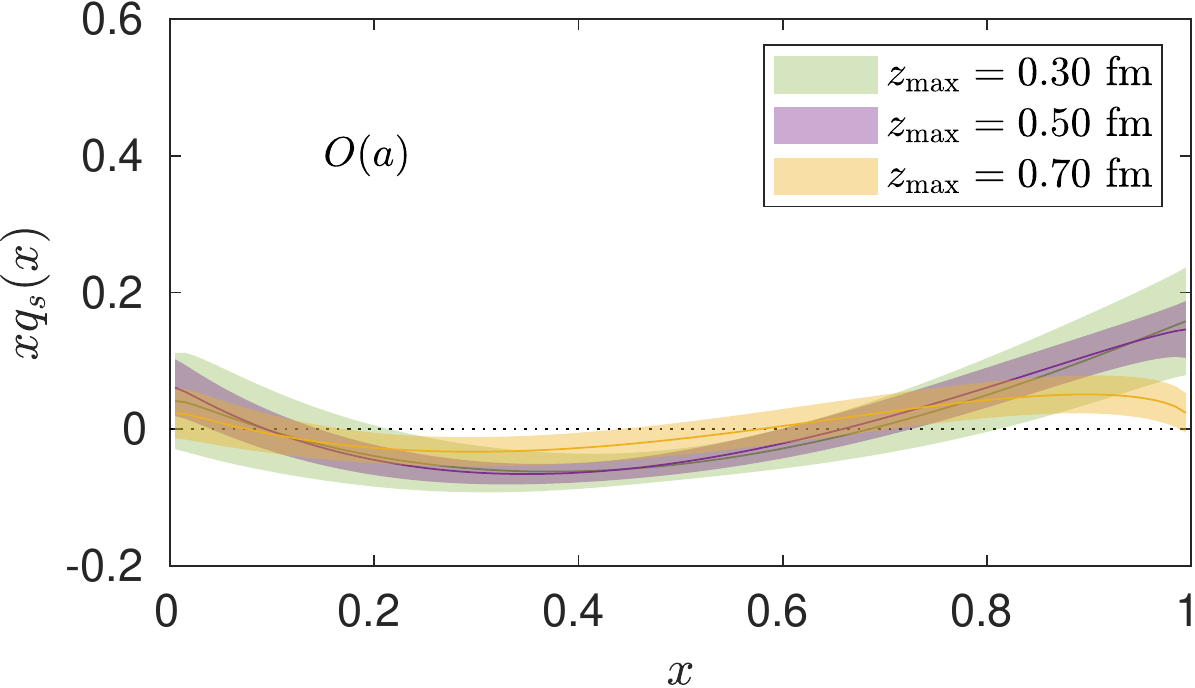}
    \includegraphics[width=0.49\textwidth]{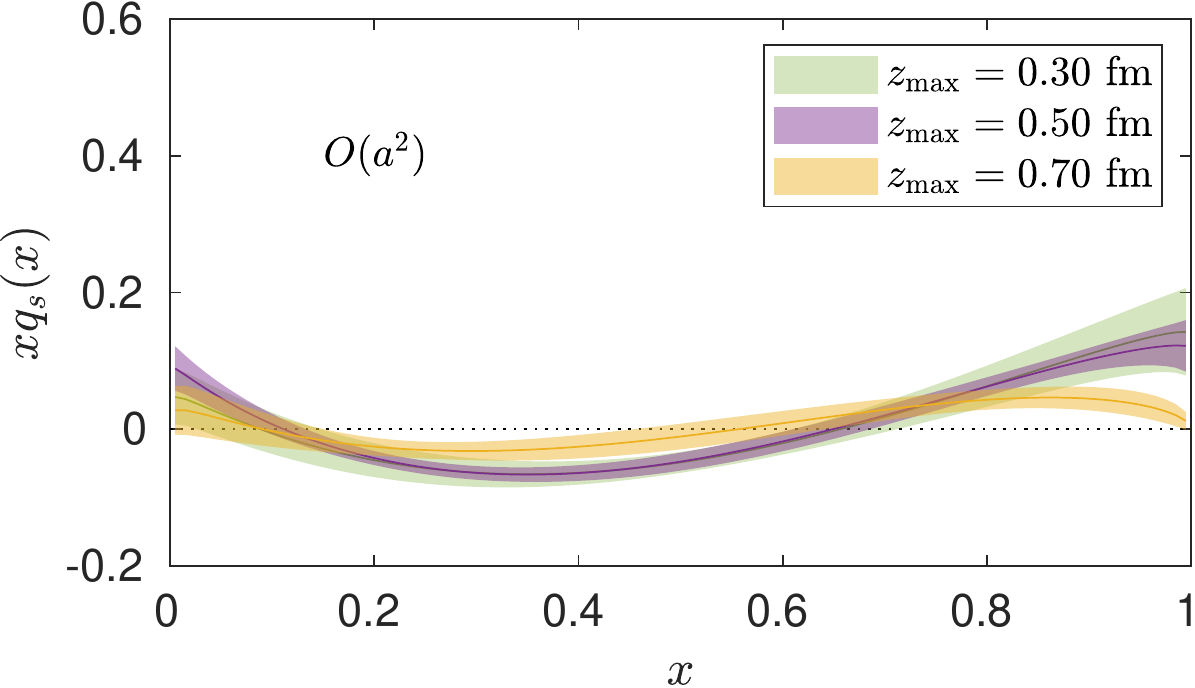}
\end{center}
\vspace*{-0.5cm} 
\caption{Comparison of continuum-extrapolated PDFs from fitting ansatz reconstruction with three $\zmax$ values, $0.3,\,0.5,\,0.7$ fm. From top to bottom: $q_v$, $q_{v2s}=q_v+2\bar{q}$, $q=q_v+\bar{q}$, $q_s=\bar{q}$. The left/right column shows results with $\mathcal{O}(a)$/$\mathcal{O}(a^2)$ extrapolation. Results are plotted for PDFs multiplied by $x$ in order to better visualize differences at large $x$.}
\label{fig:contzmax}
\end{figure*}

\begin{table*}[t!]
\begin{center}
\begin{tabular}{|c|c|c|c|c|c|c|c|c|c|c|c|c|c|}
\hline
\multirow{3}{*}{$\zmax$ [fm]} & \multicolumn{4}{c|}{A60} & \multicolumn{4}{c|}{B55} & \multicolumn{4}{c|}{D45} \\
\cline{2-13}
 & \multicolumn{2}{c|}{$q_v$} & \multicolumn{2}{c|}{$q_{v2s}$} & \multicolumn{2}{c|}{$q_v$} & \multicolumn{2}{c|}{$q_{v2s}$} & \multicolumn{2}{c|}{$q_v$} & \multicolumn{2}{c|}{$q_{v2s}$} \\
\cline{2-13}
& $\alpha$ & $\beta$ & $\alpha$ & $\beta$ & $\alpha$ & $\beta$ & $\alpha$ & $\beta$ & $\alpha$ & $\beta$ & $\alpha$ & $\beta$ \\
\hline
0.3 & -0.136(19) & 1.58(9) & -0.81(20) & 0.06(17) & -0.112(27) & 1.49(8) & -0.65(27) & 0.71(50) & -0.157(14) & 1.40(8) & -0.78(16) & 0.01(7) \\
0.5 & -0.135(15) & 1.72(9) & -0.66(29) & 0.37(31) & -0.126(11) & 1.76(7) & -0.60(27) & 0.59(31) & -0.157(11) & 1.54(8) & -0.89(20) & 0.09(21) \\
0.7 & -0.145(13) & 1.86(9) & -0.82(20) & 0.53(20) & -0.126(13) & 1.88(8) & -0.72(26) & 0.72(31) & -0.159(8) & 1.68(8) & -0.61(21) & 0.69(22) \\
\hline
\end{tabular}
\caption{Parameters $\alpha$, $\beta$ of the fitting ansatz (\ref{eq:ansatz}) for the three separate ensembles; reconstruction of $q_v$ and $q_{v2s}$ with three values of $\zmax$.}
\label{tab:fitparams}
\end{center}
\end{table*}

\begin{table*}[t!]
\begin{center}
\begin{tabular}{|c|c|c|c|c|c|c|c|c|}
\hline
\multirow{3}{*}{$\zmax$ [fm]} & \multicolumn{4}{c|}{cont.\ $\mathcal{O}(a)$} & \multicolumn{4}{c|}{cont.\ $\mathcal{O}(a^2)$}\\
\cline{2-9}
  & \multicolumn{2}{c|}{$q_v$} & \multicolumn{2}{c|}{$q_{v2s}$} & \multicolumn{2}{c|}{$q_v$} & \multicolumn{2}{c|}{$q_{v2s}$}\\
\cline{2-9}
 & $\alpha$ & $\beta$ & $\alpha$ & $\beta$ & $\alpha$ & $\beta$ & $\alpha$ & $\beta$\\
\hline
0.3 & -0.195(42) & 1.10(25) & -0.60(28) & 0.24(78) & -0.167(31) & 1.26(16) & -0.70(24) & 0.12(29)\\
0.5 & -0.199(32) & 1.27(27) & -0.79(21) & 0.04(15) & -0.172(21) & 1.45(16) & -0.90(17) & 0.07(18)\\
0.7 & -0.203(26) & 1.38(28) & -0.45(33) & 0.63(34) & -0.176(17) & 1.56(16) & -0.49(26) & 0.75(29)\\
\hline
\end{tabular}
\caption{Parameters $\alpha$, $\beta$ of the fitting ansatz (\ref{eq:ansatz}) for the $\mathcal{O}(a)$ and $\mathcal{O}(a^2)$ continuum extrapolations; reconstruction of $q_v$ and $q_{v2s}$ with three values of $\zmax$.}
\label{tab:fitparams2}
\end{center}
\end{table*}

\begin{figure*}[t!]
\begin{center}
    \includegraphics[width=\textwidth]{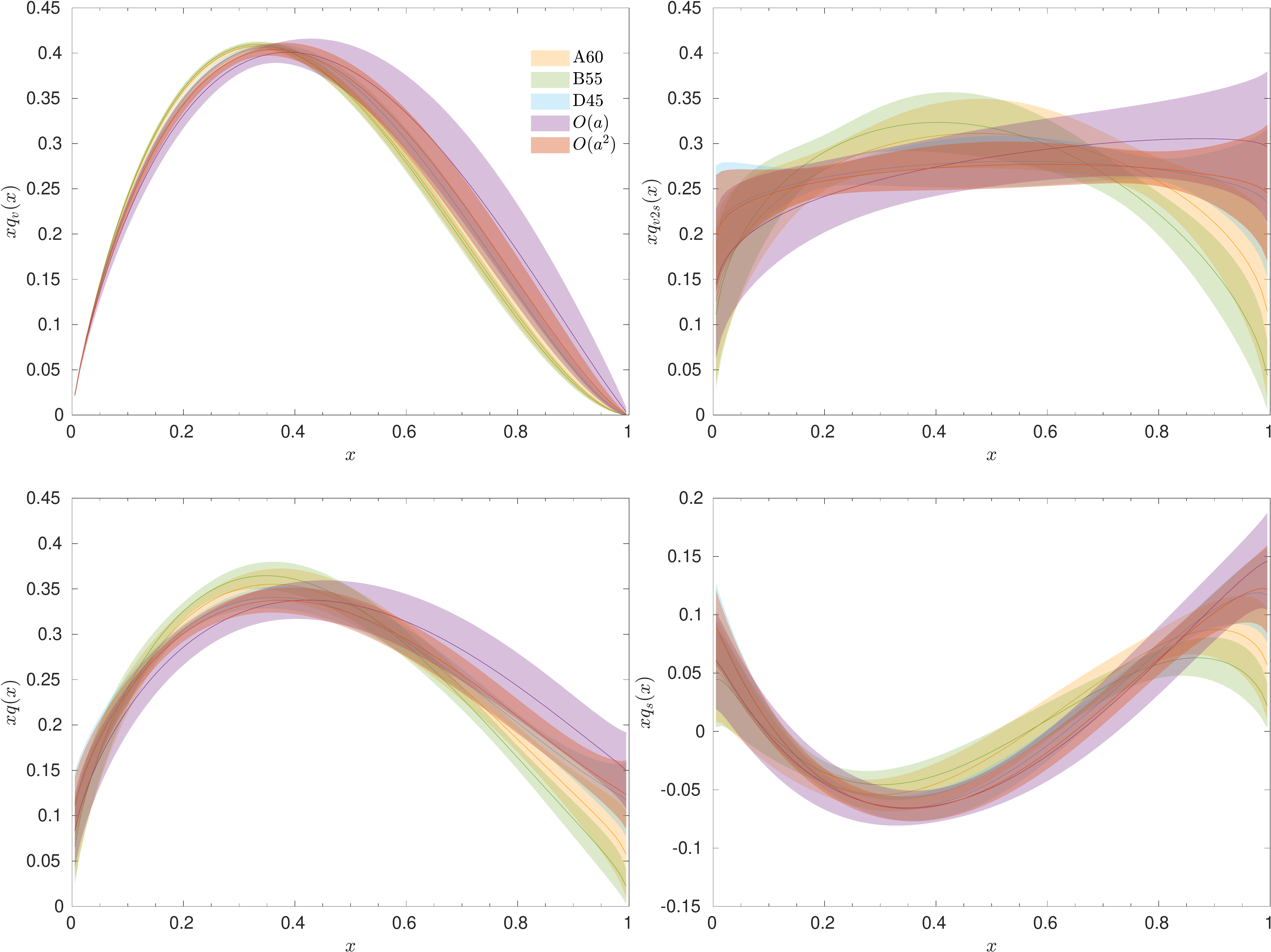}        
\end{center}
\vspace*{-0.5cm} 
\caption{PDFs from separate ensembles together with their $\mathcal{O}(a)$ and $\mathcal{O}(a^2)$ continuum limits: $q_v$ (top left), $q_{v2s}=q_v+2\bar{q}$ (top right), $q=q_v+\bar{q}$ (bottom left), $q_s=\bar{q}$ (bottom right), $\zmax=0.5$ fm. Results are plotted for PDFs multiplied by $x$ in order to better visualize differences at large $x$.}
\label{fig:cont2}
\end{figure*}

\begin{figure*}[t!]
\begin{center}
    \includegraphics[width=\textwidth]{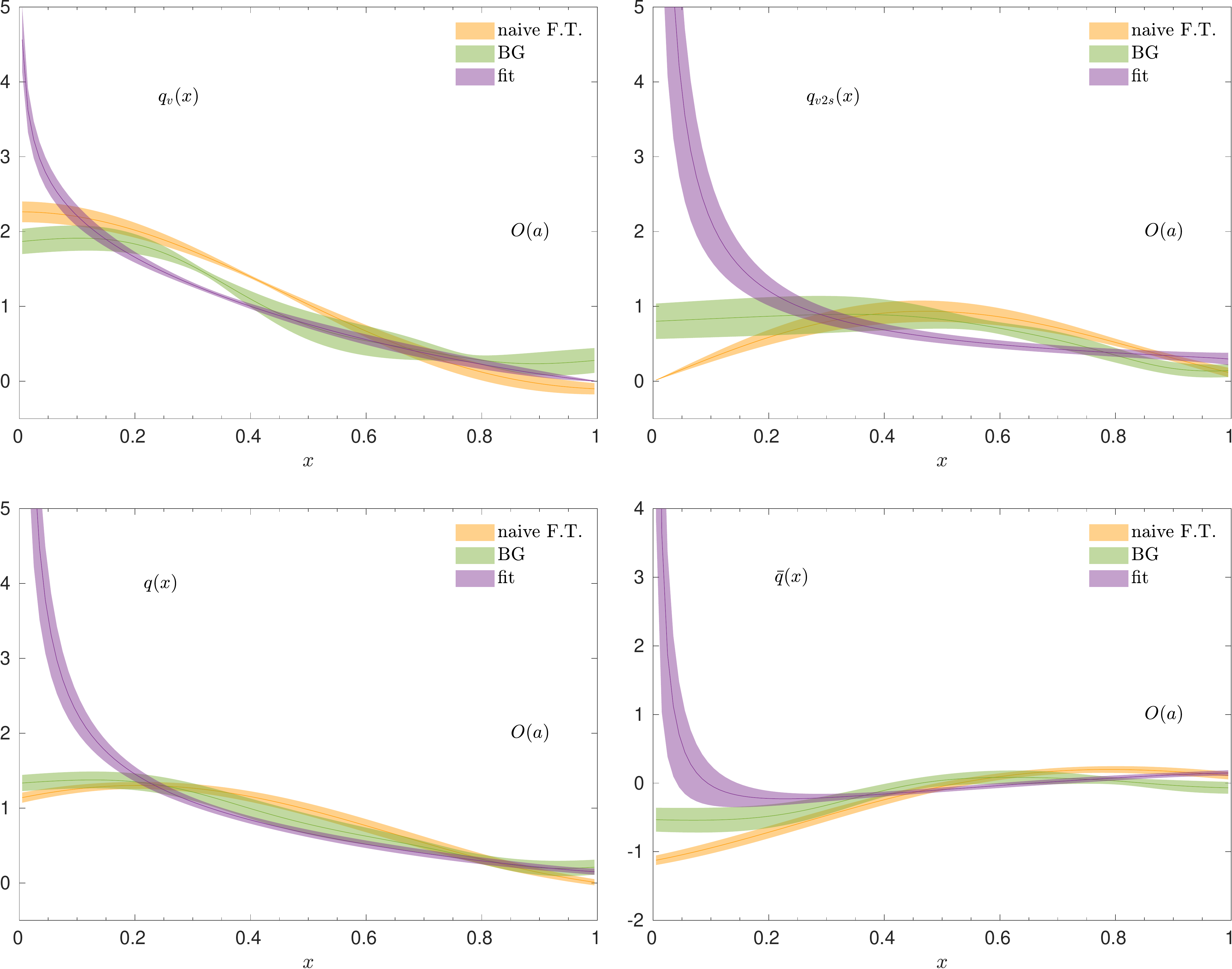}        
\end{center}
\vspace*{-0.5cm} 
\caption{Comparison of reconstruction methods for the continuum PDFs: $q_v$ (top left), $q_{v2s}=q_v+2\bar{q}$ (top right), $q=q_v+\bar{q}$ (bottom left), $q_s=\bar{q}$ (bottom right). Shown are naive Fourier transform reconstruction (orange), the Backus-Gilbert method (green) and the fitting ansatz reconstruction (purple). In all cases, $\mathcal{O}(a)$ continuum limit extrapolation from two-loop-matched ITDs is used, with $\zmax=0.5$ fm.}
\label{fig:reco}
\end{figure*}

In Fig.~\ref{fig:cont}, we show the reconstructed distributions in the continuum limit.
The continuum extrapolation is performed at the level of ITDs, thus matched continuum-extrapolated ITDs are here input to the fitting reconstruction procedure.
The inflation of errors in this extrapolation obscures any differences between one- and two-loop-matched PDFs, even at $\zmax=0.7$ fm.
In turn, the flatness of this extrapolation implies no statistically significant difference between PDFs obtained with $\mathcal{O}(a)$ and $\mathcal{O}(a^2)$ continuum fitting ansatzes.

To understand the role of $\zmax$, we take a closer look into the $\zmax$-dependence of the continuum-extrapolated, two-loop-matched PDFs.
To better illustrate the differences appearing in the large-$x$ region, we plot the PDFs multiplied by $x$ in Fig.~\ref{fig:contzmax}.
In the case of the valence distribution, PDFs reconstructed with all values of $\zmax$ are compatible with each other.
The regime of Ioffe times probed with growing $\zmax$ increases, but this has an effect only of decreasing the error in the small- to intermediate-$x$ region, without generating tension in any regime of $x$.
As we argued above, ITDs corresponding to $z>0.5$ fm may have uncontrolled HTEs and thus, the error estimate implied by $\zmax=0.5$ fm should be taken as the most reliable.
The $\zmax$-dependence in distributions involving the imaginary part of ITDs is markedly different.
While $\zmax=0.3$ fm and $\zmax=0.5$ fm cases are compatible with each other, the additional ITDs from larger Wilson line lengths influence the fits significantly.
The origin of this behavior is rather clear at the level of imaginary part of matched ITD -- $\zmax=0.5$ fm allows one to reach $\nu\approx4.7$ at $P_3\approx1.8$ GeV and the ITD reaches its maximal value around this Ioffe time.
Thus, there is considerable part of information missing on the underlying PDFs if one disregards large $z$'s for which no reliable contact can be made with the light-cone frame.
The plots of $q_{v2s}$, $q$ and $\bar{q}$ have another striking feature indicating that the reconstruction is not robust with $\zmax\leq0.5$ fm.
Namely, the PDFs are non-zero at $x=1$.
This is seemingly in contradiction with the fitting ansatz that includes the factor $(1-x)^\beta$.
However, a large subset of bootstrap samples in the fits of the imaginary part of ITDs favors a zero value for the fitting coefficient $\beta$ and thus, a non-zero value of $q_{v2s}(x=1)$.
Most likely, this is again related to reaching only the region of the maximum of Im$\,Q(\nu)$, which introduces a bias into the reconstructed PDFs.
This is indicated by the well-behaved case of $\zmax=0.7$ fm, where a clearly non-zero value of $\beta$ is preferred for all bootstrap samples.
However, distributions involving scales far beyond the perturbative regime need to be interpreted with care.
In practice, a robust reconstruction of the distributions $q_{v2s}$, $q$ and $\bar{q}$ will only be possible if the range of reliably probed Ioffe times is extended by accessing them with larger nucleon boosts and smaller values of $z$.

We provide all values of our fitting parameters of the ansatz (\ref{eq:ansatz}) in Tab.~\ref{tab:fitparams} (for separate ensembles) and Tab.~\ref{tab:fitparams2} (for continuum-extrapolated data).
Some tendencies can be observed when varying $\zmax$.
For both $q_v$ and $q_{v2s}$, $\alpha$ is largely independent of $\zmax$, while for $\beta$, there is some tendency towards its larger value when increasing $\zmax$.
The latter is particularly obvious for $q_{v2s}$, as discussed in the previous paragraph -- $\beta_{q_{v2s}}$ is consistent with zero for the two lower $\zmax$ values in continuum fits (as well as ones for the finest lattice spacing), while the additional ITD data when extending to \mbox{$\zmax=0.7$ fm} favor $\beta>0$ for all bootstrap samples.

We conclude the discussion about the $\zmax$-dependence by spelling out our choice of its preferred value.
In Fig.~\ref{fig:matched}, we have observed that ITDs pertaining to the same Ioffe time, but originating from different $(P_3,z)$ combinations, start to differ when $z$ exceeds $0.5$ fm for the real part and already $0.3$ fm for the imaginary part.
However, this observation is valid at the level of separate ensembles.
Since the continuum extrapolation inflates the errors significantly, by a factor 3-5, the criterion of compatibility of ITDs from different $(P_3,z)$ pairs can be relaxed to correspond with these increased errors.
In other words, the plausible value of $\zmax$ is such that different $(P_3,z)$ combinations with the same $P_3z$ lead to consistent ITDs in the continuum limit.
In practice, this amounts to ITDs from such combinations differing by less than around 3-$\sigma$ at the level of separate ensembles.
We apply this relaxed criterion only to the imaginary part and it allows us to justify $\zmax=0.5$ fm for $q_{v2s}$.
For the real part, we stay more conservative and do not extend $\zmax$ beyond 0.5 fm, which also allows us to have $q$ and $\bar{q}$ with the same universal $\zmax$.

Having chosen our preferred value of $\zmax=0.5$ fm, it is interesting the see the approach of the PDFs to the continuum limit at this $\zmax$, by plotting the PDFs from the separate ensembles together with their continuum limits (with two-loop matching), see Fig.~\ref{fig:cont2}.
The reconstructed valence distribution is practically identical from the ensembles with the two coarsest lattice spacings, with the one from D45 slighly below the two at small $x$ and slightly above at large $x$.
The inflated errors of the continuum PDFs, both from $\mathcal{O}(a)$ and $\mathcal{O}(a^2)$ extrapolations, imply that the latter are compatible with D45 and differ from A60/B55 up to a bit above 1-$\sigma$ in some regions of $x$.
Despite the smallness of discretization effects, we observe that the tendency is that they enhance/suppress the valence PDF at small/large $x$.
In the case of $q_{v2s}$, the situation is qualitatively similar, with this PDF being compatible between A60 and B55 and with some tensions of slightly above 1-$\sigma$ in certain $x$-ranges with respect to D45.
The latter is most susceptible to the feature mentioned above, with several bootstrap samples resulting in the vanishing of the fitting parameter $\beta$ and $q_{v2s}(x=1)>0$.
This behavior propagates also to the continuum-limit-extrapolated PDFs.
Overall, with increased errors at the stage of continuum extrapolations, we observe that discretization effects do not play a major role, with some tendencies similar to the ones in $q_v$, of certain suppression of the continuum $q_{v2s}$ at small $x$ and its enhancement at large $x$.
However, again, the latter is more indicative of not probing a large enough region of Ioffe times.
Similar conclusions can be drawn for the two remaining distributions, being linear combinations of $q_v$ and $q_{v2s}$.

Finally, again only for our preferred value of $\zmax=0.5$ fm, we compare the effects of the three reconstruction methods (Fig.~\ref{fig:reco}).
The largest differences between the fitting reconstruction and the two other methods can be seen in the small-$x$ regime.
With our relatively small $\zmax$, the ITDs corresponding to the largest probed Ioffe times, are still non-zero.
The sharp cutoff on the Ioffe time assumed in the naive reconstruction and the BG method translates to an artificially lowered value of the PDFs at small $x$.
In fact, by construction, these methods are unable to produce a divergent behavior as $x\rightarrow0$.
This feature is bypassed in the fitting reconstruction by avoiding the sharp drop of ITD values beyond Ioffe times inaccessible with $\zmax$.
While there is, obviously, no ITD data in this region, the assumption of the fitting ansatz effectively models the large-$\nu$ behavior of ITDs, with the behavior guided by data at smaller Ioffe times.
In the PDFs, this translates to an enhanced error in the small-$x$ region, as this regime of PDFs is comparatively more determined by large-$\nu$ ITDs, even if PDFs at all $x$ receive contributions from all Ioffe times.
It is clear that the errors at small $x$ can only be reduced if data at larger boosts are available.
For larger $x$, there is, in general, rather good agreement between all reconstruction methods, in particular between fitting ansatz reconstruction and BG.
The $\zmax$ cutoff in the naive Fourier transform and BG translates to a mild oscillatory behavior, especially at $x\gtrsim0.6$.
In the end, our preferred reconstruction method is the one involving the fitting ansatz.
While it has the drawback of being model-dependent, this model dependence is naturally reflected in the final errors.
In other words, at the current level of precision, we are not sensitive to corrections to the functional form of the fitting ansatz and thus, we expect that the modelling uncertainty is not significantly larger than our errors.
We view this as a temporary restriction of the approach for two reasons.
Firstly, with increased precision of the data, one can include further fitting parameters of the ansatz, making it more realistic and less model-dependent.
Secondly, if the full range of Ioffe times is probed with sufficient precision, i.e.\ the range of Ioffe times is extended such that ITDs decay to zero, all reconstruction methods should lead to compatible, model-independent results.

\begin{figure*}[p!]
\begin{center}
    \includegraphics[width=\textwidth]{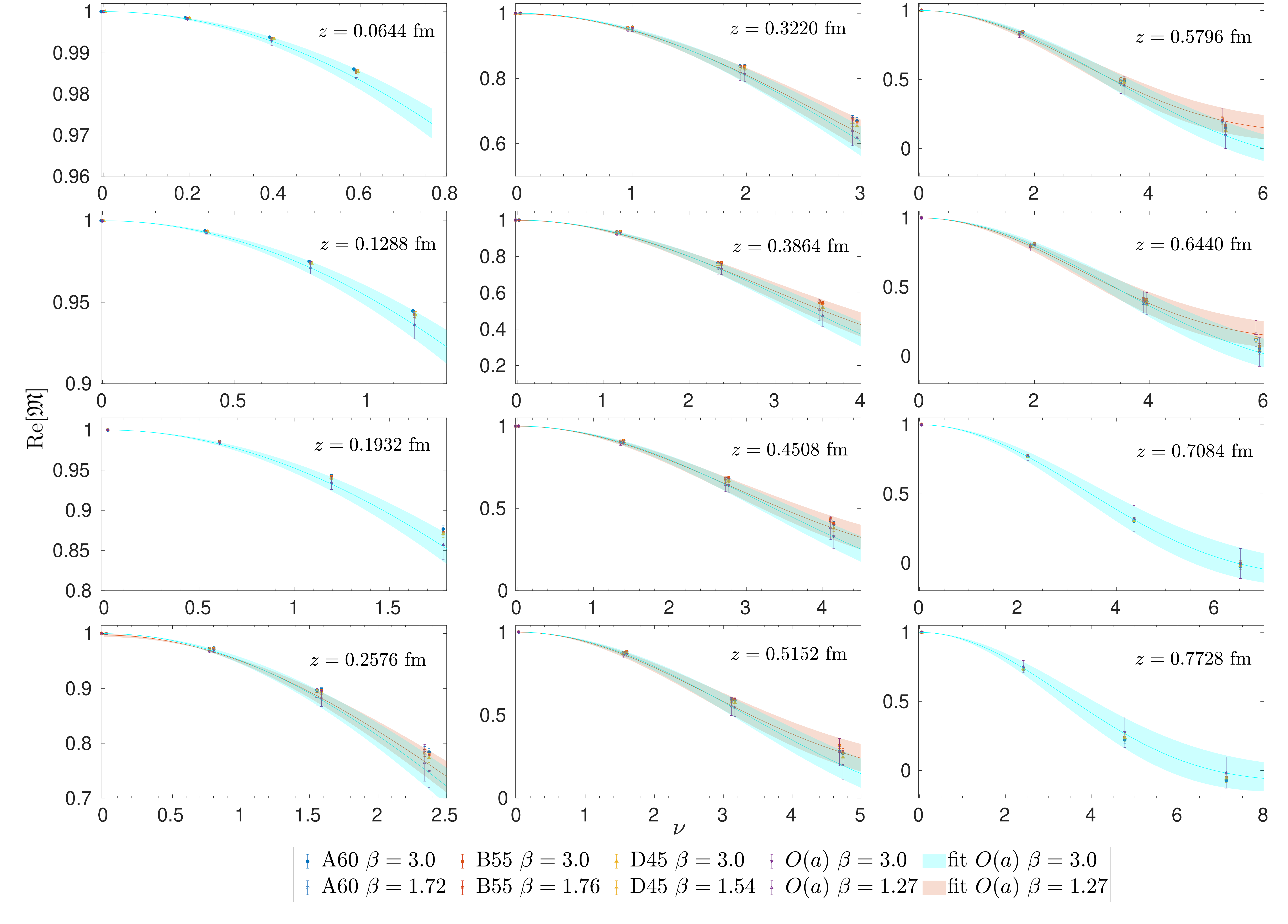}        
\end{center}
\vspace*{-0.5cm} 
\caption{Fits of $\mathcal{O}(a)$-continuum-limit-extrapolated reduced ITDs to the cosine Fourier transform of the fitting ansatz (\ref{eq:ansatz2}). Shown are also ITDs for the separate ensembles. All ITDs are interpolated to common values of $z$ (given in each panel) and $\nu=P_3z$ (see text for details) and are different for fits in the ${\bar \beta}_M^{\rm PDF fits}$ setup and with the universal ${\bar \beta}_\mathfrak{M}=3$. Small horizontal shifts are applied to the data to increase visibility, but $\nu$ is the same for each group of points.}
\label{fig:DGLAP_reduced}
\end{figure*}

\begin{figure*}[p!]
\begin{center}
    \includegraphics[width=0.49\textwidth]{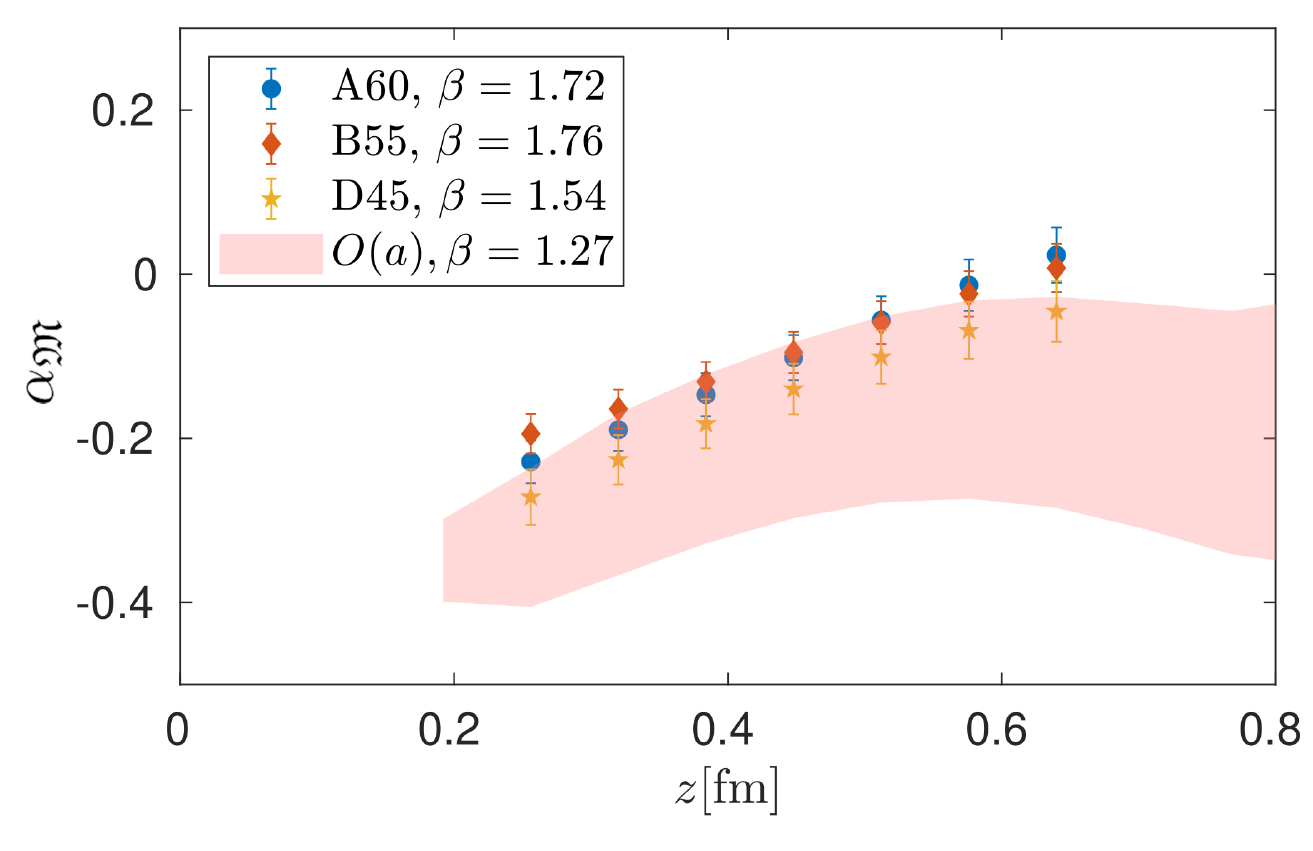}
    \includegraphics[width=0.49\textwidth]{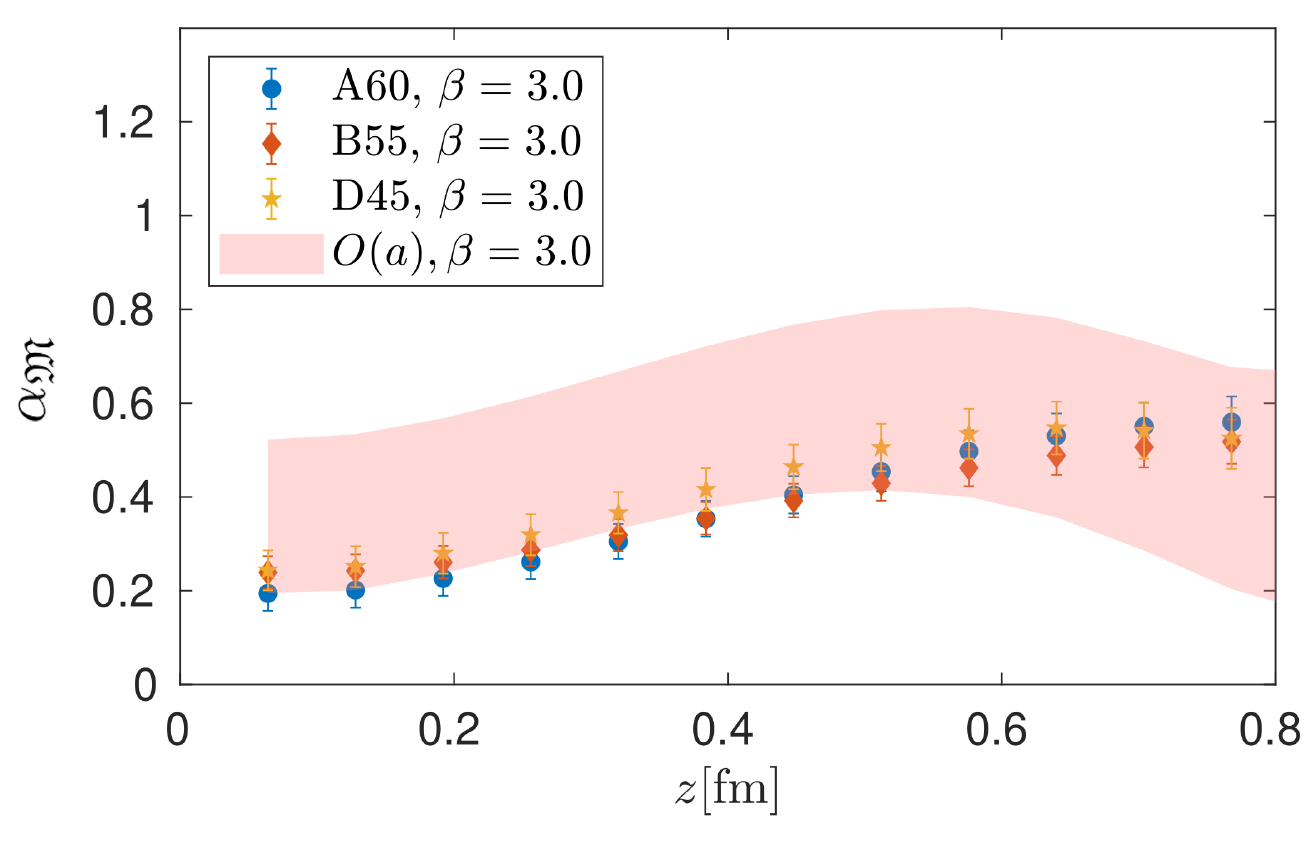}        
\end{center}
\vspace*{-0.5cm} 
\caption{Dependence of the extracted values of the fitting coefficient $\alpha_\mathfrak{M}$ on the physical distance $z$. The data points correspond to fits to the data of the separate ensembles, while the band depicts the values of $\alpha_\mathfrak{M}$ from fits to the $\mathcal{O}(a)$-continuum-limit-extrapolated data. The coefficient ${\bar \beta}_\mathfrak{M}$ is different for all ensembles and for the continuum limit in the ${\bar \beta}_M^{\rm PDF fits}$ setup (left) or fixed to 3 (right).}
\label{fig:DGLAP_reduced_a}
\end{figure*}

\begin{figure*}[p!]
\begin{center}
    \includegraphics[width=\textwidth]{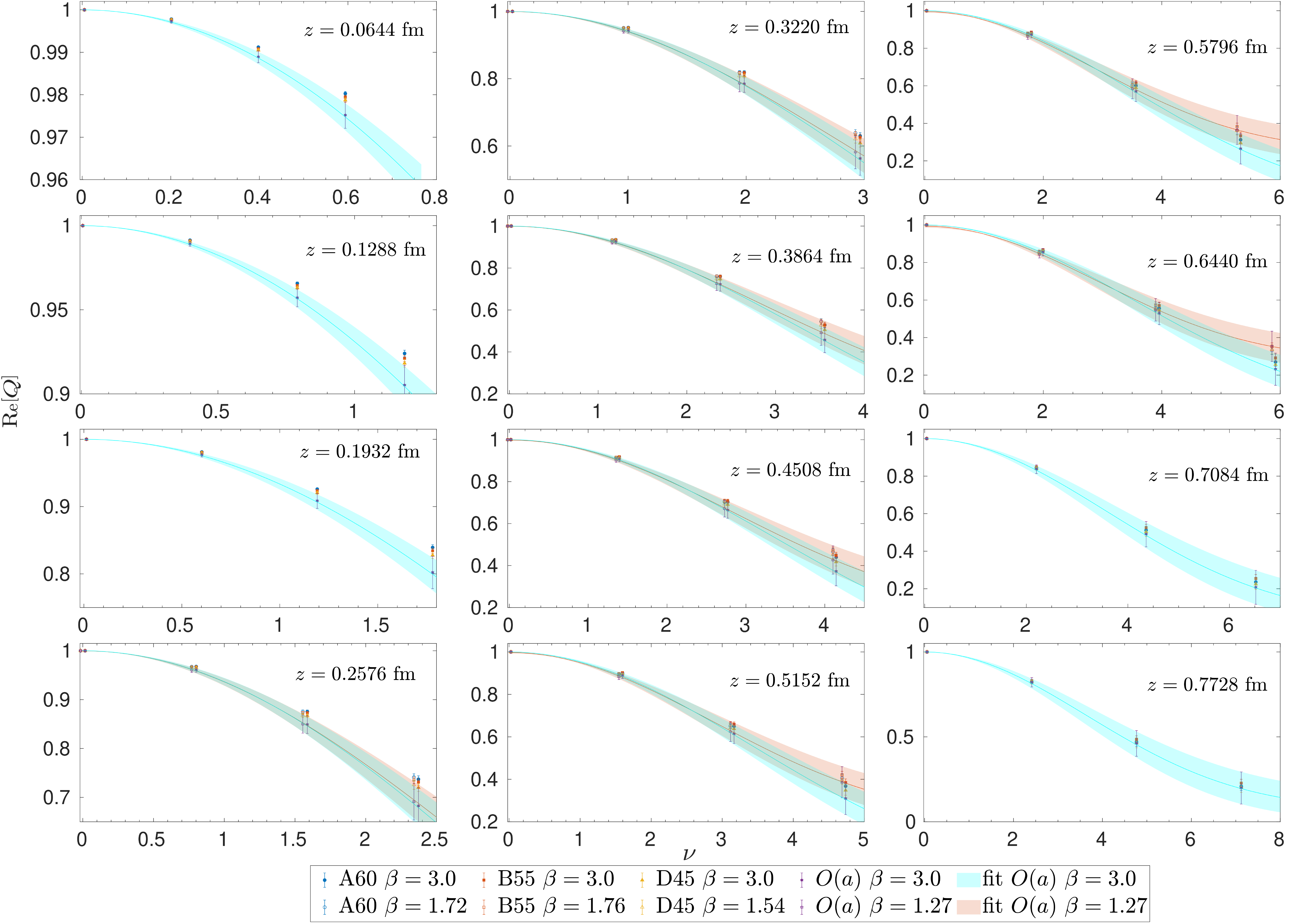}        
\end{center}
\vspace*{-0.5cm} 
\caption{Fits of $\mathcal{O}(a)$-continuum-limit-extrapolated matched ITDs to the cosine Fourier transform of the fitting ansatz (\ref{eq:ansatz2}). Shown are also ITDs for the separate ensembles. All ITDs are interpolated to common values of $z$ (given in each panel) and $\nu=P_3z$ (see text for details) and are different for fits in the ${\bar \beta}_M^{\rm PDF fits}$ setup and with the universal ${\bar \beta}_Q=3$. Small horizontal shifts are applied to the data to increase visibility, but $\nu$ is the same for each group of points.}
\label{fig:DGLAP_matched}
\end{figure*}

\begin{figure*}[p!]
\begin{center}
    \includegraphics[width=0.49\textwidth]{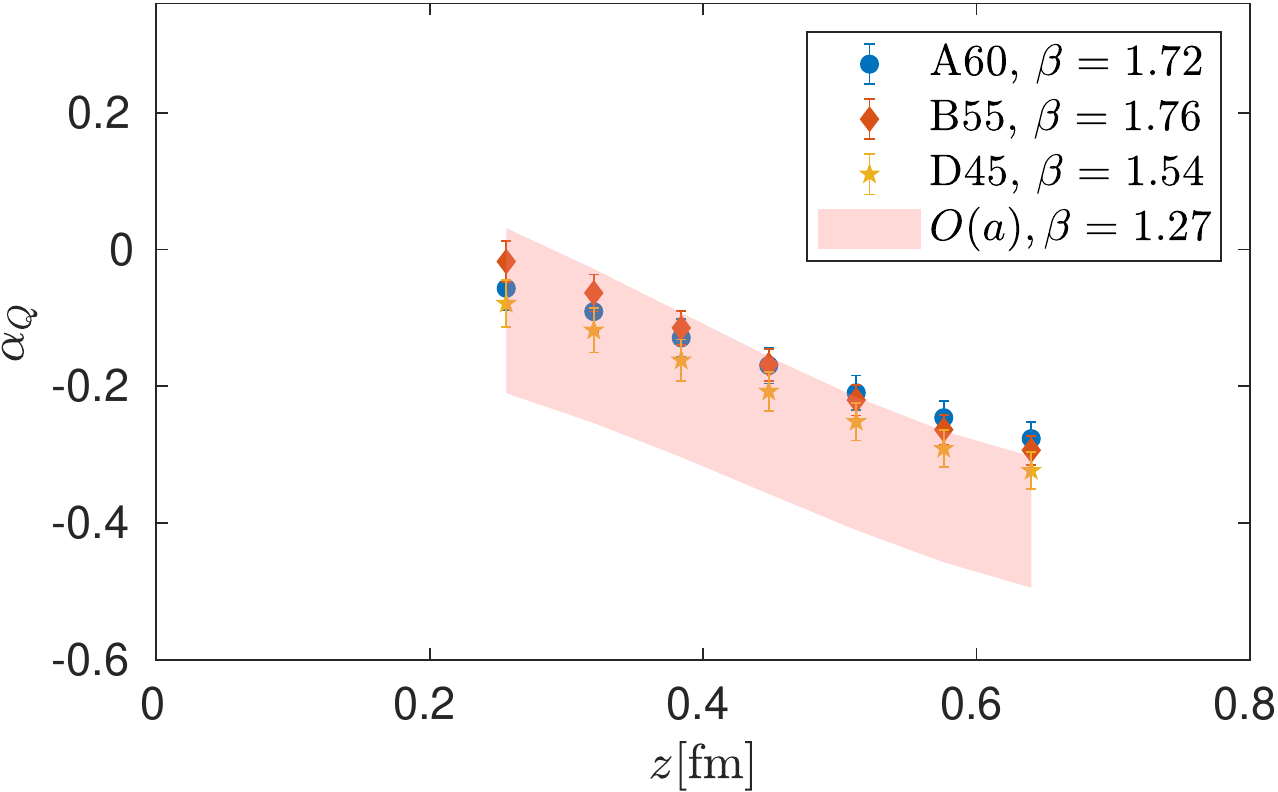}
    \includegraphics[width=0.49\textwidth]{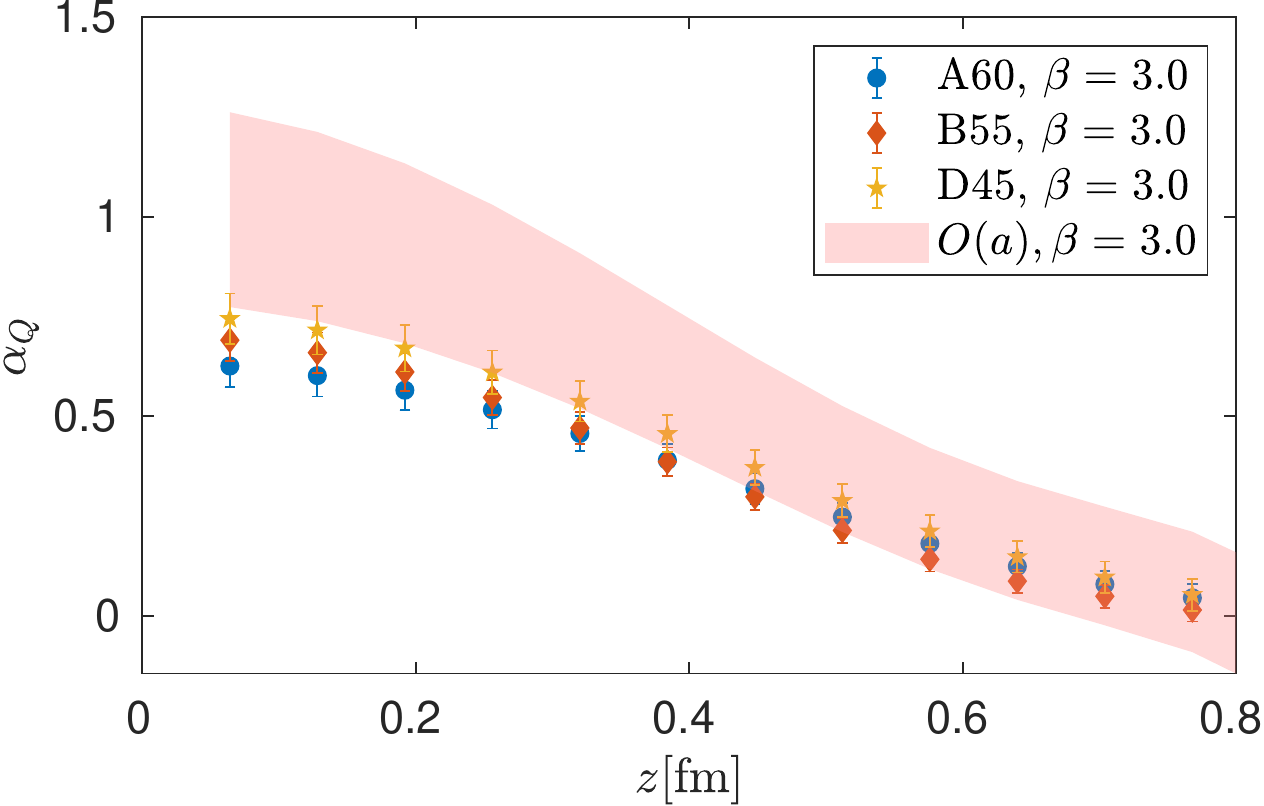}        
\end{center}
\vspace*{-0.5cm} 
\caption{Dependence of the extracted values of the fitting coefficient $\alpha_Q$ on the physical distance $z$. The data points correspond to fits to the data of the separate ensembles, while the band depicts the values of $\alpha_W$ from fits to the $\mathcal{O}(a)$-continuum-limit-extrapolated data. The coefficient ${\bar \beta}_Q$ is different for all ensembles and for the continuum limit in the ${\bar \beta}_M^{\rm PDF fits}$ setup (left) or fixed to 3 (right).}
\label{fig:DGLAP_matched_a}
\end{figure*}

\subsection{Compatibility with DGLAP evolution}
In Ref.~\cite{Egerer:2021ymv}, it was found that discretization effects lead to the violation of the DGLAP evolution of PDFs.
The authors of the aforementioned work only used a single lattice spacing, but parametrized the cutoff effects using Jacobi polynomials under the assumption that they have the form $a/|z|$ multiplied by a function of $\nu$.
Here, we can test their findings explicitly with three lattice spacings.
We recall here the methodology of this test proposed in Ref.~\cite{Egerer:2021ymv}.
It consists of fitting the real part of ITDs at fixed $z$ to a cosine Fourier transform of a phenomenologically-inspired ansatz for the valence PDF:
\begin{equation}
\label{eq:ansatz2}
q_v(x) = \frac{x^{\alpha_M} (1-x)^{{\bar \beta}_M}}{B(\alpha_M+1,{\bar \beta}_M+1)},
\end{equation}
where ${\bar \beta}_M$ is held fixed and the only fitting parameter is $\alpha_M$, where the subscript $M$ denotes the fitted ITD ($M=\mathfrak{M},\,Q$).
We consider two strategies for ${\bar \beta}_M$.
We take ${\bar \beta}_M=3$, as in Ref.~\cite{Egerer:2021ymv}, but also consider another setup with ${\bar \beta}_M$ for each ensemble taken as its fitted value in the PDF reconstruction with $\zmax=0.5$ fm, i.e. 1.72 for A60, 1.76 for B55, 1.54 for D45 and 1.27 in the continuum (see Tabs.~\ref{tab:fitparams}, \ref{tab:fitparams2}).
We refer to this setup as ${\bar \beta}_M^{\rm PDF fits}$.
The fitted values of $\alpha_\mathfrak{M}$ (fits of reduced ITDs) are expected to depend on $z$, since they are defined at different scales $1/z$.
If the DGLAP evolution is satisfied, the $z$-dependence should be considerably mildened in $\alpha_Q$ (fits of matched ITDs)

Such fits for separate ensembles are straightforward and can be performed for fixed $z/a$'s with their implied Ioffe time values.
One can also choose to fit at fixed values of $z$, e.g.\ multiplies of discrete $z$ pertinent to D45, with interpolations between neighboring $z/a$'s for A60 and B55.
However, we are also interested in such fits for continuum-extrapolated ITDs.
In this case, continuum limit extrapolations need to be performed not only at fixed $z$, but also at fixed $P_3z$.
Since the nucleon boosts of the different ensembles are slightly different, this requires an additional interpolation in Ioffe time at fixed $z$.
Thus, we use the following approach.
First, we perform fits of Eq.~(\ref{eq:ansatz2}) for all ensembles at fixed $z$ satisfying the condition $z=na_{\rm D45}$, where $n$ is integer and $a_{\rm D45}=0.0644$ fm is our finest lattice spacing.
As hinted above, this implies the need for interpolations for A60 and B55, peformed with fourth-order polynomials to the $z/a$-dependence at fixed $P_3$.
We note these fourth-order polynomials provide very good description of this dependence.
Having the ITDs at fixed $z$, fits of Eq.~(\ref{eq:ansatz2}) provide the $z$-dependence of the fitting parameter $\alpha_M$ and a parametrization of the $\nu$-dependence at fixed $z$.
The latter are used for interpolation to common $P_3z$'s required by the continuum limit extrapolation.
Peforming this interpolation, we arrive at separate-ensemble ITDs at fixed values of $z$ and $P_3z$, both being integer multiples of the values pertinent to D45.
These can be extrapolated to the continuum limit and subjected again to fits of Eq.~(\ref{eq:ansatz2}), leading to the $z$-dependence of the fitting coefficient $\alpha_M$ for continuum ITDs.

The fits of $\mathcal{O}(a)$-continuum-extrapolated reduced and matched ITDs are shown in Fig.~\ref{fig:DGLAP_reduced} and Fig.~\ref{fig:DGLAP_matched}, respectively, for both the cases of ${\bar \beta}_M^{\rm PDF fits}$ and ${\bar \beta}_M=3$.
For each value of $z$, we show ITDs of the separate ensembles, together with their continuum limit and the fitting band resulting from the ansatz of Eq.~(\ref{eq:ansatz2}).
Note that ITDs corresponding to both choices of ${\bar \beta}_M$ are different, due to the interpolations to fixed Ioffe time peformed with a different functional form.
However, in all cases, reduced and matched ITDs from interpolations with ${\bar \beta}_M^{\rm PDF fits}$ and ${\bar \beta}_M=3$ are consistent with each other.
For some distances (smaller than $0.2$ fm and greater than $0.7$ fm), description of the ITD data behavior is not possible with ${\bar \beta}_M^{\rm PDF fits}$ and then, we restrict ourselves to fits with ${\bar \beta}_M=3$.
For all the cases depicted in Figs.~\ref{fig:DGLAP_reduced}, \ref{fig:DGLAP_matched}, the fits of Eq.~(\ref{eq:ansatz2}) give good values of $\chi^2/{\rm dof}\lesssim1$.

The extracted values of the fitting coefficients $\alpha_\mathfrak{M}$ and $\alpha_Q$ are shown in Fig.~\ref{fig:DGLAP_reduced_a} and Fig.~\ref{fig:DGLAP_matched_a}, respectively.
At finite lattice spacings, we see a striking dependence of $\alpha_M$ on the distance for both choices of ${\bar \beta}_M$, implying violation of the DGLAP relation.
While reduced ITDs are defined at different scales $z$ and this is unsurprising, the violation of DGLAP can also be seen in matched ITDs, defined at a common scale of $\mu=2$ GeV.
After the continuum limit extrapolation, $\alpha_\mathfrak{M}$ becomes practically independent of the distance for reduced ITDs.
Obviously, the inflated errors of continuum ITDs imply that the actual dependence on the distance may be hidden within statistical errors.
For matched ITDs, there is still considerable deviation of $\alpha_Q$ at the smallest and the largest distances, pointing also to the fact that the observed $z$-independence of $\alpha_\mathfrak{M}$ may be accidental.
Thus, our data yield no support to the hypothesis that discretization effects are responsible for the violation of the DGLAP evolution.
However, there is also no contradiction with this hypothesis -- the visibly inflated errors upon continuum extrapolation make all values of $\alpha_Q$ for $z\lesssim0.4-0.5$ fm, depending on the scenario for $\bar{\beta}_M$, compatible with one another.
In this way, the violation of DGLAP of continuum ITDs is not seen within our errors for the distances entering the ITDs used for the reconstruction of the final PDFs.
Beyond $z\approx0.4-0.5$ fm, the observed violation may be due to enhanced HTEs of $\mathcal{O}(z^2\LambdaQCD^2)$.
These effects are bound to be present, but cannot be seen until $z\approx0.4-0.5$ fm either when comparing ITDs from different combinations of $(P_3,z)$ and the same Ioffe time or when looking at the violations of the DGLAP relation.

We remark that this study of the violation of the DGLAP relation is rather inconclusive, because of the significantly inflated errors of continuum ITDs, but also because of the rather simplified and model-dependent methodology.
More conclusive statements can only be reached with more precise data.

\begin{figure*}[t!]
\begin{center}
    \includegraphics[width=\textwidth]{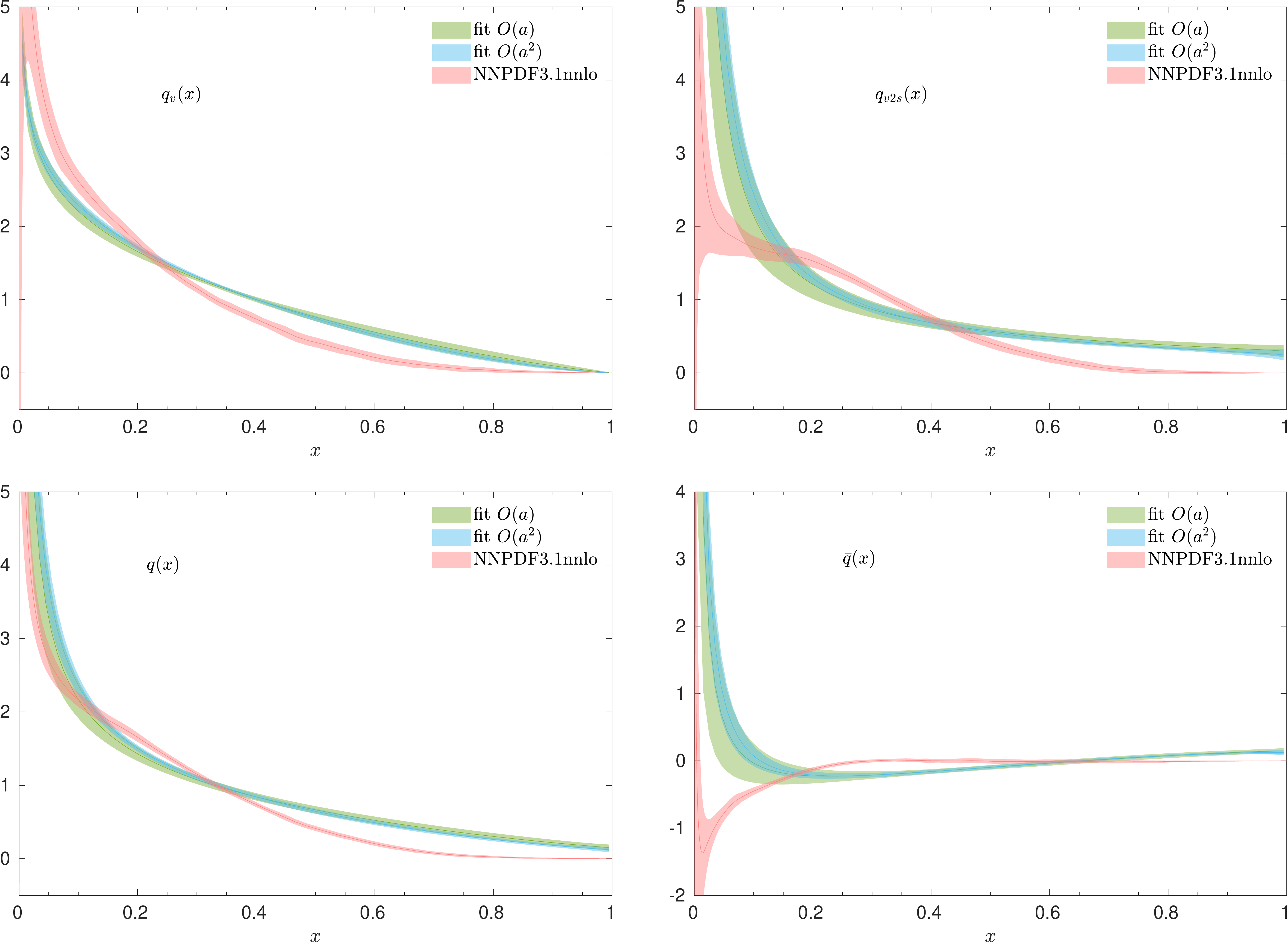}        
\end{center}
\vspace*{-0.5cm} 
\caption{Comparison of lattice-extracted PDFs with the corresponding NNPDFs (3.1, NNLO) \cite{Ball:2017nwa} for: $q_v$ (top left), $q_{v2s}=q_v+2\bar{q}$ (top right), $q=q_v+\bar{q}$ (bottom left), $q_s=\bar{q}$ (bottom right). All PDFs are reconstructed with a fitting ansatz from two-loop-matched ITDs with $\zmax=0.5$ fm, using either $\mathcal{O}(a)$ or $\mathcal{O}(a^2)$ continuum limit extrapolation.}
\label{fig:final}
\end{figure*}

\subsection{Final results}
Now, we show our final PDFs obtained in this study and we compare them to distributions coming from one of the phenomenological extractions, NNPDF3.1 at NNLO \cite{Ball:2017nwa}.
All the lattice-extracted PDFs come from ITDs calculated at three lattice spacings, extrapolated to the continuum limit at $\mathcal{O}(a)$ or $\mathcal{O}(a^2)$.
The probed range of Ioffe times extends from zero to around 4.7, with the latter value determined by the maximal length of the Wilson line, $\zmax=0.5$ fm, that can be justified to be small enough for the short-distance factorization to hold.
We emphasize again that $\zmax$ needs to be chosen in conjunction with the available precision of data.
The criterion that we advocate for is the agreement between ITDs extracted from different combinations of $(P_3,z)$ at fixed $P_3z$, which leads rather unambiguously to the conclusion that $\zmax=0.5$ fm is a safe choice.
For the real part of matched ITDs, no deviations between fixed-$\nu$ ITDs at different $(P_3,z)$ is seen even with the precise separate-ensemble data up to $z>0.5$ fm.
The situation is different for the imaginary part -- we observe inconsistencies between $(P_3,z)$ combinations already around $z=0.3$ fm, but they are small enough until $z=0.5$ fm to be hidden within statistical errors after the continuum limit extrapolation.
Thus, while $\zmax=0.5$ fm is clearly too large for the imaginary part at the 2-3\% precision level (of separate ensembles), it is beyond statistical precision in the continuum (with $\mathcal{O}(10\%)$ errors).
In turn, for the real part, HTEs are apparently smaller and invisible even with 2-3\% statistical uncertainties.
In general, $\zmax=0.5$ fm may seem rather large from the point of validity of perturbation theory ($1/\zmax\approx0.4$ GeV), but the violation of factorization is lessened to some degreee when taking the ratios of matrix elements by the partial cancellation of HTEs between the numerator and the denominator. 
The remainder of these effects is hidden in our statistical errors as long as one does not include ITDs originating from Wilson line lengths above $\zmax$, if $\zmax$ is adjusted according to the achieved level of precision.

The reduced ITDs were subjected to a two-loop matching procedure and we have shown that the two-loop correction to the standard one-loop matching used in earlier works is a small effect, thus establishing good convergence of perturbation theory in the matching.
Finally, the matched ITDs are used in a fitting reconstruction of PDFs, by employing a phenomenologically-inspired fitting ansatz.
We have argued that the implied model dependence is reflected in the errors of the PDFs, particularly at small $x$.

In the top left panel of Fig.~\ref{fig:final}, the valence PDF is compared to NNPDF.
We note that the statistical precision of the lattice result is similar to the precision of the valence NNPDF, even after the continuum limit extrapolation that inflates the errors.
However, the errors of the lattice-extracted PDF are only statistical, with some sources of systematic uncertainties unquantified.
In this work, one of the most obvious systematics of lattice computations has been evaluated, by extrapolating the data to the continuum limit.
However, the lattice ensembles that we have used involve a non-physical pion mass of around 370 MeV, which is bound to play an important role.

This role can be understood by inspecting the PDF $q_{v2s}$ (top right panel of Fig.~\ref{fig:final}), whose first moment is $\langle x\rangle_{u-d}=\int_0^1 dx\, x(q_v(x)+2\bar{q}(x))$.
Already several years ago, it was discovered that $\langle x\rangle_{u-d}$ is significantly above its phenomenological value when computed with non-physically-heavy quarks, see e.g.\ Ref.~\cite{Constantinou:2014tga}.
In particular, at $m_\pi\approx370$ MeV, $\langle x\rangle_{u-d}$ is around 40-70\% too large, for example $\langle x\rangle_{u-d}=0.270(3)$ for our ensemble B55, computed from local operators \cite{Abdel-Rehim:2015owa} at the same source-sink separation.
We can compare the latter with our values of $\langle x\rangle_{u-d}$ from the integration of the fitting-reconstructed $q_{v2s}$: $0.264(6)$ (A60), $0.254(5)$ (B55), $0.266(7)$ (D45), $0.269(25)$ ($\mathcal{O}(a)$ continuum limit, i.e.\ the $q_{v2s}$ of Fig.~\ref{fig:final}).
The larger value of $\langle x\rangle_{u-d}$ manifests itself as an enhanced value of the PDF at $x\gtrsim0.5$.
This pion-mass-related behavior is observed also in the valence distribution, which is the dominating input of $\langle x\rangle_{u-d}=\int_0^1 dx\, x(q_v(x)+2\bar{q}(x))$, the valence part giving in the continuum limit $0.263(20)$ and the sea part amounting to $2\cdot0.003(11)$.
The enhanced value of the valence PDF at intermediate and large $x$ implies, obviously, its suppressed value at small $x$.
Similar conclusions hold for $q=q_v+\bar{q}$ (bottom left panel of Fig.~\ref{fig:final}), with the most striking discrepancy with respect to NNPDF occuring in the large-$x$ regime.
This discussion allows us to speculate that the non-physical pion mass of the present study is the main systematic uncertainty responsible for the difference between our PDFs and ones from global fits.

We also emphasize again the difficulty related to distributions involving the antiquarks.
They receive contributions additionally from the imaginary part of matched ITDs ($q=q_v+\bar{q}$ and the antiquark distribution itself) or solely from it ($q_{v2s}$), which are further away from zero at our maximal Ioffe time corresponding to $\zmax=0.5$ fm.
This implies that a larger range of Ioffe times is missing in the reconstruction.
Together with the lack of the normalization condition, this translates to larger variability of the fits and significantly larger errors of the PDF.
Moreover, the data at the available Ioffe times are not enough to exclude a vanishing value of the coefficient $\beta$ of the $q_{v2s}$ fitting ansatz that governs the large-$x$ behavior of the PDF, leading to its non-vanishing value at $x=1$.
Thus, this qualitative feature signals non-robust reconstruction of this distribution and the need for probing a larger range of Ioffe times, i.e.\ ITDs obtained at larger nucleon boosts.
The antiquark distribution $\bar{q}$ (bottom right panel of Fig.~\ref{fig:final}), similarly to $q$ originating from both $q_v$ and $q_{v2s}$ and hence related to both the real and imaginary part of ITDs, is strongly suppressed and does not allow for meaningful conclusions.
At large-$x$, it is affected by the non-vanishing value of $q_{v2s}$ at $x=1$ and at small $x$, by very large errors.
Given that it probes the difference between the behavior contained in the real and imaginary parts of ITDs, its extraction seems to be the most difficult. Clearly,  a prerequisite for its robust determination is to extend the range of available Ioffe times, i.e.\ to increase the accessed nucleon boost.

\section{Summary and prospects}
\label{sec:summary}
In this paper, we tested discretization effects in partonic distributions extracted using the pseudo-distribution approach on the lattice. 
This is one of the most important systematic effects in lattice calculations in general and its quantification is necessary to obtain final meaningful results.
We concentrated on the unpolarized isovector PDF of the nucleon and we calculated the relevant matrix elements using ensembles of gauge field configurations at three lattice spacings, ranging from 0.093 fm to 0.064 fm, at a non-physical pion mass of about 370 MeV.
The bare matrix elements were produced with three or four nucleon boosts up to around 1.8 GeV and the divergences that they contain were renormalized by forming appropriate ratios.
Such ratios are functions of two Lorentz invariants, the Wilson line length ($z$) and its product with the nucleon boost, the so-called Ioffe time ($\nu\equiv P_3z$).
Thus, they are called (reduced or pseudo-) Ioffe time distributions or ITDs.
Pseudo-ITDs are Euclidean objects that describe spatial correlations in a boosted nucleon and their crucial property is that they can be perturbatively factorized into the relevant physical ITDs defined on the light front.
Until recently, this factorization was available only at one loop, but recently, the two-loop correction was calculated \cite{Li:2020xml}.
In our work, we implemented this correction for the first time in the pseudo-distribution approach to address another of the most important systematic effects, in this case unrelated to the lattice computation, namely the truncation effects in the perturbative factorization.
The matched ITDs are still coordinate-space objects and further systematics hides in their translation to momentum space of Bjorken-$x$ fractions, which we tested employing three methods of the reconstruction of the $x$-dependence.
Other systematic effects that we addressed included the influence of stout smearing of the operator insertion and the dependence of the results on the maximal length of the inserted Wilson line.

We begin our final discussion with the latter.
The choice of this maximal length, $\zmax$, is delicate.
On the one hand, at fixed maximal nucleon boost, it is profitable to have large $\zmax$ enabling access to a wide range of Ioffe times, ideally such that the ITDs have decayed to zero.
On the other hand, the pseudo-distribution approach is based on a short-distance factorization, necessitating values of $z$ in the perturbative regime.
Too large values of $\zmax$ imply the presence of uncontrolled higher-twist effects of $\mathcal{O}(z^2\LambdaQCD^2)$.
Yet the optimal value of $\zmax$ is \emph{a priori} not clear.
With the requirement of perturbation theory being applicable at the scale $1/\zmax$ taken literaly, one would be limited to values of $\mathcal{O}(0.2)$ fm.
However, this does not take into account two important aspects.
The first one is that the definition of reduced ITDs involves a ratio of matrix elements with the same $z$, in which higher-twist effects can partially cancel.
The second, in turn, is a pragmatic one -- the finite precision of lattice results implies that small effects may be hidden in statistical errors.
This fact allows us to define a practical criterion for establishing $\zmax$ by inspecting ITDs from different combinations of $(P_3,z)$ and the same product $P_3z$.
If they agree, the ITD corresponding to the largest $z$ is still statistically consistent with a ``perturbatively-safe'' ITD, i.e.\ one with $z$ that is unambiguously in the perturbative regime.
With our precision of the data, $\zmax$ is, thus, determined to be 0.5 fm for the continuum-extrapolated distributions.
We note that our previous work \cite{Bhat:2020ktg}, at one lattice spacing, but with physically-light quarks, established $\zmax=0.8$ fm to be safe in the practical sense.
In the current work, our statistical errors are significantly smaller at the non-physical pion mass, and, hence, our sensitivity to higher-twist effects is larger.
Thus, we argue that the optimal safe value of $\zmax$ is not universal and has to be established for each considered set of lattice data, with the expectation that more precise data imply the necessity of a decreased $\zmax$.

We now turn to discretization effects, the main motivation of this work.
Our main finding is that these effects are relatively small in our setup, being on the verge of statistical significance for most Ioffe times.
At the level of PDFs, they lead to tendencies that the distributions are enhanced at small $x$ and suppressed at large $x$.
However, the inflation of errors in the continuum limit extrapolations makes the final continuum distributions always consistent with the ones at the finest lattice spacing.
We tested extrapolations linear both in the lattice spacing and its square, and given the flatness of these fits, they lead to compatible results.
However, the increase of final errors with respect to those of the separate ensembles is much larger with the $\mathcal{O}(a)$ ansatz.
This increase amounts to a factor of order 3-5 for most Ioffe times, while it is around half as large for the $\mathcal{O}(a^2)$ extrapolation.
It brings about an important, although rather obvious conclusion that it is an important direction for the future to better understand the $\mathcal{O}(a)$ discretization effects and to implement a full $\mathcal{O}(a)$-improvement program in the lattice calculations of PDFs, if eventually these are to lead to precise extractions.
We note that the size of cutoff effects depends on the employed discretization -- thus, the conclusion about the comparatively small effects in our study may not be universal.
Nevertheless, for our setup, it is reassuring and excludes large systematics due to discretization effects.

Concerning truncation effects, we found that the two-loop effects are negligible at this level of precision and with our value of $\zmax$.
More precisely, separating the factorization effect into evolution from scales $1/z$ at which the matrix elements are defined and matching from the Euclidean to the physical observables, the two-loop effects in the former are quite significant already at distances of order 0.4 fm.
However, the matching part acts in the opposite direction on the ITDs, making the two-loop correction larger than our precision only at $z>\zmax$, i.e.\ it affects only ITDs that are not included in the reconstruction of PDFs.
The overall effect of evolution and matching is small and statistically insignificant in the continuum-extrapolated results.

Finally, we comment on the reconstruction of the $x$-dependence from matched ITDs.
Generically, this procedure is ill-defined, as one is trying to determine a continuous distribution from a discrete set of data truncated at some finite Ioffe time.
From this point of view, it is clear that some assumptions are needed in this step.
As hinted above, we used three reconstruction methods.
Two of these, the naive Fourier transform and the Backus-Gilbert method, implicitly assume that ITDs are zero beyond $\zmax$.
Such a sharp cutoff leads to unreliable results, manifested in suppressed small-$x$ behavior and oscillations at large $x$.
Thus, the method of choice is reconstruction with a fitting ansatz.
Although it is model-dependent, we argue that this is only seemingly a restriction at the current stage.
Given the substantial statistical error in the continuum limit, our expectation is that the model dependence is a subdominant source of uncertainty.
With more precise data and probing the full range of Ioffe times (until ITDs decay to zero), more realistic fitting ansatzes can be used and/or one can reliably use the model-independent Backus-Gilbert approach.

Having addressed the different systematic effects and having established the relative smallness of discretization effects and truncation effects in the matching, we compared our final reconstructed PDFs to ones from global fits.
There is clear qualitative agreement with the latter for all types of considered distributions, but still wide regions of $x$ values showing quantitative tension.
In this work, we eliminated some of the systematics earlier considered as likely culprits for the disagreement, particularly discretization effects and truncation effects in the matching.
Likewise, there is convincing evidence that higher-twist effects are smaller than our statistical precision.
In this way, it is most probable that the disagreement with global fits is to the largest extent induced by the non-physical pion mass of our simulations.
This is further justified by the consistency of the observed enhancement of the distributions at large $x$ values over phenomenology with lattice-calculated average momentum fraction $\langle x\rangle$ at the non-physical pion mass.

Our work leads to rather unambiguous conclusions for directions of further work.
First, such a continuum-limit study with two-loop matching should be repeated at the physical pion mass.
This would allow us to test the conjecture that the too-heavy quarks are responsible for the observed differences with respect to global fits.
Up to now, one pseudo-distribution study of unpolarized PDFs exists directly at the physical point, done by our group in 2020 \cite{Bhat:2020ktg} at a single lattice spacing $a\approx0.094$ fm.
Another calculation close to the physical pion mass and extrapolating to it (from 172, 278, 358 MeV) was published by the HadStruc Collaboration \cite{Joo:2020spy}.
In Ref.~\cite{Bhat:2020ktg}, we observed consistency with NNPDF in a wide range of Bjorken-$x$ values ($x\lesssim0.3$, $x\gtrsim0.6$) already within statistical errors.
Moreover, we admitted plausible values of unquantified systematic uncertainties, including ones from discretization effects, which allowed us to extend the consistency with phenomenology to the whole range of $x$.
The magnitude of these estimates of cutoff effects is in agreement with the present study.
Nevertheless, their explicit check at the physical pion mass is mandatory at some point.
The key challenge is to perform such a calculation with sufficient precision to reach meaningful conclusions about the size of discretization effects.
The $\mathcal{O}(a)$ continuum extrapolation of the present work inflates the errors 3-5 times with respect to the one of the separate ensembles, as mentioned above, which implies that the precision of Ref.~\cite{Bhat:2020ktg} would translate to 30-50\% errors of the physical-continuum PDFs.
Thus, reaching meaningful precision for the study of cutoff effects would necessitate simulations with much finer lattice spacings or, preferably, implementing an $\mathcal{O}(a)$-improvement program \cite{Chen:2017mie,Green:2020xco}.
Some indication of the role of $\mathcal{O}(a)$ effects is provided by the observed mild tension between 0-stout and 5-stout continuum limits upon $\mathcal{O}(a^2)$ extrapolation.

A second important direction is to be able to probe the full range of Ioffe times, i.e.\ such that matched ITDs have decayed to zero.
Currently, accessing $\nu\lesssim5$, the missing information is provided by the fitting ansatz and the large-$\nu$ ITDs values are guided by the accessible-$\nu$ behavior and the form of the ansatz, thus introducing some model dependence into the final results.
While this is reflected in the enhanced errors, particularly at small $x$, it is clear that it is desirable that this is avoided.
Probing the full range of Ioffe times would also allow for reliable usage of the Backus-Gilbert method that currently suffers from an even more severe model dependence, implied by the strong assumption that ITDs are zero beyond $\zmax$.
Moreover, the insufficient maximum Ioffe time is reflected in pathological behavior of the distribution $q+\bar{q}$, with fits for many bootstrap samples insensitive to the $1-x$ part of the fitting ansatz and implying a non-physical, non-vanishing value of this distribution at $x=1$.
However, increasing the range of Ioffe times is difficult when taking the need for a moderately small $\zmax$ into account, since it implies the necessity of simulating larger hadron boosts.
The latter leads to an exponential increase of the computational cost, caused by the decaying signal-to-noise ratio and an increased excited-states contamination, implying the need for larger source-sink separations.
The problem is severely aggravated when trying to combine both postulates, of physical-point continuum limit calculations probing the full range of Ioffe times by accessing large nucleon momenta.
In fact, the cost of this seems prohibitive at present and most likely, such computations are realistic only upon methodological improvements allowing for more favorable signal for large boosts.
Ideally, this should be combined with above discussed $\mathcal{O}(a)$-improvement of the relevant matrix elements, giving, in practice, a factor of around 2 reduction of errors of continuum distributions.

Nevertheless, the prospects of the pseudo-distribution approach and other related methods are very good. For many physical distributions, we have entered an era of laboriously quantifying several sources of systematic uncertainties.
This is a prerequisite to eventually obtaining robust results with fully reliable uncertainties.
In this work, we have shown that discretization effects and truncation effects in the matching can be fully under control, which is an important step in this quest.

\begin{acknowledgments}
  M.B., W.C.\ and K.C.\ acknowledge support by the National Science Centre
  (Poland) grant SONATA BIS no.\ 2016/22/E/ST2/00013. 
  M.C. acknowledges financial support by the U.S. Department of Energy, 
  Office of Nuclear Physics, Early Career Award under Grant No.\ DE-SC0020405. 
  J.R.G. acknowledges support from the Simons Foundation through the Simons Bridge for Postdoctoral Fellowships scheme.
  The calculations were performed at the Poznań Supercomputing and Networking Center (Eagle supercomputer) and at the J\"ulich Supercomputing Centre (JURECA \cite{jureca}), using the Grid library~\cite{Boyle:2016lbp} and the DD-$\alpha$AMG solver~\cite{Frommer:2013fsa} with twisted mass   support~\cite{Alexandrou:2016izb}. 
\end{acknowledgments}

\appendix
\section{Backus-Gilbert method}
The criterion the BG method adds to the discrete and truncated set of lattice data is to minimize the variance of the solution to the inverse problem.
This consists in maximizing the stability of the solution with respect to statistical variation of the data.
Such a condition is a model-independent assumption which chooses a unique distribution from the infinite number of solutions from a given set of input lattice ITDs.

The reconstruction procedure is performed separately for each values of $x$, which we take with a step of 0.01.
The mathematical criterion of the BG method leads to a $d$-dimensional vector $\textbf{a}_K(x)$, $d$ being the number of available discrete Ioffe times from the lattice calculation.
This vector is an approximate inverse of the Fourier kernel function $\textbf{K}(x)$, i.e.\ the cosine or the sine function for the distributions $q_v$ and $q_{v2s}$, respectively.
Thus,
\begin{equation}
\Delta(x,x')=\sum_{\nu}a_K(x)_\nu K(x')_\nu,    
\end{equation}
where $\textbf{K}(x')$ is a $d$-dimensional vector with elements $K(x')_\nu=\cos(\nu x')$ or $K(x')_\nu=\sin(\nu x')$.
The function $\Delta(x,x')$ approaches the Dirac delta function as the number of input ITDs increases.
When $d$ is finite, the $\Delta(x,x')$ approximation to $\delta(x-x')$ is the one with minimized width.
The width minimization conditions are given e.g.\ in Ref.~\cite{Karpie:2018zaz} and yield
\begin{equation}
\textbf{a}_K(x)=\frac{\textbf{M}_K^{-1}(x)\,\textbf{u}_K}{\textbf{u}_K^T\,\textbf{M}_K^{-1}(x)\,\textbf{u}_K},    
\end{equation}
with the $d\times d$-dimensional matrix elements $\textbf{M}_K(x)$ given by
\begin{equation}
M_K(x)_{\nu\nu'}=\int_0^1 dx'\,(x-x')^2 K(x')_\nu \, K(x')_{\nu'}+\rho\,\delta_{\nu\nu'}
\end{equation}
and elements of the $d$-dimensional vector $\textbf{u}_K$ are
\begin{equation}
u_{K\nu}=\int_0^1 dx'\,K(x')_\nu.    
\end{equation}
The matrix $\textbf{M}_K(x)$ can have eigenvalues arbitrarily close to zero, which need to be regularized.
We employ the Tikhonov regularization \cite{Tikhonov:1963}, which introduces a free parameter $\rho$, see also Refs.~\cite{Ulybyshev:2017szp,Ulybyshev:2017ped,Karpie:2018zaz}), which makes the matrix invertible by moving its lowest eigenvalues away from zero.
The choice of $\rho$ should be made such that the results are not biased and the resolution of the method is relatively unaffected.
We find that $\rho=10^{-3}$ is appropriate, with little effects when changing it by an order of magnitude.
In turn, much smaller values of $\rho$ introduce large oscillations in the reconstructed distributions due to the presence of very small eigenvalues of $\textbf{M}_K(x)$, while much larger values distort visibly the shapes of the final PDFs.
The latter are given by
\begin{equation}
q_{v/v2s}(x,\mu)=\sum_{\nu} a_K(x)_\nu \, {\rm Re}/{\rm Im}\, Q(\nu,\mu). 
\end{equation}

\bibliography{references.bib}

\end{document}